\documentclass[%
 reprint,
nofootinbib,
 amsmath,amssymb,
 aps,
floatfix,
]{revtex4-2}

\usepackage{orcidlink}
\usepackage{appendix}  
\usepackage{xcolor}
\usepackage{graphicx}
\usepackage{dcolumn}
\usepackage{bm}
\usepackage{hyperref}
\usepackage[mathlines]{lineno}

\usepackage{float}

\begin{document}

\preprint{APS/123-QED}

\title{Anisotropy signal of UHECRs from a structured magnetized Universe}

\author{Simone Rossoni\,\orcidlink{0009-0007-4477-8817}}
\email{simone.rossoni@desy.de}
\affiliation{
 II. Institut f\"{u}r Theoretische Physik, Universit\"{a}t Hamburg,\\ Luruper Chaussee 149, 22761 Hamburg. }

\author{G\"{u}nter Sigl\,\orcidlink{0000-0002-4396-645X}}
\email{guenter.sigl@desy.de}
\affiliation{
 II. Institut f\"{u}r Theoretische Physik, Universit\"{a}t Hamburg,\\ Luruper Chaussee 149, 22761 Hamburg. }

\date{\today}

\begin{abstract}
\noindent The surprising isotropy of the ultra-high-energy cosmic ray (UHECR) sky makes it difficult to identify their sources. Observables such as energy spectrum, mass composition and arrival directions are affected by interactions with background photon fields and by deflection in the extragalactic and galactic magnetic fields (EGMF and GMF). In this work, we simulate the propagation of UHECRs with energy above $8\,\text{EeV}$ in magnetized replicas of the local Universe, obtained from constrained simulations of the Large Scale Structure. We obtain the real magnetic deflection in structured EGMF models with realistic three-dimensional simulations. We investigate different scenarios for the UHECR source distributions and densities. The effect of the GMF can be different depending on the field model considered. In this work we consider the JF12 model by mapping the arrival directions at the edge of the galaxy to those at Earth. We study the arrival direction distribution of the propagated UHECRs, and in particular their angular power spectrum, dipole and quadrupole moments. We find that the properties of the source distribution affect the cosmic ray anisotropy more than the EGMF model considered. In particular, the low multipole components depend on both the source distribution and the density. We also find that it is difficult to simultaneously reproduce the observed dipole and quadrupole values above $8\,\text{EeV}$. In general, we predict too large a quadrupole strength, incompatible with observations.
\end{abstract}

\keywords{ultra-high-energy cosmic rays, extragalactic magnetic field, extragalactic propagation, anisotropies, dipole, quadrupole}

\maketitle

\section{Introduction}
\label{sec_intro}
In recent decades, the detection of previously unobserved, extremely energetic particles has opened up the possibility of studying acceleration mechanisms that cannot be achieved in modern accelerators. These particles, now called ultra-high-energy cosmic rays (UHECRs), consist of ionized nuclei with energy $\gtrsim 10^{18}\,\text{eV}$ ($\equiv1\,\text{EeV}$) and are detected by the most modern observatories, such as the Pierre Auger Observatory \cite{PierreAuger:2015eyc} in the southern hemisphere (Argentina) and the Telescope Array \cite{TelescopeArray:2008toq} in the northern hemisphere (USA). Measuring the arrival directions of UHECRs in the Earth's atmosphere provides an opportunity to study the possible correlation in the sky with powerful astrophysical objects. Despite the enormous experimental efforts and the large number of UHECRs observed, the sources and the acceleration mechanism of these particles remain an unsolved problem of modern astrophysics.
\par In the last 10 years we have made considerable progress in our understanding of the distribution of UHECR arrival directions. A large-scale dipole anisotropy in the UHECR sky above $8\,\text{EeV}$ was discovered by the Pierre Auger collaboration \cite{PierreAuger:2017pzq,PierreAuger:2018zqu}. Recent analyses of the UHECRs detected during Phase 1 of the Pierre Auger Observatory's operation show a dipole with a strength of $\sim 7\%$ for UHECR energy above $8\,\text{EeV}$ with a significance of $6.9\sigma$ \cite{PierreAuger:2023fcr,PierreAuger:2024fgl}, pointing more than $\sim100^\circ$ away from the galactic center of the Milky Way, suggesting an extragalactic origin of the UHECR flux at this energy. Furthermore, the significance of the dipole amplitude for the energy range $8-16\,\text{EeV}$ reached $5.7\sigma$. The search for large scale anisotropy by the Telescope Array collaboration also led to the observation of a dipole component. However, due to the limited statistics, this result is compatible with both Pierre Auger results and an isotropic distribution \cite{TelescopeArray:2020cbq}. In contrast, the observed quadrupole and higher multipole moments are not statistically significant and consistent with isotropy in the combined Pierre Auger and Telescope Array analysis \cite{PierreAuger:2023mvf}. Correlation analyses were performed with several powerful astrophysical sources, such as Active Galactic Nuclei (AGN) and Starburst Galaxies (SBG), to identify the most plausible candidate source class of the UHECRs \cite{PierreAuger:2018qvk,PierreAuger:2022axr}. In particular, in \cite{PierreAuger:2022axr} the SBG catalogue shows the most statistically significant deviation from isotropy above $40\,\text{EeV}$, with a deviation of $\sim4\sigma$. 
\par In the northern hemisphere, Telescope Array has also reported the presence of hotspots that deviate from the isotropy \cite{TelescopeArray:2014tsd,Kim:2021mcf,TelescopeArray:2021dfb,TelescopeArray:2023bdy,Kim:2023ksw}. In equatorial coordinates $\sim(144^\circ,41^\circ)$ an excess of events with energy $\gtrsim60\,\text{EeV}$ with $4.8\sigma$ Li-Ma significance \cite{Li:1983fv} was observed at $\sim20^\circ$ away from the supergalactic plane.  The presence of a new hotspot of events with energies between $\sim30-60\,\text{EeV}$ has also recently been observed in the direction of the Perseus-Pisces supercluster. The origin of the observed UHECR hotspots has been studied from a theoretical point of view in \cite{Taylor:2023qdy}, where the most massive galaxies in the Local Group (referred to as \textit{Council of Giants}) have been considered. In particular, an early UHECR pulse from Centaurus A, scattered by the other galaxies, could give rise to multiple UHECR excess signals in the sky, while late echoes propagate out into the local extragalactic space. Recently, a hotspot associated with the Sombrero galaxy was claimed to have a global significance of $3.3\sigma$ \cite{He:2024vnm}.
\par The first difficulty in identifying UHECR sources is that the information obtained from particles observed on Earth must be processed to account for extragalactic energy losses and mass compositional changes. At the EeV energy scale, cosmic rays can trigger photohadronic interactions with cosmic photon fields, such as the Cosmic Microwave Background (CMB) and the Extragalactic Background Light (EBL), through the so-called GZK effect \cite{Greisen:1966jv,Zatsepin:1966jv}. Protons interact with cosmic background photons through Bethe-Heitler pair production \cite{Bethe:1934za} ($p+\gamma \rightarrow p + e^+ + e^-$) and photopion production ($p+\gamma\rightarrow p+\pi^0$ or $p+\gamma\rightarrow n+\pi^+$), and heavy nuclei also through photodisintegration. Considering the UHECR energy between $10-100\,\text{EeV}$, the typical energy loss length for protons and iron nuclei is of the order of a few times $100\,\text{Mpc}$, while for intermediate nuclei such as nitrogen it is even less than $10\,\text{Mpc}$ \cite{Allard:2011aa}. The decrease of the UHECR horizon and the increase of the dipole with increasing energy observed by the Pierre Auger Observatory \cite{PierreAuger:2021dqp} provide further evidence for the extragalactic origin of CR anisotropies above $8\,\text{EeV}$.
\par The second major obstacle in the search for UHECR sources is the presence of galactic and extragalactic magnetic fields (GMF and EGMF). The spatial structure and strength of the EGMF is characterized by various structures such as clusters, filaments and cosmic voids. However, the EGMF is still poorly understood and could be one of the main sources of uncertainty in UHECR propagation studies (see \cite{AlvesBatista:2019tlv} for a general review). Two main magnetogenesis scenarios can be identified. In the first scenario, called \textit{primordial}, the present EGMF is the result of the evolution of a seed field produced in the early Universe. The second scenario, called \textit{astrophysical}, considers the case where the magnetic field is generated by the feedback of astrophysical sources in the extragalactic space. Therefore, UHECR anisotropies can also be used to constrain EGMF parameters, as done in \cite{Lee:1995tm,Bray:2018ipq}.
\par The uncertainties present in the various models of the EGMF have a direct impact on UHECR simulations. In \cite{Sigl:2003ay,Sigl:2004yk,Armengaud:2004yt} unconstrained simulations of the local cosmic web have been used to study the anisotropy signal of UHECRs above $10\,\text{EeV}$, and show that large deflections up to $\sim20^\circ$ are expected even above $100\,\text{EeV}$. This implies that UHECR astronomy would not really be feasible even at such high energies.  Large deflections have also been obtained in \cite{Dundovic:2017vsz}, where an angular spread of a single nearby source normalized to the isotropic background of $\sim50^\circ$ was found. On the other hand, in \cite{Dolag:2004nt,Dolag:2004kp}, simulations of cosmic ray propagation in constrained replicas of the local Universe have shown that large deflections occur only when particles pass through magnetised clusters, but these consist of a very small fraction of the volume. UHECRs above $40\,\text{EeV}$ are expected to be deflected less than $\sim1^\circ$ in most cases, opening up the possibility of cosmic ray astronomy. Similar results were also found in \cite{Das:2008vb}, where above $60\,\text{EeV}$ the deflections are small enough for cosmic ray astronomy. 
\par Recent constrained cosmological simulations developed in  \cite{Hackstein:2016pwa,Hackstein:2017pex} have also shown that the anisotropy signal associated with propagating UHECRs with energy $\gtrsim1\,\text{EeV}$ is not affected by different magnetogenesis scenarios and is nearly compatible with ballistic propagation. However, the distribution of the sources considered and the position of the cosmic ray observer within the simulation volume are important factors in distinguishing the observed anisotropies from the isotropic prediction. In particular, sources within $\sim50\, \text{Mpc}$ mostly dominate the anisotropy signal and show a deviation from isotropic scenarios when their position is inferred from the local matter distribution. Similarly, in the scenarios considered in \cite{AlvesBatista:2017vob}, propagation in highly magnetised cosmic voids $\sim1\,\text{nG}$ shows small deflections in most of the sky.
\par The UHECR anisotropy signal from local (distance $\lesssim25\,\text{Mpc}$) AGNs and SBGs has been studied in \cite{deOliveira:2021ckh,deOliveira:2023oir} for different source luminosities and EGMF scenarios from \cite{Hackstein:2017pex}, showing that the interpretation of the large-scale anisotropy signal and the presence of cosmic ray hotspots in the sky observed by the Pierre Auger and Telescope Array observatories depends on the astrophysical scenario considered. In addition, a dipolar and quadrupolar signal for UHECRs above $32\,\text{EeV}$, produced by isotropic and uniform CR sources, but limited to $100\,\text{Mpc}$ from Earth in structured extragalactic magnetic fields, has also been obtained in \cite{deOliveira:2022wfn}, where a pure proton composition was considered. These anisotropies could be caused by the trapping of the cosmic rays in the magnetic filaments of the primordial magnetic field model. The EGMF models of \cite{Hackstein:2017pex} were also considered in \cite{deOliveira:2020cte}, where the authors showed how the arrival directions of secondary neutral particles, such as photons and neutrinos, from Centaurus A could also be affected by different EGMF scenarios. 
\par The GMF is now better understood and constrained than the EGMF, but there are still uncertainties that can affect the propagation of cosmic rays. One of the most comprehensive and widely used GMF models in UHECR studies is that developed by Jansson and Farrar in \cite{Jansson:2012pc,Jansson:2012rt}, hereafter referred to as JF12. Galactic deflection of UHECRs with the JF12 model was studied in \cite{Farrar:2017lhm}, where the variety and complexity of possible deflection patterns depending on cosmic ray energy and source direction were discussed. In particular, below rigidities of $\sim10\,\text{EeV}$, deflections can also reach $\sim90^\circ$. In \cite{dosAnjos:2018ind} the energy and mass dependence of the elongation of cosmic ray hotspots in the sky due to the GMF for $E>40\,\text{EeV}$ from different candidate sources was studied. There are several models of the GMF in the literature, based on different observations of synchrotron radiation and Faraday rotation (see \cite{Jaffe:2019iuk} for a review). The propagation of cosmic rays in the GMF can also lead to a modification of the observed anisotropy signal at relatively low \cite{Mollerach:2022aji} and very high energies \cite{Bakalova:2023rgy,Korochkin:2025ugg}. In particular, \cite{Bakalova:2023rgy} has shown how different models of GMF introduce dipole amplitude suppression and dipole direction shifts.  It has been shown in \cite{Shaw:2025ykm} how large multipoles, such as the quadrupole moment, can be affected by the propagation of UHECR in the galactic halo. Recent new modelling of the GMF can also be found in \cite{Unger:2023lob,Korochkin:2024yit}.
\par The final quantity that has an important effect on the prediction of the CR anisotropies is the source number density (i.e. the number of sources per unit volume) and the underlying distribution from which a source catalogue is extracted. In particular, it has been shown in \cite{deOliveira:2023kvu} that an isotropic but inhomogeneous source distribution can give rise to anisotropies due to the presence of a magnetic field. The most general and realistic assumption about the distribution of the UHECR sources is that they follow the local baryon distribution, also called the Large Scale Structure (LSS) \cite{diMatteo:2017dtg,Wittkowski:2017nfd,Globus:2018svy,Ding:2021emg,Allard:2021ioh,Bister:2023icg,Allard:2023uuk,PierreAuger:2024fgl}. The interplay of source number density and EGMF is particularly important in defining the multipole structure of the UHECR arrival direction distribution. In \cite{Bister:2023icg} this problem has been studied in detail and it was shown that for a small magnetic deflection (i.e. magnetization parameter\footnote{The magnetization parameter is defined as $\beta_\text{EGMF}=B_\text{rms}\sqrt{\lambda_\text{c}}$, where $B_\text{rms}$ is the root mean square magnetic strength and $\lambda_\text{c}$ is the coherence length of the magnetic field. See also Sec.~\ref{subsec_magnetic_fields} and Eq.~\eqref{deflection_angle} for the correlation between $\beta_\text{EGMF}$ and magnetic deflection.}$\beta_\text{EGMF}\lesssim0.1\,\text{nG}\,\text{Mpc}^{1/2}$) a source density $n_s>10^{-3}\,\text{Mpc}^{-3}$ is required to reproduce the observed multipole signal. Furthermore, scenarios characterized by $\beta_\text{EGMF}\geq3\,\text{nG}\,\text{Mpc}^{1/2}$ or a homogeneous source distribution are also excluded. A similar study considering local star-forming galaxies as UHCER sources, was performed in \cite{vanVliet:2021deg}, where upper limits on the source density $n_s\lesssim9\cdot10^{-2}\,\text{Mpc}^{-3}$ and $\beta_\text{EGMF}\lesssim20\,\text{nG}\,\text{Mpc}^{1/2}$ were found. In \cite{Bister:2024ocm} different source densities and newly developed GMF models were considered to reproduce the observed dipole and quadrupole. The authors have shown that the continuous source scenario ($n_s=\infty$) cannot reproduce the observed dipole, and a source density of $n_s\simeq10^{-4}\,\text{Mpc}^{-3}$ seems to be favored. Recent studies have also considered the scenarios where transient sources emit UHECRs in extragalactic space \cite{Marafico:2024qgh}, finding that the turbulent magnetic field strength in the Local Sheet should be of $0.5-20\,\text{nG}$ and with a coherence length of $10\,\text{kpc}$, and identifying long gamma-ray bursts as possible source candidates. Interestingly, it was also shown in \cite{Mbarek:2025xvg} that the EGMF-induced time delay in the propagation of UHECRs from AGNs can be comparable to or even larger than the source duty cycle. This would make the correlation of the arrival directions of UHECRs and their sources very difficult. 
\par Given the uncertainties in the number density of UHECR sources and the limited knowledge of the EGMF, we consider several scenarios in this work to account for several possible UHECR trajectory scenarios to Earth. We consider two source density values and three EGMF models together with the ballistic case, as discussed in Sec.~\ref{subsec_magnetic_fields} and Sec.~\ref{subsec_crpropa}, respectively. The analysis of the simulation outputs is described in Sec.~\ref{subsec_analysis} and the results are presented in Sec.~\ref{sec_results}. The discussion of the results and conclusions are given in Sec.~\ref{sec_discussion_conclusions}.

\section{UHECR simulations}
\label{sec_uhecr_sim}
As they propagate through extragalactic space, the trajectories of UHECRs are subject to two types of interaction. As described in the previous section, UHECRs are energetic enough to trigger photohadronic interactions with cosmic photon fields (CMB and EBL). These interactions sensitively modify the injected energy spectrum and composition (and hence the magnetic rigidity). Photohadronic interactions also define the interaction horizon. Secondly, the presence of EGMFs changes the arrival directions of UHECRs with respect to the line of sight of the source, and low energy nuclei emitted by distant sources could also be deflected to such an extent that they would never reach the Earth within the age of the Universe. For example, protons with energy $E=8\,\text{EeV}$ coming from a source at $100\,\text{Mpc}$ are deflected by $\theta_\text{rms} \sim 40^\circ$ by a magnetic field with root mean square strength $B_\text{rms}=1\,\text{nG}$ and coherence length $\lambda_\text{c}=1\,\text{Mpc}$ (see Sec.~\ref{subsec_magnetic_fields} for the definition of $\theta_\text{rms}$). Magnetic deflection could give rise to a magnetic horizon which would further limit our maximum observable distance. In the following sections, we will describe the interaction and the magnetic parameters we have considered in this work (a brief introduction can be found in \cite{Rossoni:2023lgq,Rossoni:2023bef}). The combination of all these effects is crucial for the calculation and interpretation of UHECR anisotropies.

\subsection{Cosmic magnetic fields}
\label{subsec_magnetic_fields}
The spatial configuration and local strength of the EGMF is a topic still under debate and investigation. Under the simple assumption of a statistically homogeneous EGMF, characterized by its root mean square (rms) strength $B_\text{rms}$ and coherence length $\lambda_\text{c}$, the typical deflection angle can be estimated. In the absence of energy loss, the rms deflection angle $\theta_\text{rms}$ over a distance $r$ greater than the Larmor radius of the particle $r_\text{L}$ reads\footnote{We recall that the Larmor radius of a charged particle with energy $E$ and charge number $Z$ in a magnetic field of rms strength $B_\text{rms}$ is given by $r_L\simeq1. 1\,\left(\dfrac{E/Z}{1\,\text{PeV}}\right)\left(\dfrac{B_\text{rms}}{1\,\mu\text{G}}\right)^{-1}\,\text{pc}$} 
\begin{linenomath}
\begin{align}
\label{deflection_angle}
\theta_\text{rms} \left(E,r \right) & \simeq 0.8^\circ \cdot  Z \left( \dfrac{E}{100\,\text{EeV}}\right)^{-1} \left( \dfrac{r}{10\,\text{Mpc}}\right)^{1/2} \nonumber \\
& \,\,\,\,\,\,\,\,\,\,\,\,\,\,\,\,\,\,\,\,\,\,\,\,\,\,\,\, \times \left( \dfrac{\lambda_\text{c}}{1\,\text{Mpc}}\right)^{1/2} \left( \dfrac{B_\text{rms}}{1\,\text{nG}}\right) \nonumber \\
& = 0.8^\circ \cdot  Z \left( \dfrac{E}{100\,\text{EeV}}\right)^{-1} \left( \dfrac{r}{10\,\text{Mpc}}\right)^{1/2} \nonumber \\
& \,\,\,\,\,\,\,\,\,\,\,\,\,\,\,\,\,\,\,\,\,\,\,\,\,\,\,\, \times \left( \dfrac{\beta}{1\,\text{nG}\,\text{Mpc}^{1/2}}\right)\, ,
\end{align}
\end{linenomath}
where $\beta$ is defined in Sec.~\ref{sec_intro} and $Z$ and $E$ are the charge number and the energy of the particle, respectively \cite{Waxman:1996zn}. The above expression immediately shows that even over small distances (but still larger than the Larmor radius) where a magnetic field of $B_\text{rms}\sim 0.1\,\mu\text{G}$ is present, high-energy protons can be deflected by many degrees. 
\begin{figure}[t]
\centering
\begin{minipage}{8cm}
\centering
\includegraphics[scale=0.45]{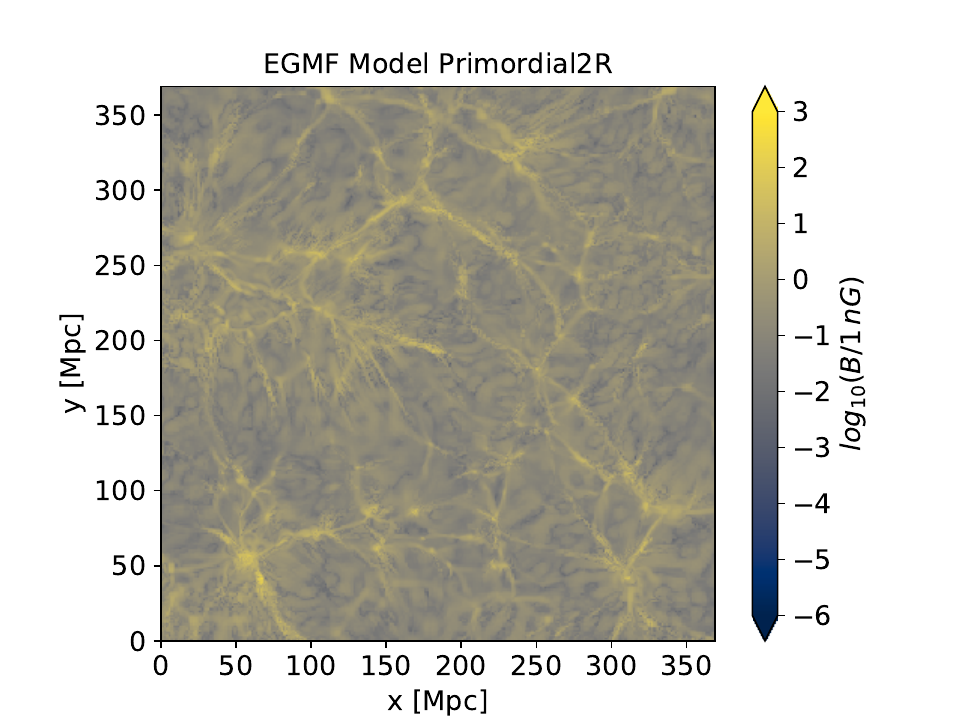}
\end{minipage}
\begin{minipage}{8cm}
\centering
\includegraphics[scale=0.45]{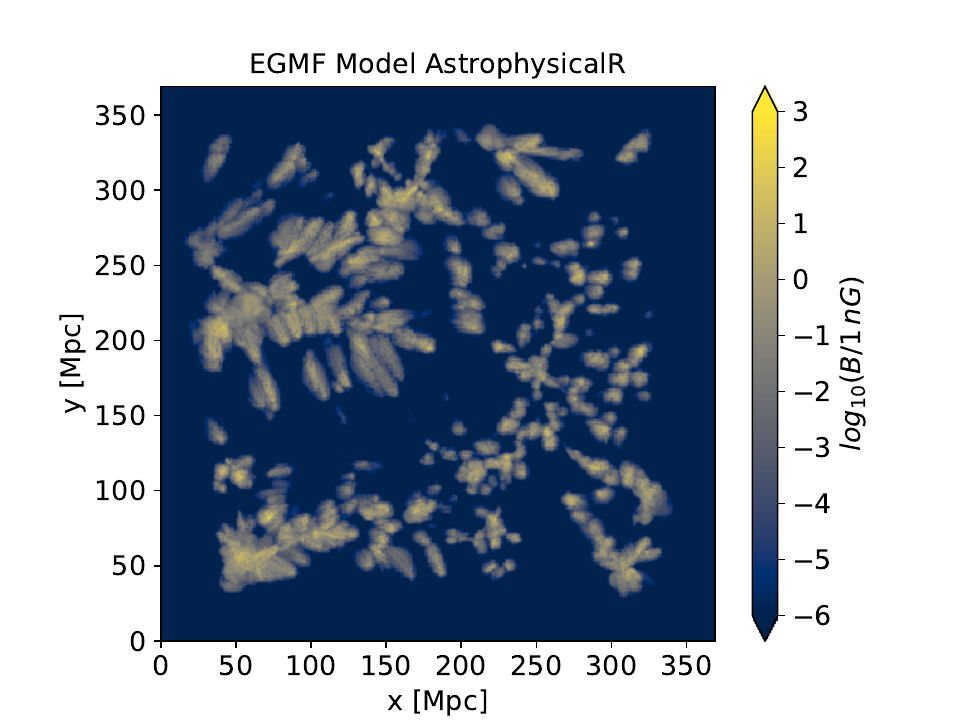}
\end{minipage}
\begin{minipage}{8cm}
\centering
\includegraphics[scale=0.45]{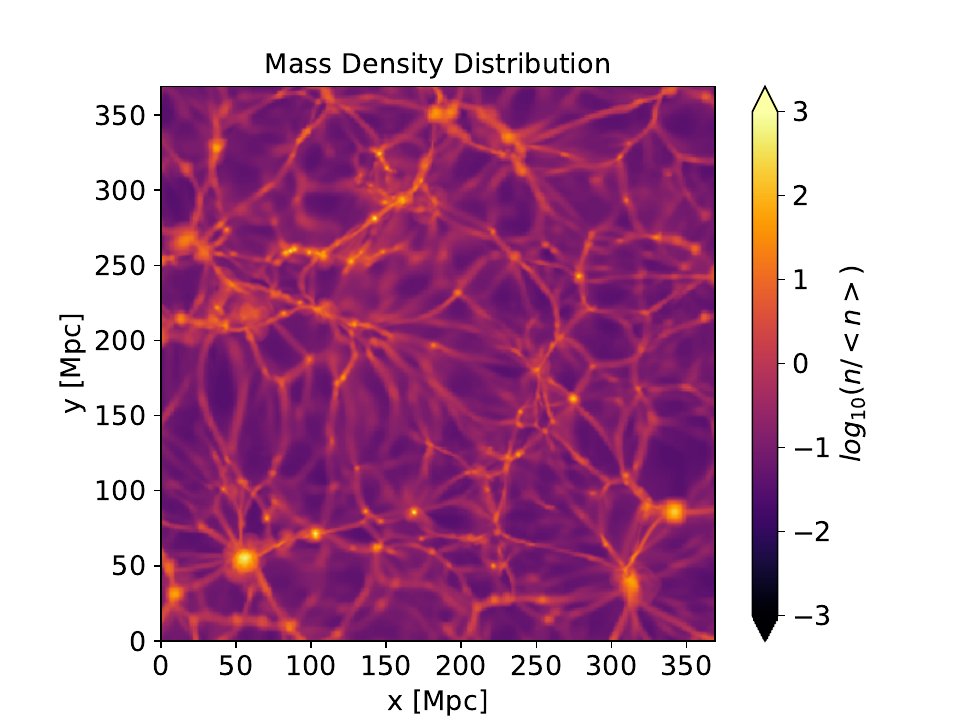}
\end{minipage}
\caption{Magnetic field strength in nG (upper and middle panels) and mass number density $n$ normalized to the average value in the simulation volume $\left< n\right>$ (lower panel) in super-galactic coordinates at redshift $z=0$. The panels show the super-galactic plane orthogonal to the Z-axis including the Milky Way at the center. EGMF models \textit{primordial2R} (upper panel), \textit{astrophysicalR} (middle panel) and mass density distribution from \cite{Hackstein:2017pex}.}
\label{egmf_baryon_maps}
\end{figure}
\par The deflection angle in Eq.~\eqref{deflection_angle} can be used to estimate how the values of $B_\text{rms}$ and $\lambda_\text{c}$ affect the multipole distribution. Since the multipole parameter $l$ is related to the angular scale $\theta\sim180^\circ/l$ (see Appendix~\ref{appendix_pixels}), we obtain that, for cosmic ray energies $E\geq8\,\text{EeV}$, the magnetization parameter must be such that
\begin{equation}
\beta \gtrsim 2 \cdot
Z^{-1}\left( \dfrac{10}{l}\right) \left( \dfrac{r}{10\,\text{Mpc}}\right)^{-1/2} \,\text{nG}\,\text{Mpc}^{1/2} \, .
\end{equation}
Above $8\,\text{EeV}$, we expect that the UHECR flux is mostly produced by sources within a distance of $r\lesssim100\,\text{Mpc}$. Furthermore, the average observed UHECR mass composition in this energy range is $\left<A\right>\sim 10$ \cite{PierreAuger:2023bfx,PierreAuger:2023xfc}. Under these conditions, we obtain that
\begin{equation}
\beta \gtrsim 10^{-1} \left(\dfrac{10}{l}\right)\,\text{nG}\,\text{Mpc}^{1/2} \, .
\end{equation}
For $l=1,2$ (i.e. for the dipole and quadrupole components), this is equal to
\begin{equation}
\label{magnetic_field_impact_d_Q}
B_\text{rms} \gtrsim 1 \cdot \left(\dfrac{\lambda_\text{c}}{1\,\text{Mpc}}\right)^{-1/2}\,\text{nG} \, .
\end{equation}
\par In this work, we consider constrained MagnetoHydroDynamics (MHD) simulations of the local cosmic web to investigate realistic magnetic deflections. In our UHECR simulations we have implemented the evolved EGMF models described in \cite{Hackstein:2017pex}, obtained with the cosmological grid code \texttt{ENZO} \cite{ENZO:2013hhu}. In these simulations, a numerical replica of the local Universe at redshift $z=0$ is obtained by initializing a volume of $(500\,\text{Mpc}/h)^3$ (with the Hubble parameter is $h=0.677$) at $z=60$, where only the internal subregion with the Milky Way in the center is constrained. The evolution of the cosmic baryonic structures and the EGMF are simulated together\footnote{The evolved  EGMF models and the baryon distribution used in this work can be downloaded from \href{https://crpropa.github.io/CRPropa3/pages/AdditionalResources.html}{https://crpropa.github.io/CRPropa3/pages/AdditionalResources.html}}. 
\par Among the models presented in \cite{Hackstein:2017pex}, we consider those called \textit{primordial2R} and \textit{astrophysicalR}. We consider these two models to be representative of two general magnetogenesis scenarios mentioned in Sec.~\ref{sec_intro}. The \textit{primordial2R} scenario consists of an EGMF obtained from a power-law spectrum with spectral index $n_\text{B}=-3$ at $z=60$. The normalization of this model is such that the rms value of the field is $1\,\text{nG}$ at $z=0$. In the other model the EGMF is of astrophysical origin, and the magnetic energy is assumed to be $50\%$ of the injected thermal energy. The magnetic field is released as a dipole around halos where the physical gas density exceeds the value $n_\text{crit}=10^{-2}\,\text{cm}^{-3}$. These two EGMF models are shown in the upper and middle panels of Fig.~\ref{egmf_baryon_maps} in super-galactic coordinates within a length scale of $250\,\text{Mpc}/h$ each side (the plane orthogonal to the Z-axis including the Milky Way is shown). Note that the volume of the MHD simulations described above was $(500\,\text{Mpc}/h)^3$ in order to remove numerical effects due to periodic boundary conditions. For this reason, UHECR simulations are performed within a smaller internal volume of $(250\,\text{Mpc}/h)^3$ (for more details, see \cite{Hackstein:2017pex}). 
\par Both models show a non-trivial spatial distribution, characterised by cosmic structures such as clusters, filaments and voids. The structure of the EGMF is clearly correlated with the baryon density distribution at $z=0$, shown in the lower panel of Fig.~\ref{egmf_baryon_maps}. The different magnetization of extragalactic voids is given by the assumed field origin. In \textit{astrophysicalR} the EGMF is only produced where the matter density if different from zero. Therefore, this model is characterized by magnetized clusters and filaments, but voids do not show any magnetization. On the other hand, the seed field of the \textit{primordial2R} model is uniformly present throughout the volume at $z=60$. During the evolution of cosmic structures, the EGMF correlates with the matter density, giving rise to the structures observed in Fig.~\ref{egmf_baryon_maps}, and also keeping magnetised \textit{baryon-empty} regions of the volume. 
\par The result in Eq.~\eqref{magnetic_field_impact_d_Q} for the effect of the EGMF on the dipole and quadrupole moments cannot be applied to the structured field models described above. In these cases, the $B_\text{rms}$ and $\lambda_\text{c}$ can only be defined locally within cosmic structures (such as clusters, filaments, and voids), and they do not describe the properties of the magnetic field in the entire simulation volume. However, in both structured models, large deflections are expected to occur at most within cosmic structures. 
\par To better understand the effects due to the spatial distribution of the EGMF, we also consider two other propagation scenarios: the \textit{ballistic} scenario and the \textit{statistical} scenario. In the \textit{ballistic} simulations, the propagation of the cosmic rays is purely ballistic. In the \textit{statistical} scenario the EGMF is given by a statistically homogeneous Kolmogorov-like power-law spectrum with rms field strength $B_\text{rms}=1\,\text{nG}$ and coherence length $\lambda_\text{c}=1\,\text{Mpc}$. We chose these parameters because of the result on the effect of the EGMF on the dipole and quadrupole moments in Eq.~\eqref{magnetic_field_impact_d_Q}. The \textit{ballistic} and \textit{statistical} scenarios give us the opportunity to study the effects not only of the presence of an EGMF, but also of the influence of a structured magnetic field configuration.

\begin{figure*}[t]
\centering
\begin{minipage}{7cm}
\centering
\includegraphics[scale=0.4]{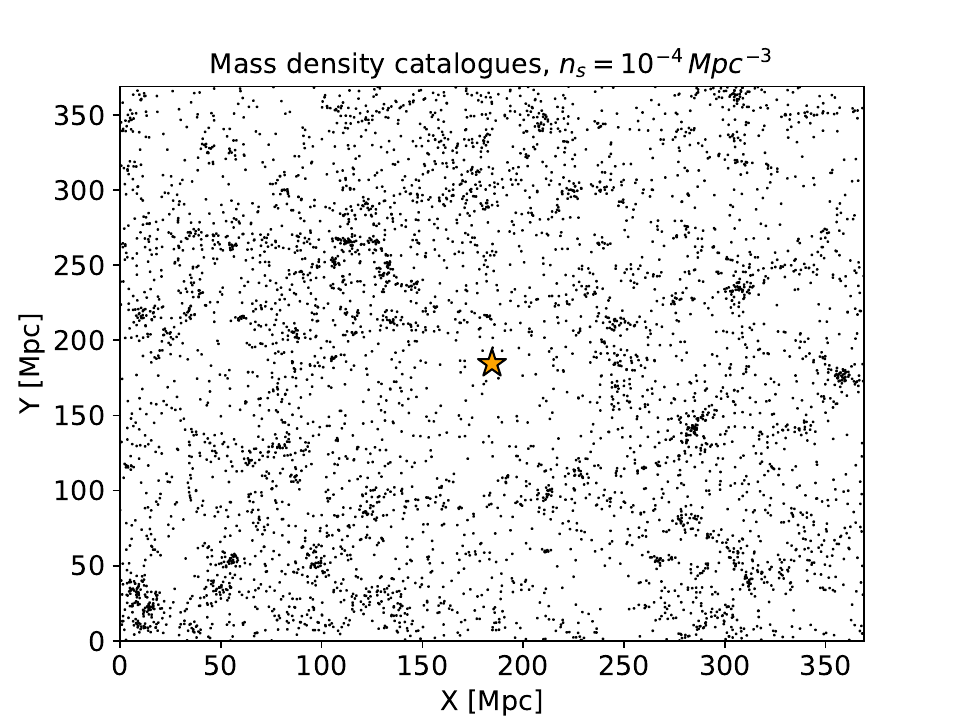}
\end{minipage}
\begin{minipage}{7cm}
\centering
\includegraphics[scale=0.4]{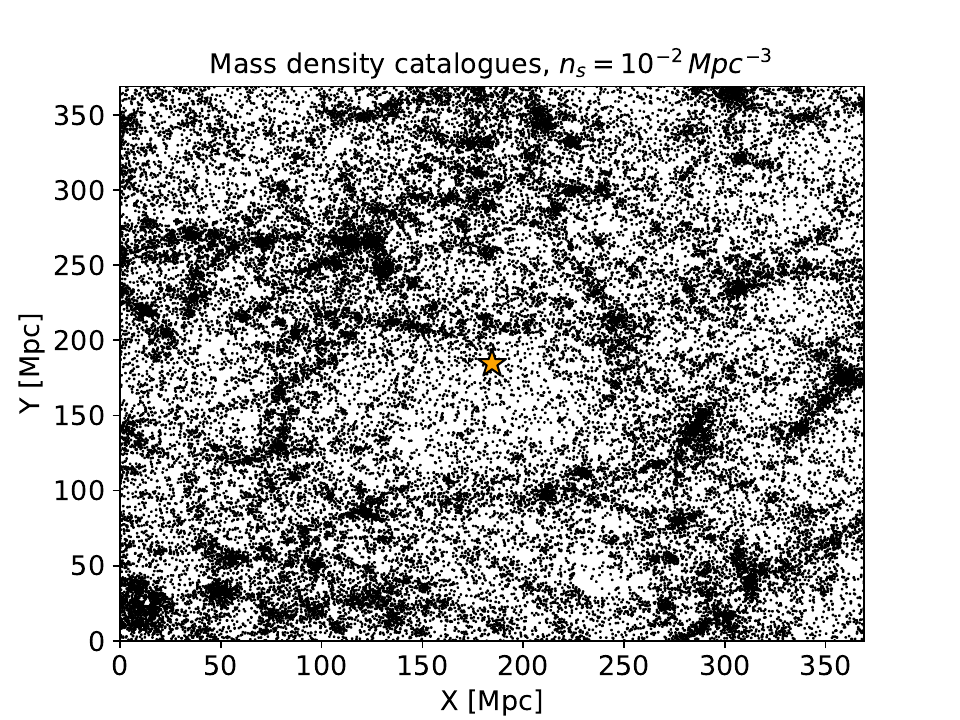}
\end{minipage}
\begin{minipage}{16cm}
\centering
\includegraphics[scale=0.4]{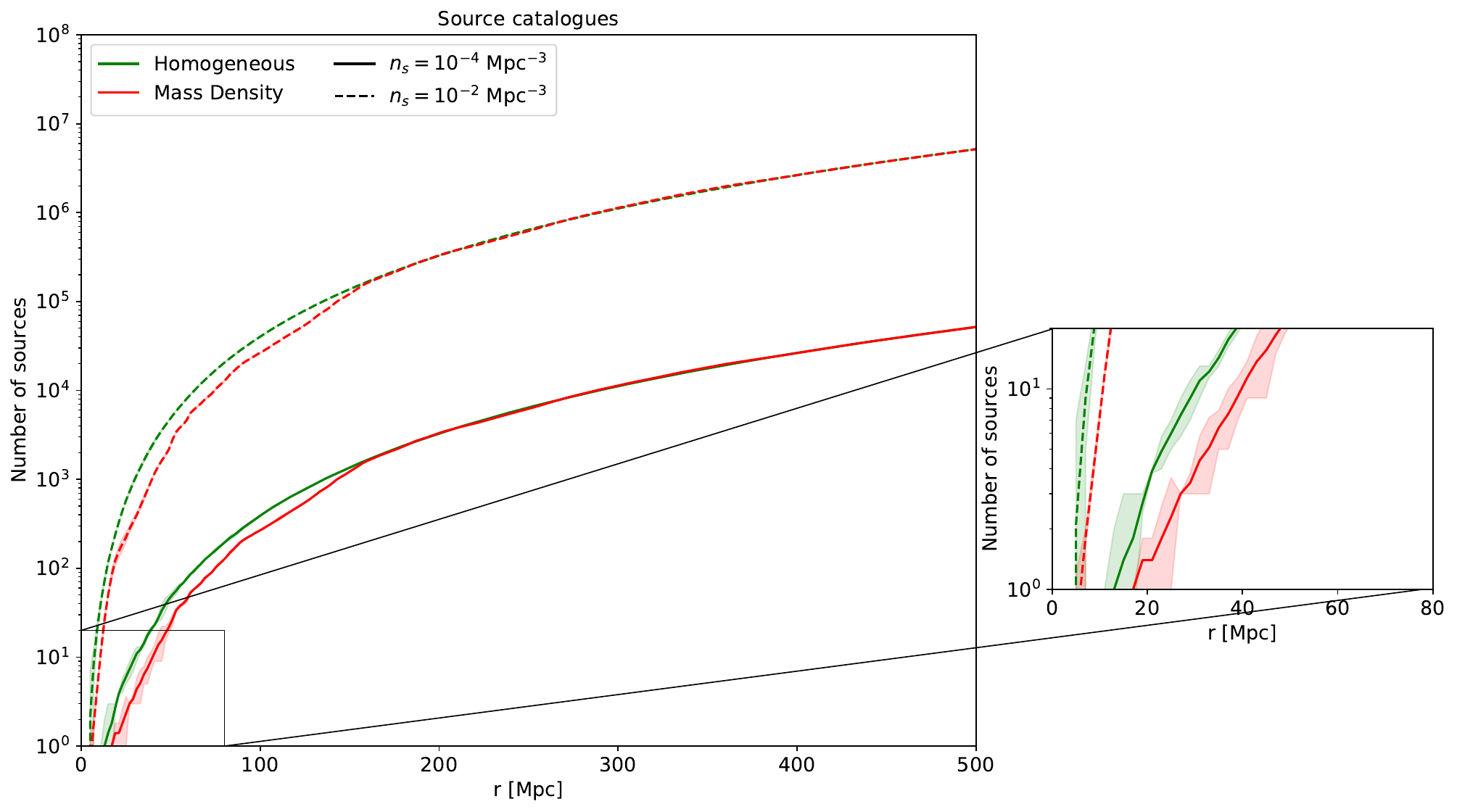}
\end{minipage}
\caption{Upper panels: scatter plot of a catalogue of source locations in the low density (left) and high density (right) scenarios. Both panels show the positions of the sources projected onto the supergalactic plane orthogonal to the Z axis, including the Milky Way (yellow star) in the center. In the high density scenario, only $10\%$ of the sources in the catalogue are shown. Lower panel: Number of sources as a function of the distsourceance from the observer for the two catalogues considered in this work. The \textit{homogeneous} catalogues are in green and the \textit{mass density} catalogues are in red. Solid lines correspond to the source number density $n_s=10^{-4}\,\text{Mpc}^{-3}$ and dashed lines correspond to $n_s=10^{-2}\,\text{Mpc}^{-3}$. The shaded areas represent the cosmic variance multiplied by three. A zoom is shown for $r\leq 80\,\text{Mpc}$.}
\label{source_distance}
\end{figure*}
\subsection{CRPropa simulations}
\label{subsec_crpropa}
As described in the previous sections, the extragalactic propagation of UHECRs is perturbed by several processes. To fully reproduce all the physical effects during the propagation, in this work we use the Monte Carlo code \texttt{CRPropa 3}\footnote{\href{https://crpropa.desy.de}{https://crpropa.desy.de}} \cite{AlvesBatista:2016vpy}, in particular we consider the released version \texttt{CRPropa 3.2} \cite{AlvesBatista:2022vem}. The equation of motion of a UHECR is numerically integrated in a three-dimensional volume, and the propagation is simulated by calculating the actual extragalactic magnetic deflection, and by taking into account interactions with the background photon fields. Photohadronic interactions with the Cosmic Microwave Background (CMB) and the Extragalactic Background Light (EBL) from \cite{Gilmore:2011ks} are considered. Photohadronic interactions include Bethe-Heitler pair production, pion production, and photodisintegration. Nuclear decay of unstable species and adiabatic energy loss due to the expansion of the Universe are also considered. However, since the adiabatic energy loss rate is 3 to 4 times smaller than the total photohadronic energy loss rate at $8\,\text{EeV}$ (see Appendix~\ref{appendix_ELL}), cosmological effects are not larger than other uncertainties of the model.
\par We define a cubic simulation volume of $250\,\text{Mpc}/h$ on each side with periodic boundary conditions (see also Sec.~\ref{subsec_magnetic_fields}). In particular, when a particle leaves the simulation volume, its trajectory is continued on the opposite side, and the injection position is changed accordingly. In this way, the maximum observed source distance does not depend on the size of the simulation volume, but it is automatically determined by the physical processes considered in the simulations (i.e. magnetic deflection and interactions). Periodic boundary conditions also allow us to account for the contribution of sources at distances larger than the sources in the catalogues selected in the local box around the observer. 
\par In order to terminate the propagation of a single cosmic ray particle, two potential \textit{end-conditions} have been introduced. Given out interest in anisotropies in the UHECR sky with energy $E\ge 8\,\text{EeV}$, we terminate the simulation when the cosmic ray energy falls below this value, as the event can no longer contribute. The second condition is based on the fact that the EGMF models in \cite{Hackstein:2017pex} used in this work are produced with constrained cosmological simulations. We can then identify the position of the Milky Way in the center of the simulation volume, where we place our UHECR observer. A spherical surface with radius $R_\text{O}=1\,\text{Mpc}$ is located in the center of the volume; if a particle intersects the sphere, it is recorded and the simulation ends. We correct the recorded arrival directions for the finite size of the observer as discussed in Appendix~\ref{appendix_ad_correction}.  We would also like to emphasise that the use of periodic boundary conditions corresponds to the position of a spherical observer in each virtual replica of the simulation volume. 
\par Together with the EGMF model, the distribution of the source locations can strongly influence the observed anisotropies \cite{Bister:2023icg}. This becomes important in particular at the highest energies due to the presence of the interaction horizon. We consider two different scenarios for the source number density\footnote{We emphasise that our source catalogues do not aim to reproduce any realistic astrophysical object. In this study we call all injection positions \textit{source}, without taking into account the presence of a real astrophysical accelerator at the same coordinates.}: in the first scenario we consider a synthetic catalogue of $5\cdot10^3$ point-like sources corresponding to a source number density of $n_s=10^{-4}\,\text{Mpc}^{-3}$, while the second scenario is given by a synthetic catalogue of $5\cdot10^5$ point-like sources corresponding to a source number density of $n_s=10^{-2}\,\text{Mpc}^{-3}$ \cite{vanVliet:2021deg,Bister:2023icg,Bister:2024ocm}. These density values can be related to real candidate source classes. In particular, $n_s\simeq10^{-2}\,\text{Mpc}^{-3}$ corresponds to the number density of Milky Way-like galaxies \cite{Conselice:2016zid}, while infrared starburst galaxies (SBGs) are characterized by a number density of $n_s\simeq10^{-4}\,\text{Mpc}^{-3}$\cite{Gruppioni:2013jna,Condorelli:2022vfa,Coleman:2022abf}. For these two source densities we generate two different catalogues of source locations, called \textit{homogeneous} and \textit{mass density}. In the first catalogue, we obtain the source positions with a homogeneous sampling in the simulation volume. On the contrary, the \textit{mass density} catalogue is computed by sampling the source positions with a probability density proportional to the baryon density distribution from \cite{Hackstein:2017pex}, shown in Fig.~\ref{egmf_baryon_maps}. Following the idea already introduced in Sec.~\ref{subsec_magnetic_fields}, the purpose of two source distribution models is to study the effect on the arrival directions induced by a structured source distribution. 
\par A single sample of a source catalogue from these models represents only one possible discrete realization of a continuous source distribution. Since the location of the UHECR sources is still unknown, the results could be affected by the cosmic variance associated with the synthesised source catalogue. For this reason, we consider $N_\text{R}=10$ different realizations of the two classes of source catalogues: the number of sources as a function of the distance from the observer is shown in the lower panel of Fig.~\ref{source_distance}. The \textit{homogeneous} catalogues are shown in green and the \textit{mass density} catalogues are shown in red. The solid lines correspond to source number density $n_s=10^{-4}\,\text{Mpc}^{-3}$ and the dashed lines correspond to $n_s=10^{-2}\,\text{Mpc}^{-3}$. The shaded red and green areas represent the cosmic variance (here estimated by simply calculating the sample variance) multiplied by three. The spatial distributions of one of the source catalogues for each number density are also shown in the upper panels of Fig.~\ref{source_distance} in super-galactic coordinates. In both panels, the positions of the sources are projected onto the supergalactic plane orthogonal to the Z axis, including the Milky Way (yellow star) in the center. It can be seen that the number of sources in the two classes of catalogues differs when the distance is $r\lesssim150\,\text{Mpc}$. In particular, the \textit{homogeneous} catalogue contains more sources at small distances than the \textit{mass density}. This is a direct consequence of the complex structure of the baryon density in the local Universe, where sources are more concentrated in regions of high density (i.e. clusters and filaments). At a distance of $r\gtrsim 200 \,\text{Mpc}$, the difference between the two catalogues is no longer relevant, as predicted by the cosmological principle. Both of them show a trend like $N_\text{source}\propto r^3$, and the cosmic variance gradually becomes negligible. Furthermore, we can see that the cosmic variance of both source catalogues becomes negligible when a very high density of sources is considered.
\par The propagation simulations are initiated with five different nuclear species: protons p, helium $^4$He, nitrogen $^{14}$N, silicon $^{28}$Si and iron $^{56}$Fe nuclei. For all species considered, the initial energy is extracted following a power-law injection rate of the form $Q_A(E) \propto E^{-1}$, between $8\,\text{EeV}$ and $10^3\,\text{EeV}$. This energy spectrum at the sources does not correspond to a real physical spectrum at the escape from the accelerator, but it was only used to increase the probability of recording high energy particles at the end of the simulation. The initial momentum direction is chosen to be isotropic at the injection position. We then created a simulation database containing all the combinations of EGMF models, source catalogues and injected compositions discussed above. If $N_\text{B}$ is the number of EGMF models, $N_\text{S}$ is the number of source models, $N_\text{A}$ is the number of injected species and $N_\text{R}$ is the number of source catalogue realisations, then the total number of simulated scenarios in our database is $N_\text{B}\times N_\text{S}\times N_\text{R}\times N_\text{A}=4\times 4\times 10\times 5=800$ ($400$ simulations in low density scenarios and $400$ in high density scenarios). For each simulated scenario in our database, we obtained $N=5\cdot10^4$ observed events\footnote{In the high source density scenario the number of sources present in a catalogue is $5\cdot10^5$, which is one order of magnitude larger than the final number of events collected by the observer. However, the propagation of UHECRs is simulated in a three-dimensional volume, and to obtain such a statistic, we had to inject $4\cdot10^7$ particles for each species, for a total of $2\cdot10^8$ simulated particles for each combination of source distribution and EGMF model.} above $8\,\text{EeV}$, as in \cite{PierreAuger:2024fgl}.

\subsection{Output analysis}
\label{subsec_analysis}
\par As discussed above, we simulate the extragalactic propagation of five different nuclear species separately, all injected with an emissivity of the form $Q_A(E)\propto E^{-1}$. To reproduce a realistic injection scenario, we will re-weight our simulated cosmic rays. We consider the scenario where the UHECR emissivity at the escape from the source\footnote{In this work, we do not study the possible change in the emissivity of accelerated UHECR due to \textit{in-source} interactions. Therefore, we do not distinguish between accelerated and escaped cosmic rays from the source environment.} is described by a power-law with a broken exponential cutoff following the Peters cycle \cite{Peters:1961mxb}, as already described in \cite{Wittkowski:2017okb,PierreAuger:2022atd,PierreAuger:2023htc,PierreAuger:2024hlp}. Therefore, the injection rate takes the form
\begin{equation}
\label{best_fit_1}
Q_A(E) = Q_0\, a_A \left( \dfrac{E}{1\,\text{EeV}}\right)^{-\gamma} f_{\text{cut},A}(E)\, ,
\end{equation}
where the cutoff function is the broken exponential 
\begin{equation}
\label{best_fit_2}
f_{\text{cut},A}(E)=
\begin{cases}
1 \,\,\,\,\,\,\,\,\,\,\,\,\,\,\,\,\,\,\,\,\,\,\,\,\,\,\,\,\,\,\,\,\,\,\,\,\,\,\,\,\,\,\,\,\,\,\,\,\,\,\,\,\,\,\,\,\,\, E< Z_A R_\text{cut}\,; \\
\exp \left(1-\dfrac{E}{Z_A R_\text{cut}}\right) \,\,\,\,\,\,\,\,\,\, E\geq Z_A R_\text{cut}\, ,
\end{cases}
\end{equation}
where $Q_0$ is a normalization constant, $\gamma$ is the spectral index, $R_\text{cut}$ is the maximum rigidity, $Z_A$ is the atomic number of the injected species with mass number $A$ and $a_A$ is the fraction of injected particles of the same species (i.e. the coefficients $a_A$ are such that $\sum_A a_A =1$). We assume that the injection rate of all the considered sources is characterized by the same maximum rigidity $R_\text{cut}$. This is justified by the results obtained in \cite{Ehlert:2022jmy}, where it was shown that a very small variance of $R_\text{cut}$ is expected in order to reproduce the available data. The set of coefficients $a_A$ can also be expressed as integral functions $I_A=a_A Z_A^{2-\gamma}/\sum_{A'}a_{A'}Z_{A'}^{2-\gamma}$. The integral functions $I_A$ express the emitted energy fractions of the element $A$. Details about the re-weighting procedure can be found in Appendix~\ref{appendix_weight}. To find the set of injection parameters $\gamma$, $R_\text{cut}$ and $I_A$ that reproduces the observed energy spectrum and mass composition above $8\,\text{EeV}$, we define the variable $\chi^2$ as the sum of the spectral component $\chi^2_J$ and the mass component $\chi^2_{\left<A\right>}$
\begin{linenomath}
\begin{align}
\label{chi_sq_def}
\chi^2 = \chi^2_J +\chi^2_{\left<A\right>} = & \sum_i \left(\dfrac{J_{i,\text{Auger}}-J_{i,\text{sim}}}{\sigma^{J}_{i,\text{Auger}}} \right)^2\nonumber \\
& + \sum_j \left(\dfrac{\left< A \right>_{j,\text{Auger}}-\left< A \right>_{j,\text{sim}}}{\sigma^{\left< A \right>}_{j,\text{Auger}}} \right)^2 \, ,
\end{align}
\end{linenomath}
where the summations are over different energy bins, and $J_{i,\text{sim}}$ and $\left< A \right>_{j,\text{sim}}$ refer to the energy spectrum and average mass number of a given  simulated scenario. The quantities obtained from real data are indicated with the subscript \textit{Auger}: the energy spectrum above $8\,\text{EeV}$ is taken from \cite{PierreAuger:2021hun} and the mass composition from \cite{PierreAuger:2023bfx,PierreAuger:2023xfc} for the hadronic interaction model EPOS-LHC \cite{Pierog:2013ria}. The values $\sigma^{J}_{i,\text{Auger}}$ and $\sigma^{\left< A \right>}_{j,\text{Auger}}$ indicate the statistical variance of $J_{i,\text{Auger}}$ and $\left< A \right>_{j,\text{Auger}}$, respectively. We therefore find the set of injection parameters that minimize Eq.~\eqref{chi_sq_def} for our simulated scenarios.
\par The final modification to the UHECR trajectory is given by the magnetic deflection due to the GMF. The field strength of the GMF is $\sim 1-10\,\mu\text{G}$ and depends strongly on the arrival direction of the UHECR at the edge of the galaxy (i.e. on the galactic halo) \cite{Farrar:2017lhm}. Understanding the galactic deflection is a non-trivial task and it can be very large (see Sec.~\ref{sec_intro}). We assume that the observer's surface represents the edge of the Milky Way halo, and we apply a magnetic lens to the observed cosmic ray sky (after correcting for the finite size of the observer, as described in Appendix~\ref{appendix_ad_correction}). In practice, the sky at the edge of the galaxy is mapped to the sky at Earth pixel by pixel (for more details on the magnetic lensing technique, see \cite{Bretz:2013oka}). Several magnetic lenses are implemented in \texttt{CRPropa 3}: in this work we use the magnetic lens based on the JF12 model of the GMF \cite{Jansson:2012pc,Jansson:2012rt}, which includes the large-scale random component and the small-scale turbulent component.
\par For each simulated scenario, the recorded UHECRs are used to compute the energy spectrum, mass composition, mean angular deflection, sky map, angular power spectrum and multipoles (dipole vector and quadrupole tensor). In addition, sky distributions are constructed at the edge of the galaxy (i.e. at the observer's surface in the trajectory simulations) and at Earth (i.e. by applying the galactic lensing). We compute the relevant observables for each realization of the source catalogue, we calculate the cosmic average, and we estimate the cosmic variance by calculating the sample variance.
\par The calculation of anisotropy related observables involves functions defined on the unit sphere. The normalized arrival direction of an observed event is given by $\hat{\bold{n}}_i$. We use these vectors (corrected to account for the finite size of the observer) to compute the sky map distribution $\phi(\hat{\bold{n}})$ of the UHECR arrival directions (i.e. the probability density function of observing a UHECR from the direction $\hat{\bold{n}}$). We consider a finite number of pixels in the sky centered in the direction $\hat{\bold{p}}_i$, where $i=1,...,N_p$, and $N_p$ is the total number of pixels. Details of the pixellation parameters used in this study can be found in the Appendix \ref{appendix_pixels}. Given the unit vector $\hat{\bold{n}}$ and a sequence of $N$ events with arrival directions $\left( \hat{\bold{n}}_1,\,\hat{\bold{n}}_2,\,...,\,\hat{\bold{n}}_N\right)$, the sky map distribution $\phi(\hat{\bold{n}})$ is given by 
\begin{equation}
\label{signal_function_text}
\phi (\hat{\bold{n}}) = \dfrac{1}{\mathcal{N}} \sum_{i=1}^{N_p} \mathcal{N}_i \,\Delta (\hat{\bold{n}}-\hat{\bold{p}}_i) \, ,
\end{equation}
where $\mathcal{N}= \sum_{i=1}^{N_p}\mathcal{N}_i$ and $\mathcal{N}_i$ is defined as the sum of the particle weights $\omega_i$ (independent of the arrival direction) within the pixel $i$. The function $\Delta (\hat{\bold{n}}-\hat{\bold{p}}_i)$ is defined as 
\begin{equation}
\Delta (\hat{\bold{n}}-\hat{\bold{p}}_i)= \left\{ \begin{aligned} 
  & \dfrac{1}{\Delta \Omega} \,\,\,\,\,\,\,\,\,\,\,\,\,\,\,\,\,\,\,\,\,\,\, \text{for $\hat{\bold{n}}$ in the pixel $\hat{\bold{p}}_i$}; \\
  & 0 \,\,\,\,\,\,\,\,\,\,\,\,\,\,\,\,\,\,\,\,\,\,\,\,\,\,\,\,\,\,\,\,  \text{otherwise}, 
\end{aligned} \right.
\end{equation}
where $\Delta \Omega$ is the angular size of the pixel.
\par We analyse the sky map distribution in Eq.~\eqref{signal_function_text} by using the \textsc{Python} package \texttt{HEALPY}\footnote{\href{https://healpy.readthedocs.io/en/latest/index.html}{https://healpy.readthedocs.io/en/latest/index.html}} \cite{Zonca:2019vzt}, based on the Hierarchical Equal Area isoLatitude Pixellization (HEALPix\footnote{\href{https://healpix.sourceforge.io}{https://healpix.sourceforge.io}}) scheme \cite{Gorski:2004by}. In particular, the sky map of the observed events $\phi(\hat{\bold{n}})$ can be decomposed as
\begin{equation}
\label{spherical_decomp}
\phi(\hat{\bold{n}}) =\sum_{l=0}^{+\infty}\sum_{m=-l}^l a_{lm} Y_{lm}(\hat{\bold{n}})\, ,
\end{equation}
where the $Y_{lm}(\hat{\bold{n}})$ are the spherical harmonics and the $a_{lm}$ are the spherical coefficients (for details on the normalisation used in this work, see Appendix \ref{appendix_spherical_harmonics}). We compute the angular power spectrum of the observed sky maps, defined as
\begin{equation}
\label{C_l-def}
\mathcal{C}_l = \dfrac{1}{2l+1} \sum_{m=-l}^l \left| a_{lm}\right|^2 \, .
\end{equation}
We compare the obtained $\mathcal{C}_l$ with the ones expected for an isotropic distribution $\Psi_{\text iso}$ in the case of a finite number of recorded particles and pixels in the sky. These are given by 
\begin{equation}
\label{C_l_iso_text}
 \left< \mathcal{C}_l\right>_{\Psi_\text{iso}} = \left\{ \begin{aligned} 
  & \dfrac{1}{4\pi} \,\,\,\,\,\,\,\,\,\,\,\,\,\,\,\,\,\,\,\,\,\,\,\,\,\,\,\,\,\,\,\,\,\,\,\,\,\,\,\,\,\,\,\,\,\,\,\,\,\,\,\,\,\,\,\,\,\,\,\,\,\,\,\,\,\,\,\,\,\,\,\,\,\,\,\,\,\,\,\,\,\,\,\,\,\,\,\,\,\,\,\,   l=0\, ; \\
  &\dfrac{1}{2l+1}\dfrac{\Delta\Omega}{4\pi}\dfrac{\sum_{j=1}^N \omega_j^2 }{\mathcal{N}^2}\sum_{m=-l}^l\sum_{i=1}^{N_p} \left| f_{lm,i} \right|^2 \,\,\,\,\, l\neq0\,, 
\end{aligned} \right.
\end{equation}
where $\omega_j$ is the weight factor of the particle, the coefficient $\mathcal{N}$ can be written as $\sum_{j=0}^N \omega_j$, with $N$ the number of particles observed, and $N_p$ the number of pixels in the sky with angular size $\Delta\Omega$. The factors $f_{lm,i}$ are defined as 
\begin{equation}
f_{lm,i}=\dfrac{1}{\Delta\Omega}\int_{\Delta\Omega_i}d\hat{\bold{n}}\,Y_{lm}(\hat{\bold{n}}) \, ,
\end{equation}
and correspond to the average value of the spherical harmonic $Y_{lm}(\hat{\bold{n}})$ within the pixel in the direction $\hat{\bold{n}}_i$. The full derivation of Eq.~\eqref{C_l_iso_text} is discussed in Appendix~\ref{appendix_aps_estimation}. We also compute the statistical variance of the multipole moments $\mathcal{C}_l$ as done in \cite{Hackstein:2016pwa,Campbell:2014mpa,Anchordoqui:2003bx}, and it is equal to
\begin{equation}
\label{stat_var_C_l}
\sigma_{\mathcal{C}_l}=\sqrt{\dfrac{2}{2l+1}}\,\mathcal{C}_l \, .
\end{equation}
\par Given the normalization used in this work, the monopole component of the angular power spectrum is $\mathcal{C}_0= (4\pi)^{-1}$. To be independent of the chosen normalization, we will consider the angular power spectrum normalized to the monopole component $\mathcal{C}_l/\mathcal{C}_0$. The quantity $\mathcal{C}_l/\mathcal{C}_0$ is also a more appropriate choice for studying dipole and quadrupole amplitudes. In particular, the sky map $\phi(\hat{\bold{n}})$ can be decomposed as 
\begin{equation}
\label{multipole_decomp}
\phi(\hat{\bold{n}})= \phi_0 \left(1+\bold{d}\cdot \hat{\bold{n}}+\dfrac{1}{2}\,\hat{\bold{n}}\cdot\mathbf{Q}\cdot\hat{\bold{n}}+...\right),
\end{equation}
where $\phi_0$ is the mean value of $\phi(\hat{\bold{n}})$ on the sky, $\bold{d}$ is the dipole vector and $\mathbf{Q}$ is the symmetric quadrupole tensor. From the two relations \eqref{spherical_decomp} and \eqref{multipole_decomp} it can be seen that $\phi_0$, $\bold{d}$ and $\mathbf{Q}$ are functions of the spherical coefficients. In particular, it can be shown that the monopole corresponds to
\begin{equation}
\phi_0 = \dfrac{a_{00}}{4\pi}\, ,
\end{equation}
the dipole vector components to
\begin{equation}
\label{dipole_vector}
d_x = -\sqrt{6}\,\dfrac{\Re(a_{11})}{a_{00}} \,\,\, , \,\,\, d_y  = \sqrt{6}\,\dfrac{\Im(a_{11})}{a_{00}} \,\,\, , \,\,\, d_z  = \sqrt{3}\,\dfrac{a_{10}}{a_{00}} \, , 
\end{equation}
and the quadrupole tensor components to
\begin{linenomath}
\begin{align}
\label{quadrupole_tensor}
Q_{xx} & = -\dfrac{\sqrt{10}\,a_{20}-2\sqrt{15}\,\Re(a_{22})}{\sqrt{2}\,a_{00}} \, , \nonumber \\
Q_{xy} & = -\dfrac{2\sqrt{15}\,\Im(a_{22})}{\sqrt{2}\,a_{00}}\, , \,\,\, Q_{xz} = -\dfrac{2\sqrt{15}\,\Re(a_{21})}{\sqrt{2}\,a_{00}}\, , \nonumber \\
Q_{yy} & = -\dfrac{\sqrt{10}\,a_{20}+2\sqrt{15}\,\Re(a_{22})}{\sqrt{2}\,a_{00}} \, ,  \nonumber \\
Q_{yz} & = \dfrac{2\sqrt{15}\Im(a_{21})}{\sqrt{2}\,a_{00}}\, , \,\,\, Q_{zz} = \dfrac{2\sqrt{10}\,a_{20}}{\sqrt{2}\,a_{00}} \, ,  
\end{align}
\end{linenomath}
where $\Re(z)$ and $\Im(z)$ are the real and imaginary parts of $z$. It can be shown that the dipole and quadrupole amplitudes are given by 
\begin{equation}
\label{dip_amplitude}
d = \sqrt{\sum_{i} d_i^2}=3\sqrt{\frac{\mathcal{C}_1}{\mathcal{C}_0}} \, ,   
\end{equation}
\begin{equation}
\label{quad_amplitude}
Q = \sqrt{\sum_{i,j}\dfrac{Q_{ij}^2}{9}}=\sqrt{\dfrac{50}{3}\,\dfrac{\mathcal{C}_2}{\mathcal{C}_0}} \, . 
\end{equation}
For the definition of the quadrupole strength $Q$, see also \cite{PierreAuger:2024fgl}. From Eqs.~\eqref{dip_amplitude} and~\eqref{quad_amplitude} it can be seen that dipole and quadrupole amplitudes only depend on the ratio $\mathcal{C}_{1,2}/\mathcal{C}_0$.

\begin{table*}
\centering
\bgroup
\def\arraystretch{1.25}
\begin{tabular}{c|c|c|c|c|c|c|c|c}
& \textit{ballistic} & \textit{astrophysicalR} & \textit{primordial2R} & \textit{statistical} & \textit{ballistic} & \textit{astrophysicalR} & \textit{primordial2R} & \textit{statistical} \\
\hline
\hline
$\gamma$ & $-2.81$ & $-1.53$ & $-2.53$ & $-3.03$ & $-1.88$ &$-2.50$ &$-2.18$ & $-1.76$\\
$\log (R_\text{cut}/1\,\text{EV})$ & $0.07$ & $0.30$ &$0.12$ & $0.08$ & $0.26$&$0.18$ & $0.11$& $0.29$\\
$I_{p}$ & $1.3\cdot 10^{-1}$ & $8.6\cdot 10^{-2}$ & $6.6\cdot 10^{-2}$& $1.5\cdot 10^{-1}$ & $2.3\cdot 10^{-1}$& $1.1\cdot 10^{-1}$& $3.4\cdot 10^{-2}$& $1.8\cdot 10^{-1}$\\
$I_{He}$ & $1.2\cdot 10^{-1}$ & $9.1\cdot 10^{-2}$& $8.4\cdot 10^{-2}$& $0.0$ & $0.0$& $0.0$& $1.6\cdot 10^{-1}$&$0.0$ \\
$I_{N}$ & $3.7\cdot 10^{-1}$& $3.5\cdot 10^{-1}$& $4.4\cdot 10^{-1}$ & $5.3\cdot 10^{-1}$ & $2.6\cdot 10^{-1}$ & $4.6\cdot 10^{-1}$& $4.0\cdot 10^{-1}$& $3.2\cdot 10^{-1}$\\
$I_{Si}$ & $3.3\cdot 10^{-1}$& $3.8\cdot 10^{-1}$&$3.3\cdot 10^{-1}$ & $2.6\cdot 10^{-1}$ &$5.0\cdot 10^{-1}$ &$4.2\cdot 10^{-1}$ &$2.7\cdot 10^{-1}$& $4.4\cdot 10^{-1}$\\
$I_{Fe}$ & $5.1\cdot 10^{-2}$ &$8.8\cdot 10^{-2}$ &$8.3\cdot 10^{-2}$ & $6.0\cdot 10^{-2}$  & $3.8\cdot 10^{-3}$&$1.8\cdot 10^{-2}$ & $1.3\cdot 10^{-1}$& $5.9\cdot 10^{-2}$\\
\hline
& \multicolumn{4}{c|}{$n_s=10^{-4}\,\text{Mpc}^{-3}$} & \multicolumn{4}{c}{$n_s=10^{-2}\,\text{Mpc}^{-3}$} \\
\end{tabular}
\egroup
\caption{Minimization results of Eq.~\eqref{chi_sq_def} for the \textit{mass density} source scenarios. Different columns correspond to different EGMF models. The source number density of a given model is indicated at the bottom of the table.}
\label{table_density}
\end{table*}
\begin{figure*}[t]
\centering
\begin{minipage}{8.5cm}
\centering
\includegraphics[scale=0.475]{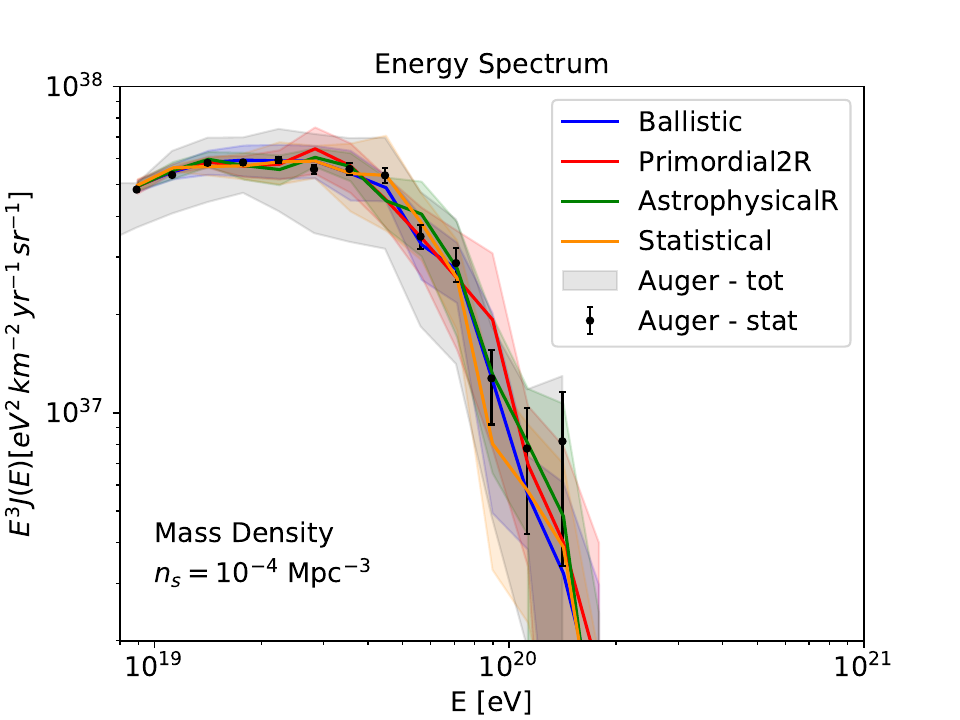}
\end{minipage}
\begin{minipage}{8.5cm}
\centering
\includegraphics[scale=0.475]{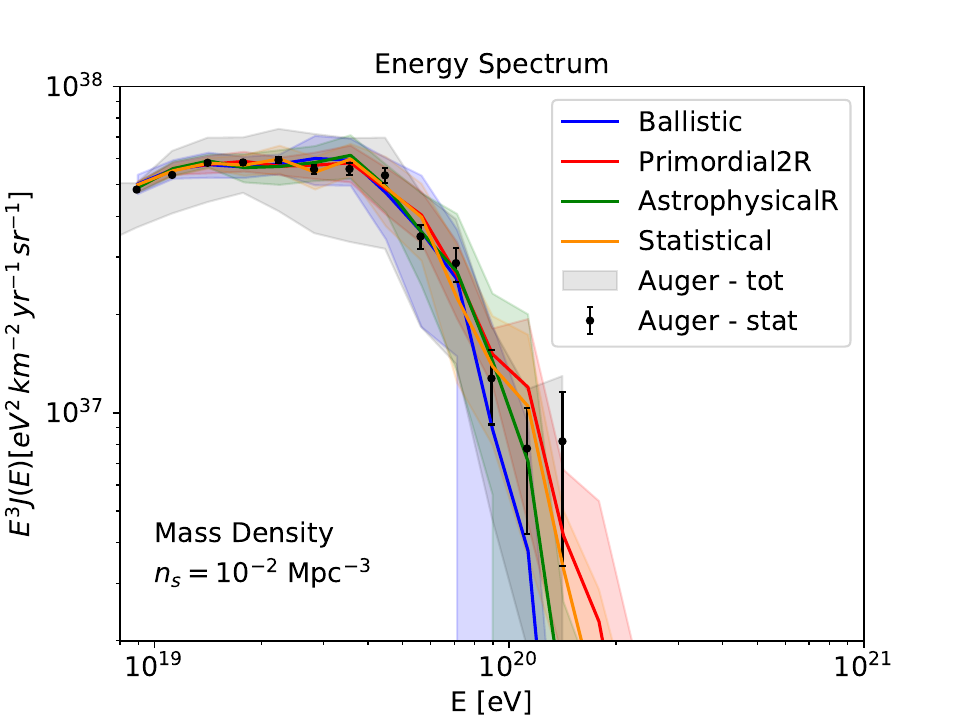}
\end{minipage}
\begin{minipage}{8.5cm}
\centering
\includegraphics[scale=0.475]{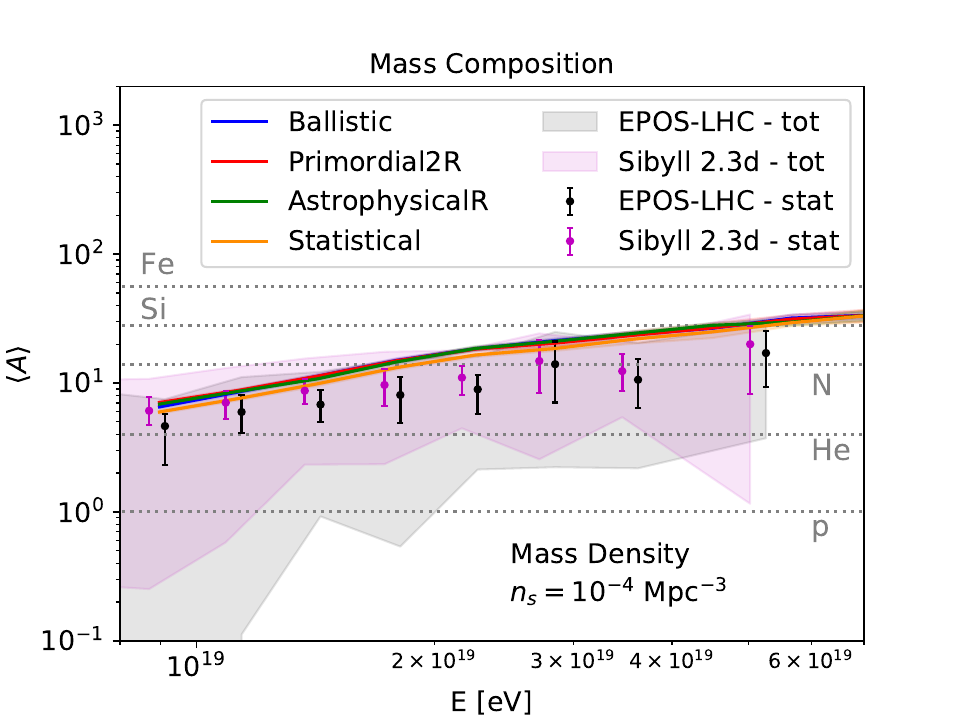}
\end{minipage}
\begin{minipage}{8.5cm}
\centering
\includegraphics[scale=0.475]{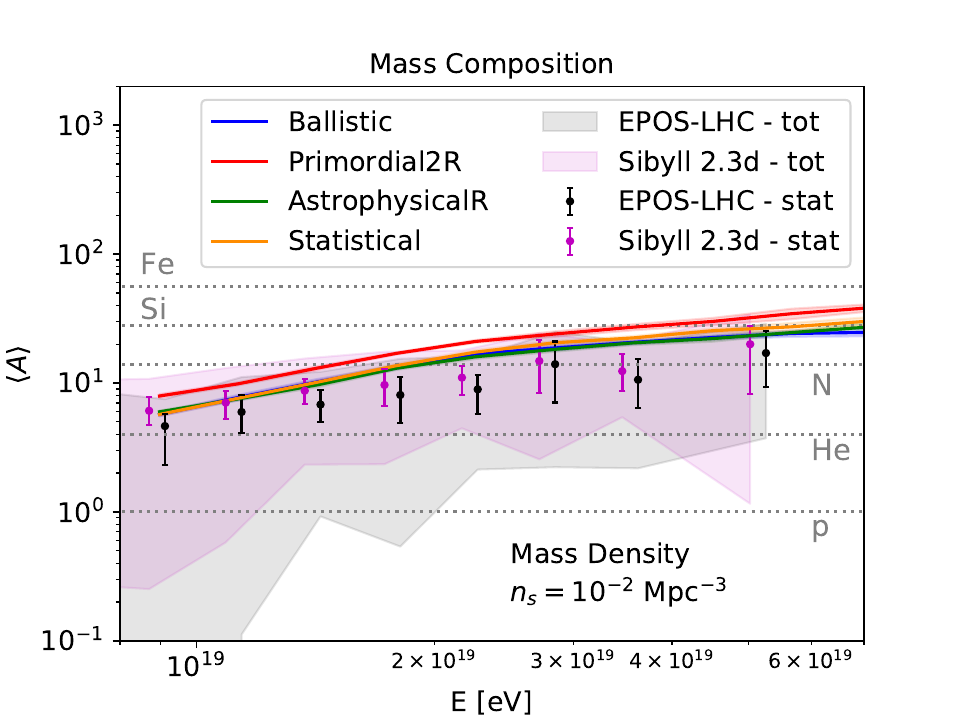}
\end{minipage}
\caption{Simulated energy spectrum (first row) and mass composition (second row) at Earth for the \textit{mass density} catalogues. Different colors correspond to different EGMF models and the shaded areas indicate the cosmic variance of the model. The simulated energy spectra are compared with the observed energy spectrum above $8\,\text{EeV}$ from \cite{PierreAuger:2021hun} shown as black points: the black error bars indicate the statistical variance, and the gray shaded areas correspond to the total variance (statistical and systematic) of the data. The simulated mass compositions are compared with the observed composition from \cite{PierreAuger:2023bfx,PierreAuger:2023xfc} shown as black points for the hadronic interaction model EPOS-LHC \cite{Pierog:2013ria}, and as purple points for Sibyll 2.3d \cite{Riehn:2019jet}: Error bars and shaded areas indicate the same uncertainties as for the energy spectrum. The left column refers to the source density $n_s=10^{-4}\,\text{Mpc}^{-3}$ and the right column to $n_s=10^{-2}\,\text{Mpc}^{-3}$.}
\label{spectrum_composition_den}
\end{figure*}
\section{Results}
\label{sec_results}
As discussed above, the observed level of anisotropy is given by the interaction of several parameters. In the absence of magnetic fields, the UHECR arrival direction distribution is given by the combination of the spatial source distribution, and stochastic interactions with background photon fields. The presence of galactic and extragalactic magnetic fields will introduce deviations that could suppress or enhance the contribution of different multipoles to the overall UHECR arrival direction distribution. In the following, we will explicitly study the effects of all these components that determine the final anisotropy signal for the \textit{mass density} source catalogues. The \textit{homogeneous} source distribution scenarios are discussed in Appendix~\ref{appendix_homo_source}. We recall that, given the observation of a dipole component in the UHECR sky above $8\,\text{EeV}$ \cite{PierreAuger:2023fcr,PierreAuger:2024fgl}, we will only focus on this energy range. 
\begin{figure*}[t]
\centering
\begin{minipage}{7.5cm}
\centering
\includegraphics[scale=0.5]{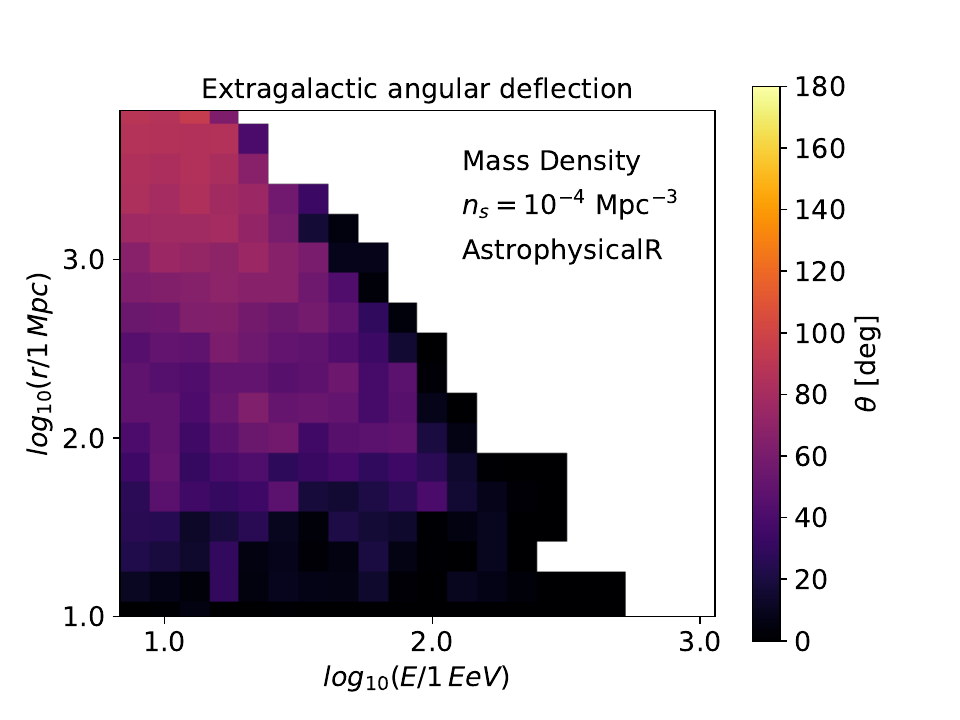}
\end{minipage}
\begin{minipage}{7.5cm}
\centering
\includegraphics[scale=0.5]{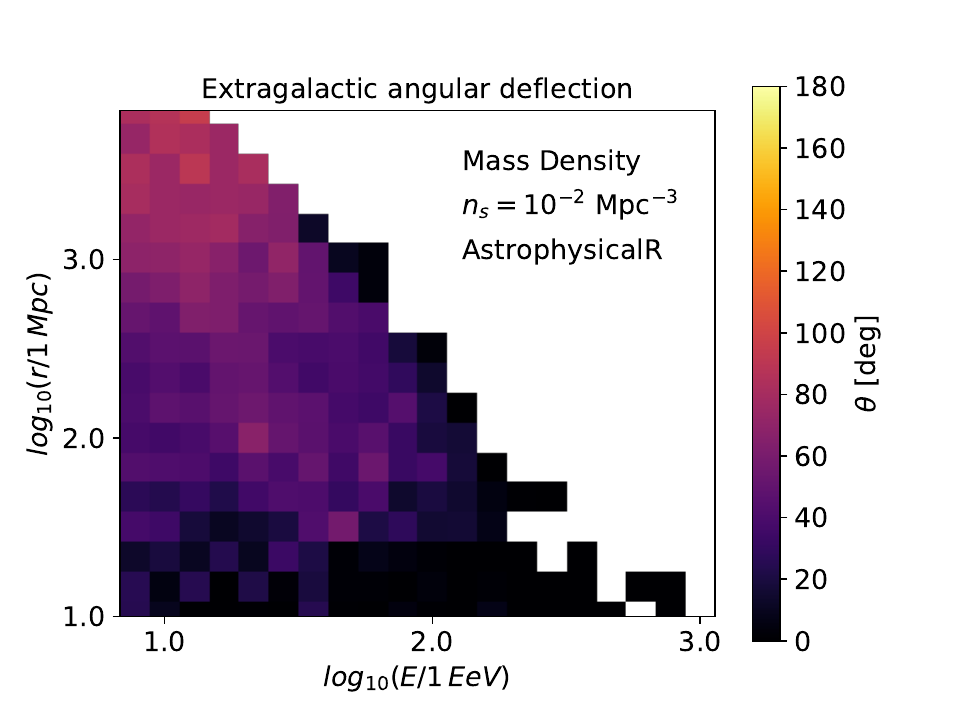}
\end{minipage}
\begin{minipage}{7.5cm}
\centering
\includegraphics[scale=0.5]{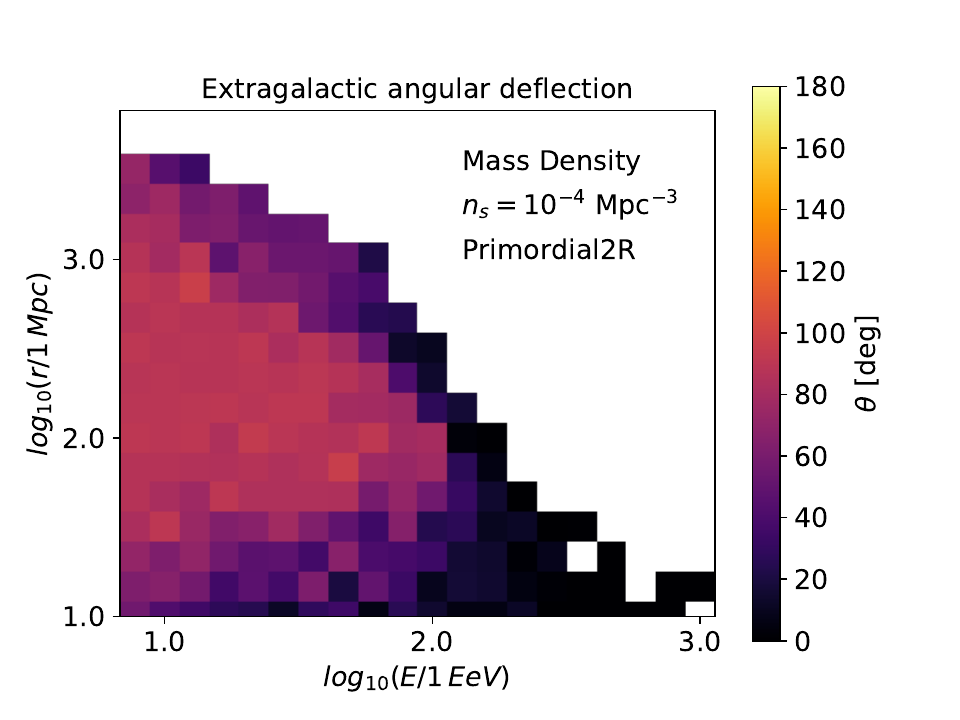}
\end{minipage}
\begin{minipage}{7.5cm}
\centering
\includegraphics[scale=0.5]{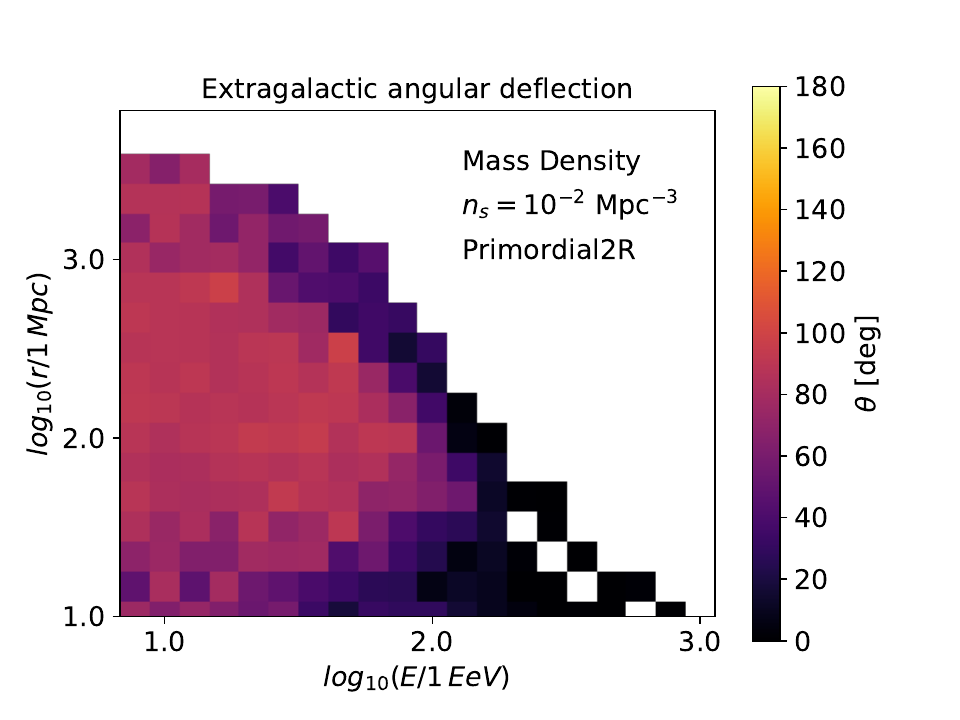}
\end{minipage}
\begin{minipage}{7.5cm}
\centering
\includegraphics[scale=0.5]{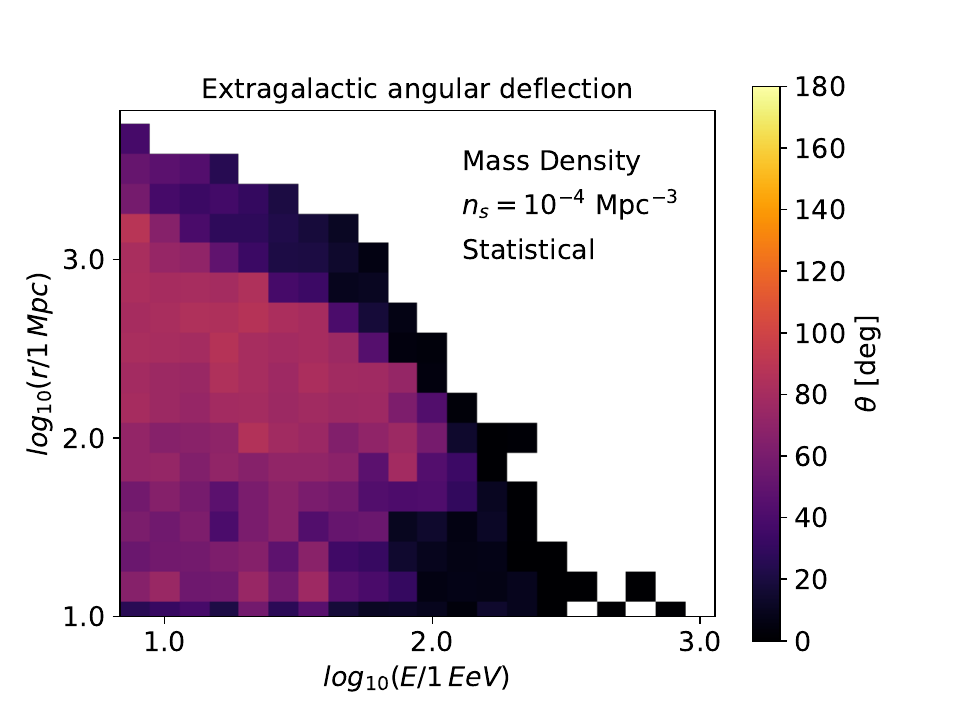}
\end{minipage}
\begin{minipage}{7.5cm}
\centering
\includegraphics[scale=0.5]{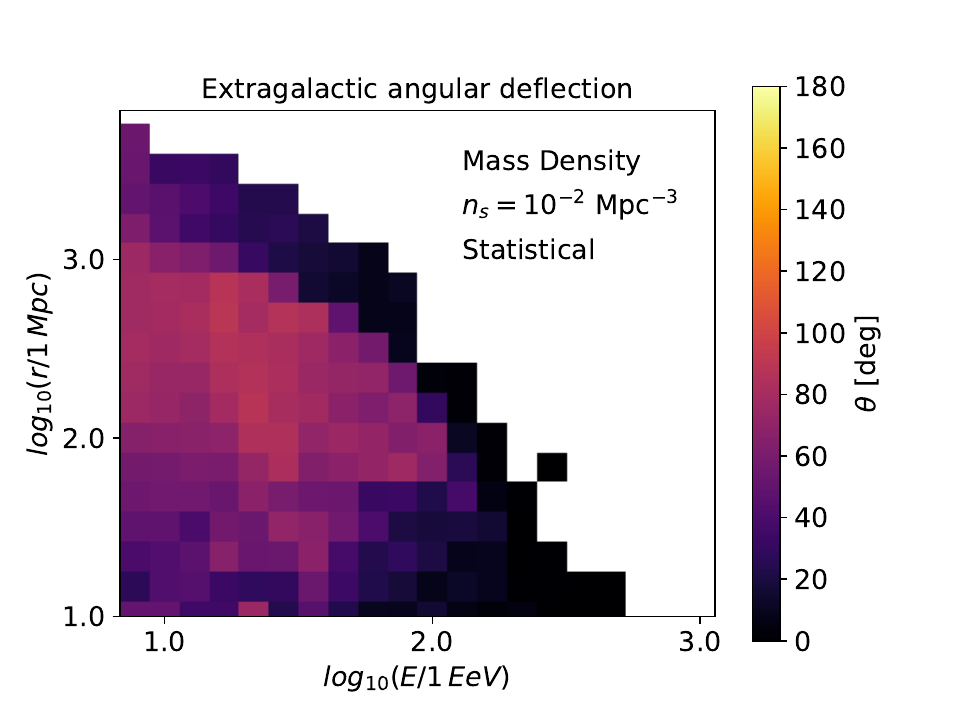}
\end{minipage}
\caption{Cosmic average of the total angular deflection distribution of the observed UHECRs above $8\,\text{EeV}$ for the \textit{mass density} source catalogues. The \textit{astrophysicalR} (first row), \textit{primordial2R} (second row) and \textit{statistical} (third row) EGMF models are shown. The deflection angle is defined as the angle between the injected and observed momentum vectors, and is given in degrees in the color bar. The axes show the observed energy (x-axis) and the source distance (y-axis). The left column refers to the source density $n_s=10^{-4}\,\text{Mpc}^{-3}$ and the right column to $n_s=10^{-2}\,\text{Mpc}^{-3}$.}
\label{deflection_den}
\end{figure*}
\par The minimization results of Eq.~\eqref{chi_sq_def} for the \textit{mass density} scenarios are given in Table~\ref{table_density} for both the source number densities considered. We found that a hard spectral index up to $\gamma\simeq-3$ is required to reproduce the observed spectrum and composition, and a rigidity cutoff of $\log (R_\text{cut}/1\,\text{EV})\simeq0.2$ on average, as also found in \cite{PierreAuger:2022atd,PierreAuger:2023htc,Bister:2023icg,PierreAuger:2024hlp}. We emphasize again that the spectral index $\gamma$, defined in the injection rate in Eq.~\eqref{best_fit_1}, characterizes the UHECR spectrum at the escape from the source environment. Due to \textit{in-source} interactions and propagation, the spectral index at the acceleration would generally be different \cite{Unger:2015laa}. The mass composition is dominated by the nitrogen and silicon components in all the scenarios. This is due to the fact that the lightest components are already suppressed above $8\,\text{EeV}$ ($E_{\text{cut},p}=Z_p R_\text{cut}\simeq1.5\,\text{EeV}$ and $E_{\text{cut},He}=Z_{He} R_\text{cut}\simeq3\,\text{EeV}$), while a large iron component would result in a too heavy mass composition at the highest energy.
\par In Fig.~\ref{spectrum_composition_den} we show the propagated energy spectra and mass composition for the \textit{mass density} scenarios obtained with the parameters in Table~\ref{table_density}. The colored lines and shaded areas correspond to the cosmic average and cosmic variance of the simulated scenarios. We also show the observed energy spectrum above $8\,\text{EeV}$ from \cite{PierreAuger:2021hun} as black points. The statistical variance is indicated by black error bars, while the total variance (statistical and systematic) is indicated by the gray shaded area. The simulated mass compositions are also compared with the observed composition from \cite{PierreAuger:2023bfx,PierreAuger:2023xfc}. We consider the hadronic interaction models EPOS-LHC \cite{Pierog:2013ria} (black points) and Sibyll 2.3d \cite{Riehn:2019jet} (purple points). The error bars and shaded areas denote again the statistical and total variance, respectively. The left column refers to the source density $n_s=10^{-4}\,\text{Mpc}^{-3}$ and the right column to $n_s=10^{-2}\,\text{Mpc}^{-3}$. We can see that our simulated scenarios are compatible with the observations within the current systematic uncertainties and the cosmic variance of our source catalogues. However, we do not find any systematic influence of the assumed source density or EGMF model on spectrum and composition.
\par The propagation of UHECRs in a magnetic field is characterized by deflections, that can affect the maximum observable source distance. In particular, low energy particles may be deflected out of the line of sight of the source and never reach the observer's sphere. We compute the UHECR angular deflection as the angle between the momentum of the cosmic ray at the injection and at the observation at the observer's surface. In order to study the effect due to the EGMF model alone, the galactic deflection is not taken into account. The distributions of the angular deflection for the \textit{mass density} source catalogues are shown in Fig.~\ref{deflection_den} for all EGMF models. In the left column of Fig.~\ref{deflection_den} the source density $n_s=10^{-4}\,\text{Mpc}^{-3}$ is considered and in the right column $n_s=10^{-2}\,\text{Mpc}^{-3}$. We consider the distribution of the deflection angle as a function of the observed cosmic ray energy and the source distance. Very large deflection angles are observed in all EGMF models, but different magnetization scenarios of the extragalactic space correspond to different angular distributions. In the \textit{astrophysicalR} model, magnetic deflections are expected only in regions characterized by high baryonic density (see Fig.~\ref{egmf_baryon_maps}). In the first row of Fig.~\ref{deflection_den} it can be seen that larger deflections than $\gtrsim90^\circ$ are obtained for cosmic rays emitted by sources at distance $r\gtrsim1\,\text{Gpc}$. This can be explained by the fact that in order to be deflected to such an extent, the UHECRs must pass through several magnetized regions separated by large un-magnetized cosmic voids. In the \textit{primordial2R} and \textit{statistical} scenarios, the magnetization of the cosmic voids is such that large deflections are observed even for sources at $\sim100\,\text{Mpc}$. It can also be seen that as the magnetization of the voids increases, the maximum distance of the observed sources decreases. This is due to the fact that particles from very distant sources suffer greater energy losses and are therefore deflected more strongly, until they enter the diffusive regime and are not able any more to reach the observer within the maximal propagation time. This leads to a magnetic horizon which occurs for both source densities considered.
\begin{figure}[t]
\centering
\begin{minipage}{8.5cm}
\centering
\includegraphics[scale=0.5]{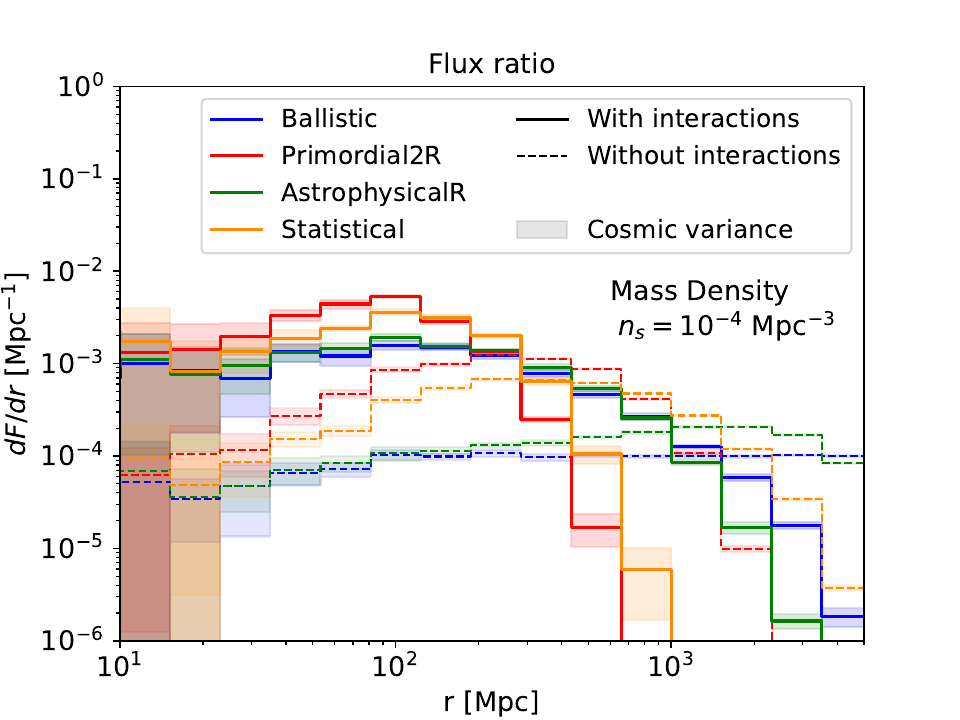}
\end{minipage}
\begin{minipage}{8.5cm}
\centering
\includegraphics[scale=0.5]{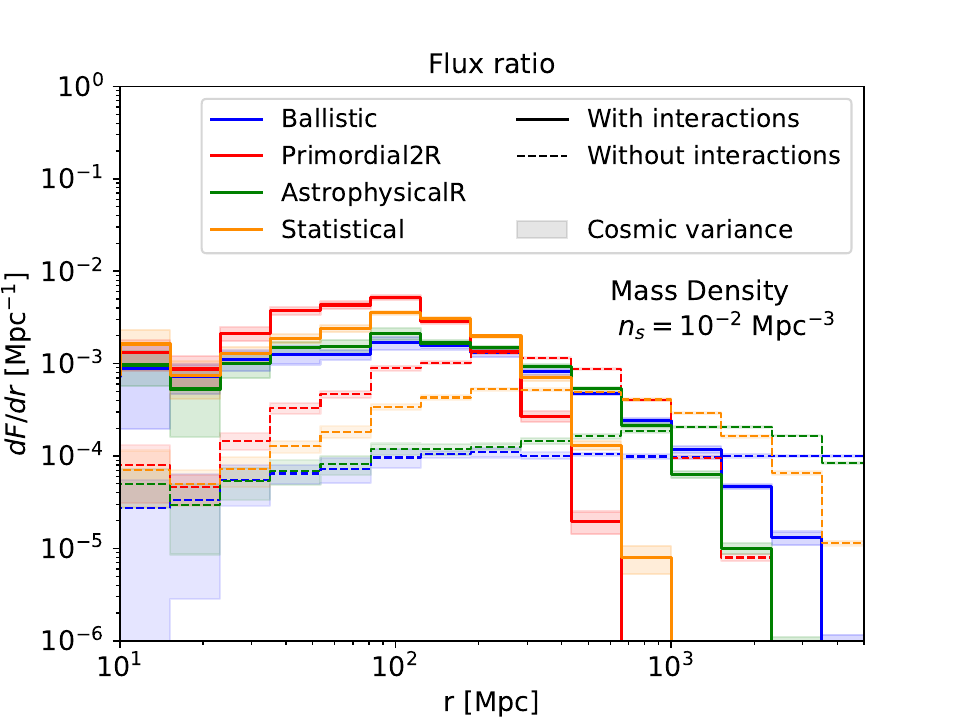}
\end{minipage}
\caption{Distribution of the relative UHECR flux contribution $dF$ above $8\,\text{EeV}$ from sources within distance interval $dr$ with distance to the observer $r$, for the \textit{mass density} source catalogues in the low density (upper) and high density scenarios (lower). Different colors correspond to different EGMF models.
The cases without stochastic interactions are shown as dashed lines. The shaded areas represent the cosmic variance.}
\label{number_sources_den}
\end{figure}
\begin{figure*}[t]
\centering
\begin{minipage}{8.5cm}
\centering
\includegraphics[scale=0.2]{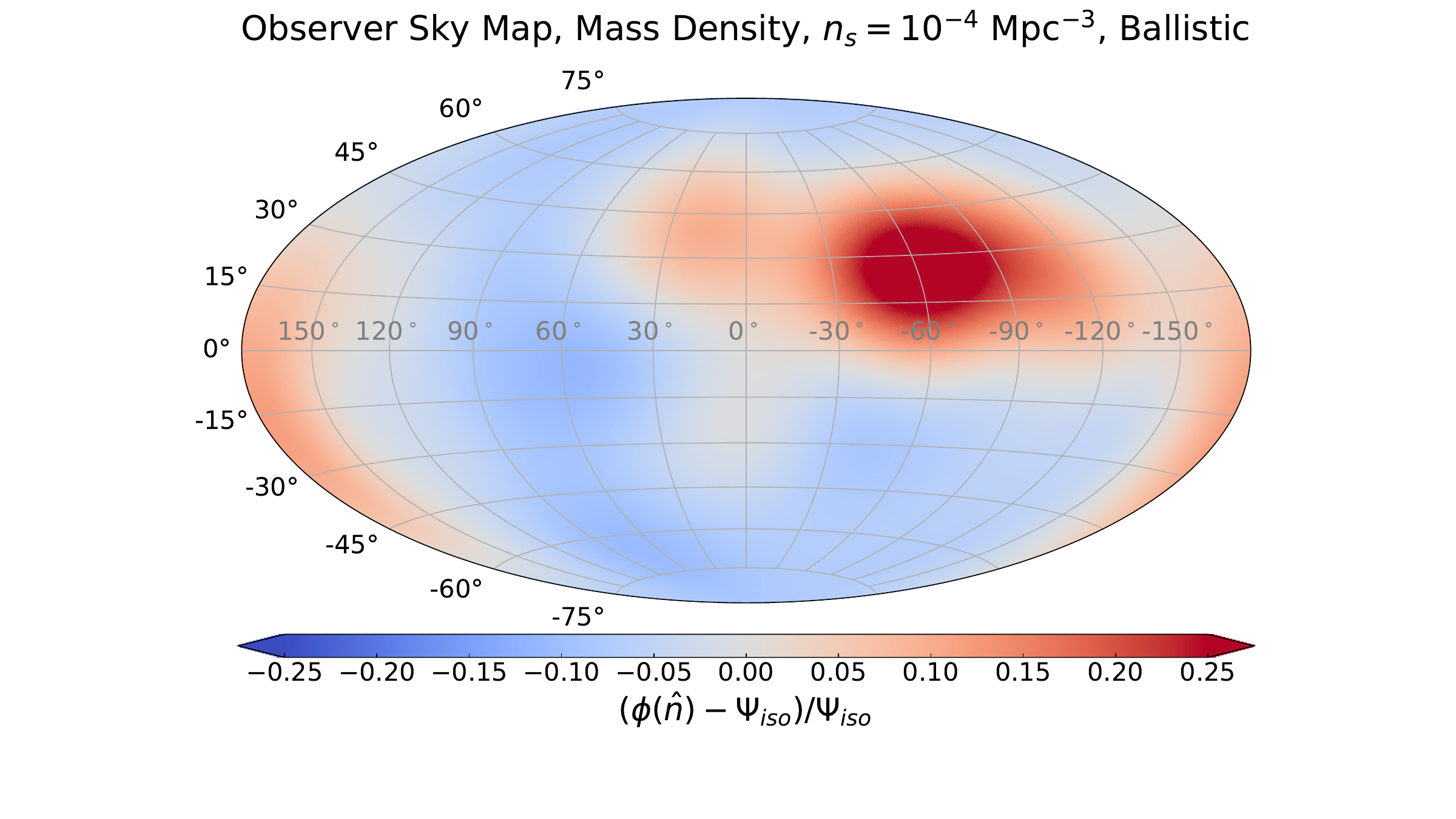}
\end{minipage}
\begin{minipage}{8.5cm}
\centering
\includegraphics[scale=0.2]{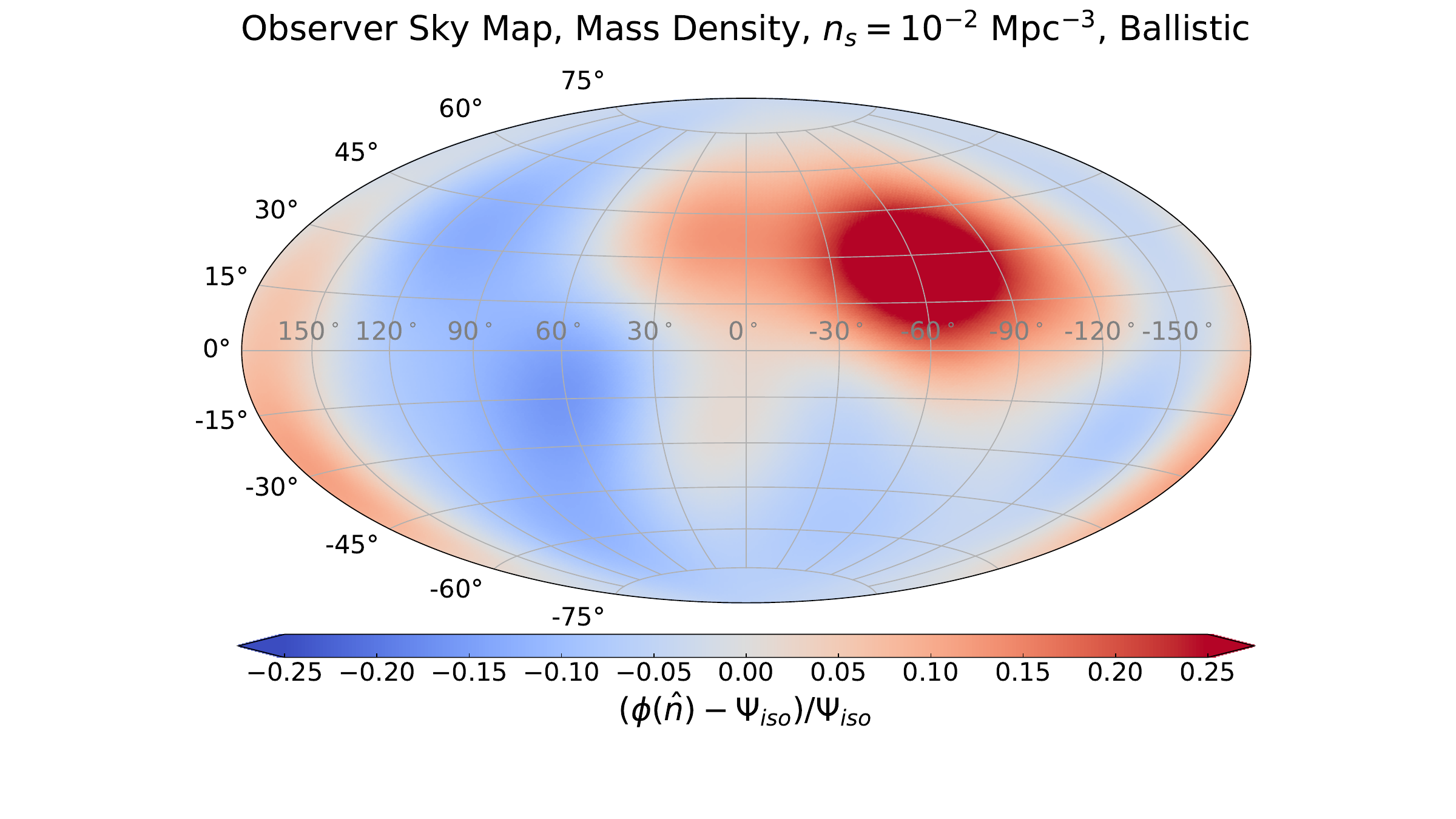}
\end{minipage}
\begin{minipage}{8.5cm}
\centering
\includegraphics[scale=0.2]{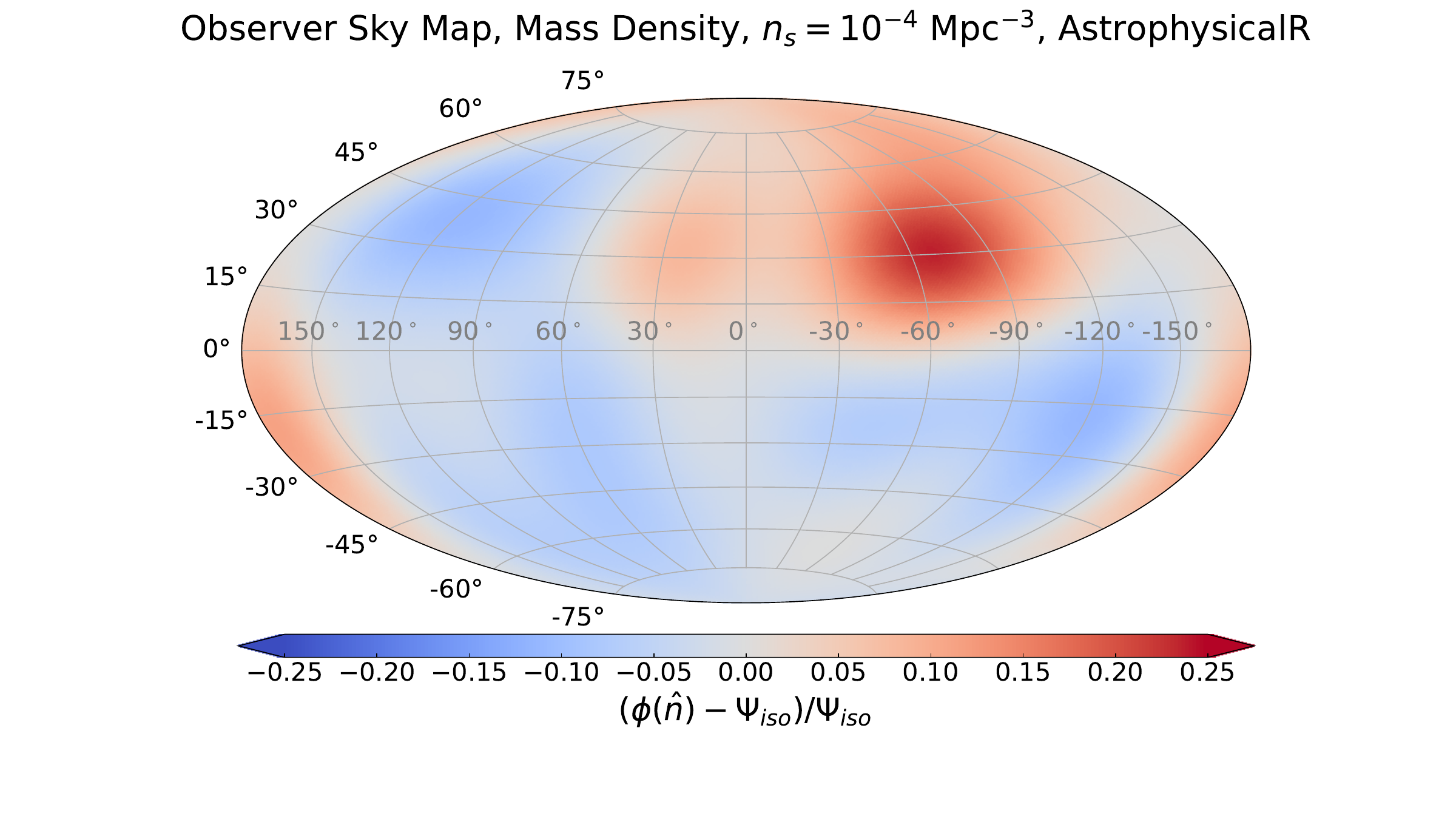}
\end{minipage}
\begin{minipage}{8.5cm}
\centering
\includegraphics[scale=0.2]{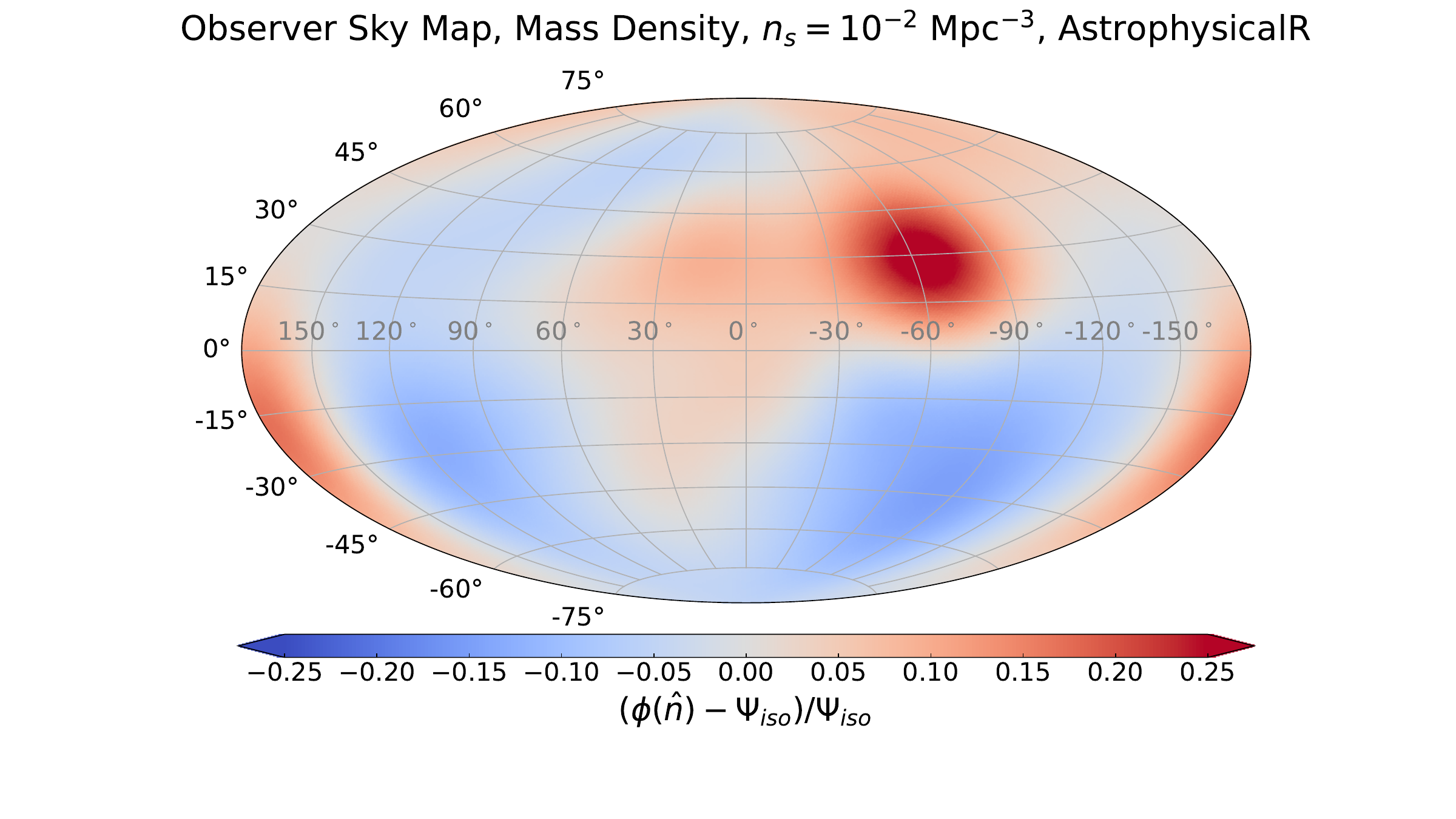}
\end{minipage}
\begin{minipage}{8.5cm}
\centering
\includegraphics[scale=0.2]{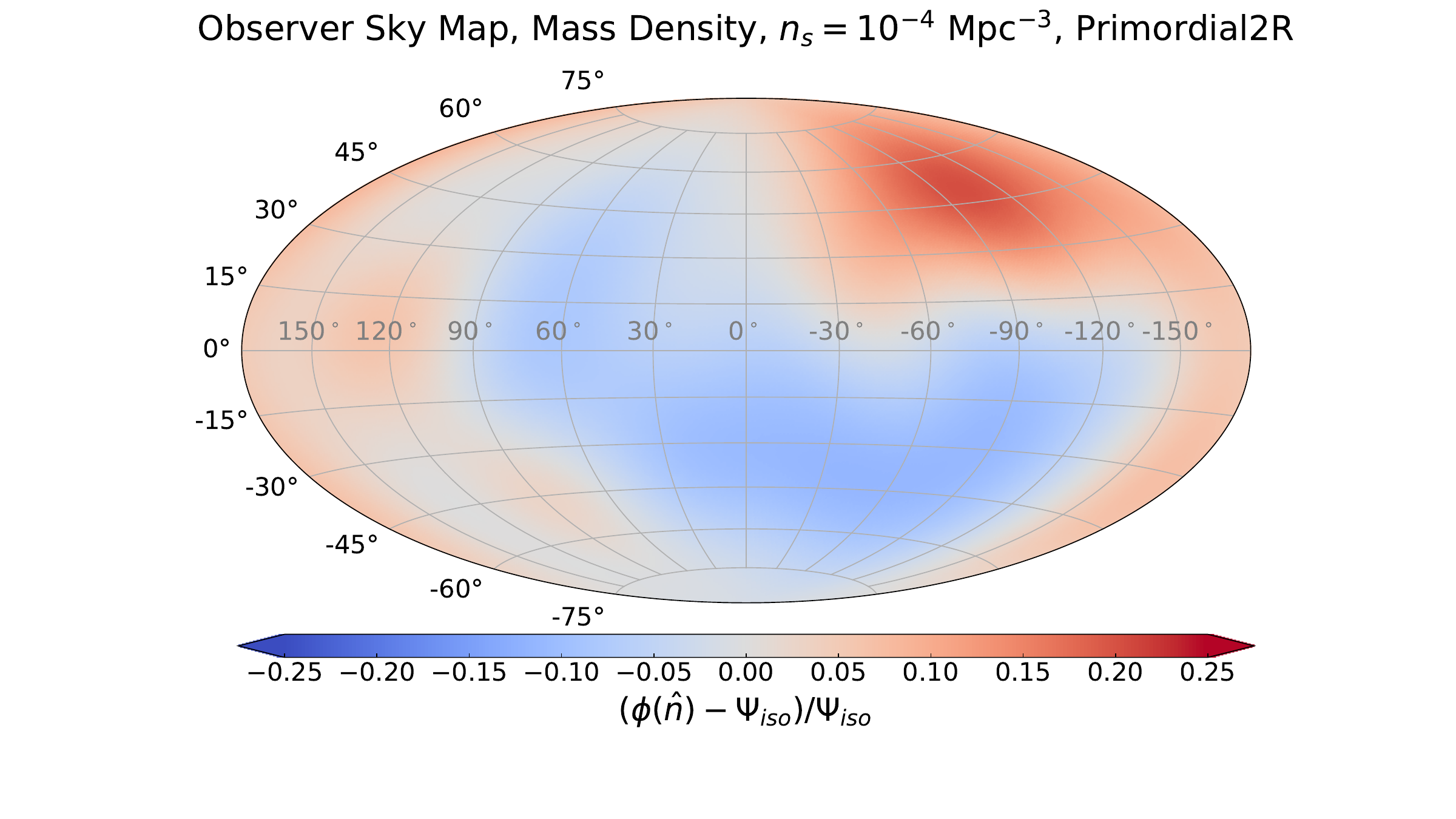}
\end{minipage}
\begin{minipage}{8.5cm}
\centering
\includegraphics[scale=0.2]{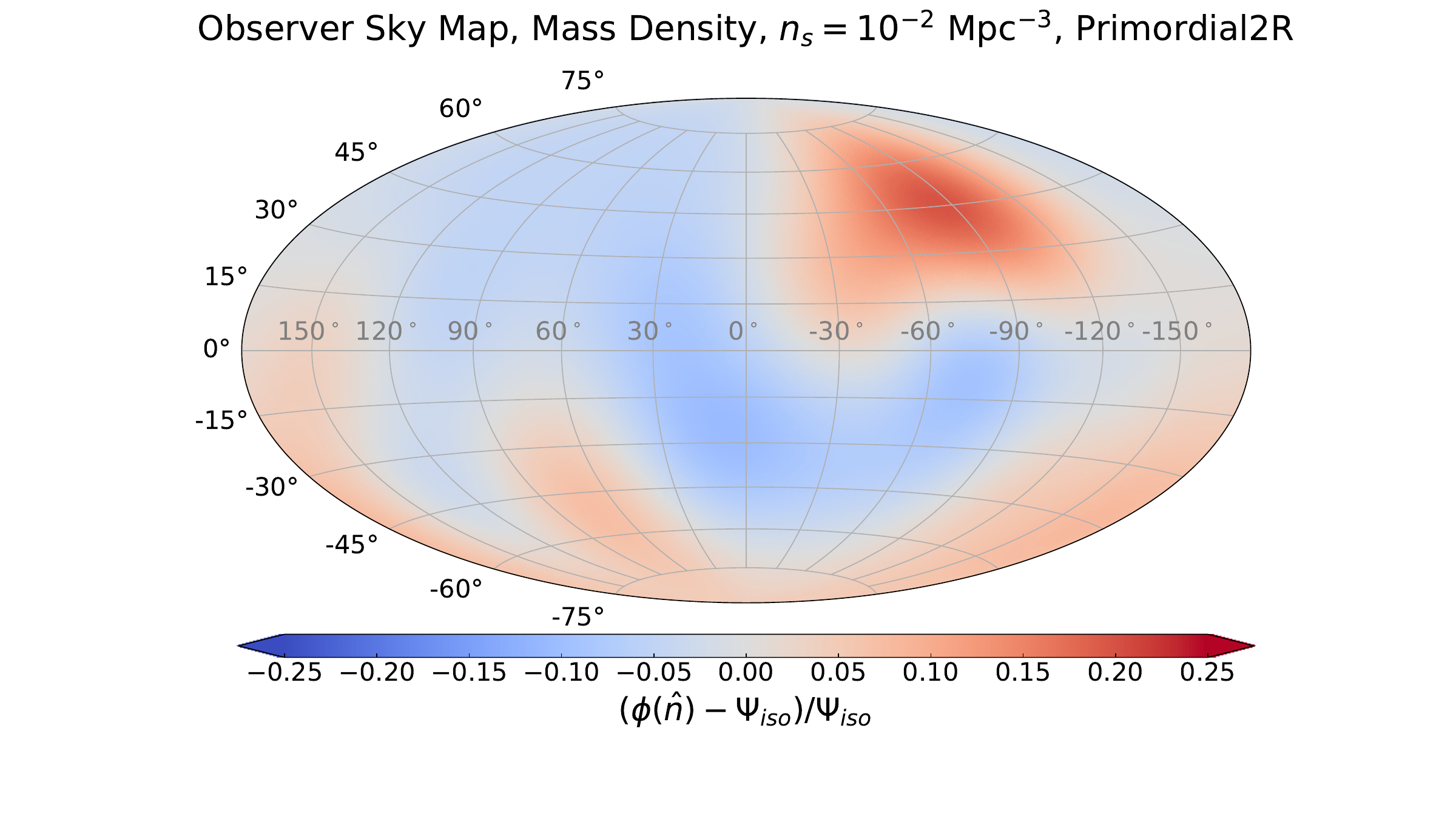}
\end{minipage}
\begin{minipage}{8.5cm}
\centering
\includegraphics[scale=0.2]{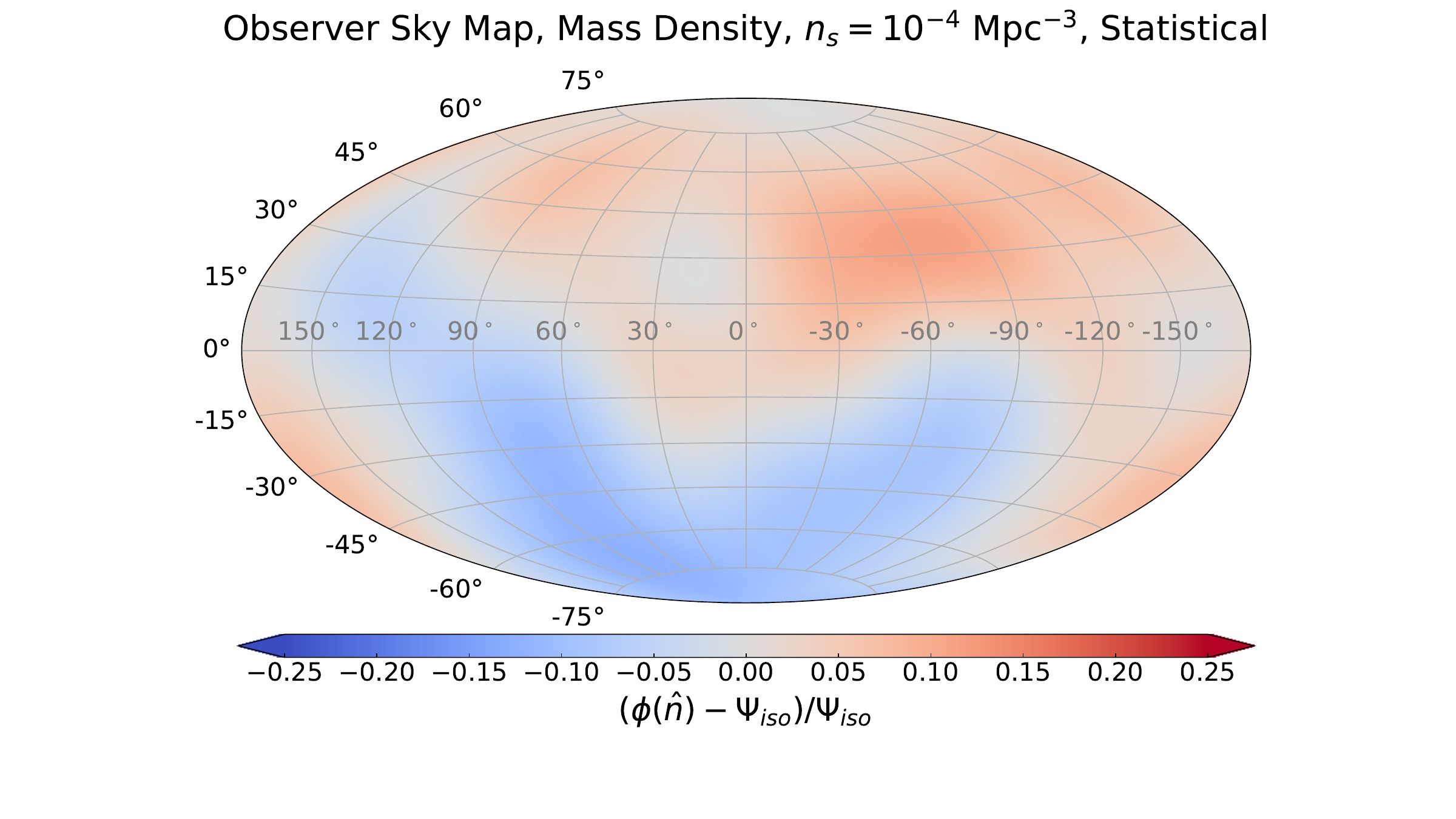}
\end{minipage}
\begin{minipage}{8.5cm}
\centering
\includegraphics[scale=0.2]{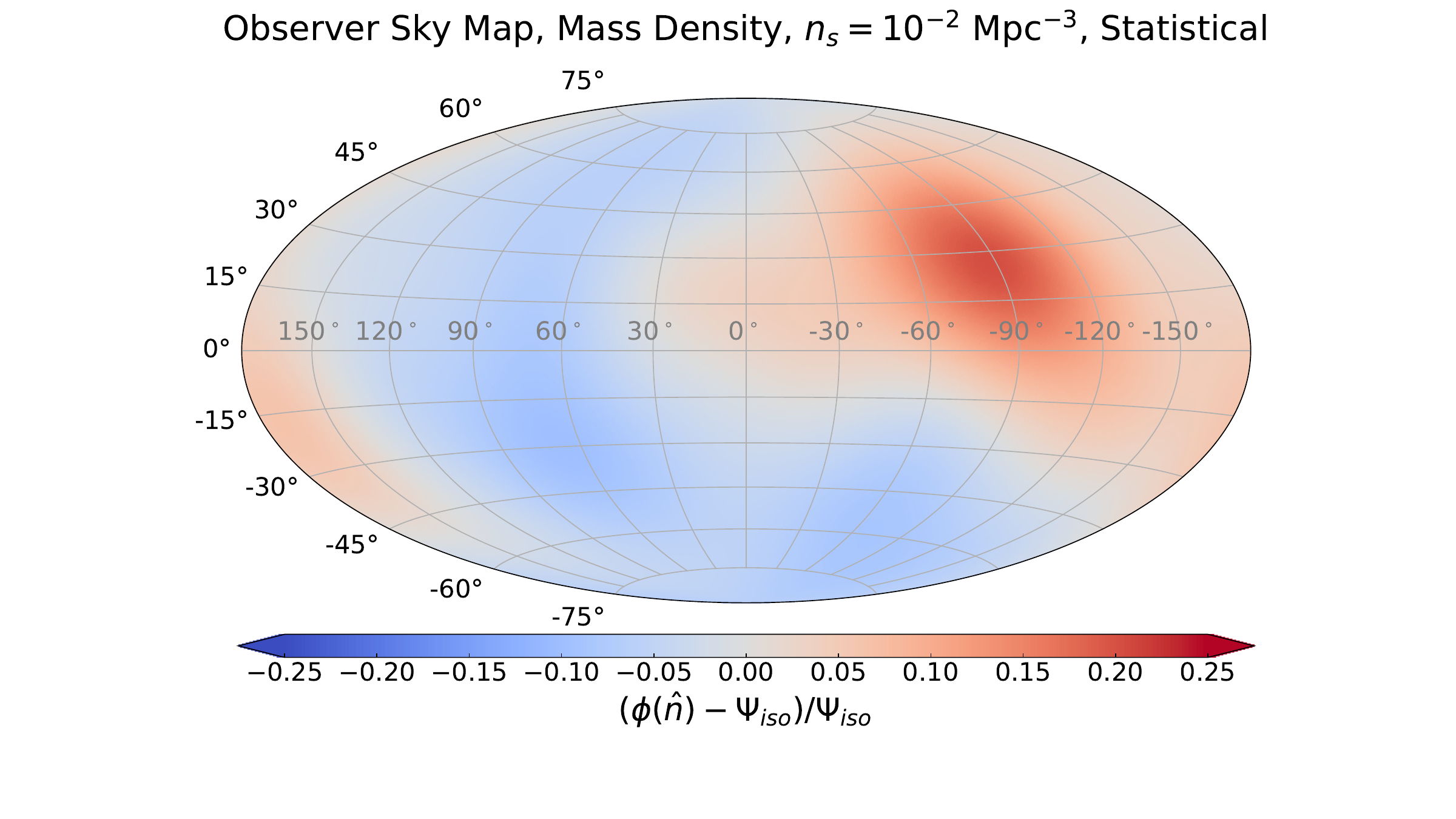}
\end{minipage}
\caption{Average maps of the fractional deviation from isotropy \eqref{frac_dev} of UHECRs above $8\,\text{EeV}$ obtained with the $10$ realizations of the \textit{mass density} source catalogues. Sky maps obtained at the surface of the observer (i.e. at the edge of the galaxy) in galactic coordinates. The \textit{astrophysicalR} (second row), \textit{primordial2R} (third row) and \textit{statistical} (fourth row) EGMF models are shown together with the ballistic case (first row). The left column refers to the source density $n_s=10^{-4}\,\text{Mpc}^{-3}$ and the right column to $n_s=10^{-2}\,\text{Mpc}^{-3}$. We apply a Gaussian smoothing to the sky maps with a full width at half maximum (FWHM) of $45^\circ$. The color bar indicates the fractional deviation from isotropy from $-0.25$ to $0.25$.}
\label{sky_maps_den}
\end{figure*}
\par To better quantify the dependence of the magnetic horizon on the EGMF, in Fig.~\ref{number_sources_den} we show the relative contribution $dF$ to the UHECR flux per source distance interval $dr$ as a function of the distance $r$ to the observer for all the EGMFs (colored lines). The low density and high density scenarios are shown in the upper and lower panels of Fig.~\ref{number_sources_den}, respectively. We also show the $dF/dr$ obtained with simulations where UHECR interactions with background photon fields are neglected (dashed lines). These lines correspond to the magnetic horizon suppression given by the EGMF alone. The shaded areas correspond to the cosmic variance of the source catalogue realizations. As expected, the scenarios without both interactions and magnetic field show the same contribution from near and distant sources (i.e. flux contributions per distance bin that are roughly constant). The small suppression obtained at small distances is due to the fixed finite value of the source densities which gives rise to a minimal source distance. We see that the more magnetized models introduce a suppression of the maximum source distance, which becomes stronger as the magnetization of the voids increases (e.g. in the \textit{primordial2R} model). This effect is even more pronounced when interactions are considered in the simulations, due to the production of low rigidity cosmic rays during the extragalactic propagation. In particular, the maximum distance obtained in the ballistic case of few Gpc is reduced to less than $1\,\text{Gpc}$. As a result, an increased contribution from sources at distances $10\,\text{Mpc}\lesssim r \lesssim 100\,\text{Mpc}$ is observed, which is almost independent of the source distribution model or density considered. Furthermore, for distances $r\lesssim20\,\text{Mpc}$ the cosmic variance of the source catalogues dominates the distribution of the observed sources when $n_s=10^{-4}\,\text{Mpc}^{-3}$.
\par The expected arrival direction probability distribution of a perfectly isotropic sky is given by $\Psi_\text{iso}=(4\pi)^{-1}$, while the obtained arrival direction distribution of our simulated scenarios is $\phi(\hat{\bold{n}})$, as defined in Eq.~\eqref{signal_function_text}. We then evaluate the deviation from isotropy by considering the fractional deviation of the arrival direction distribution $\phi(\hat{\bold{n}})$ from the isotropic prediction $\Psi_\text{iso}$, defined as
\begin{equation}
\label{frac_dev}
\delta_\phi (\hat{\bold{n}})=\dfrac{\phi (\hat{\bold{n}})-\Psi_\text{iso}}{\Psi_\text{iso}} \, .
\end{equation}
In Fig.~\ref{sky_maps_den} the average map of the fractional deviations $\delta_\phi(\hat{\bold{n}})$ on the observer surface obtained with the $10$ realizations of the \textit{mass density} scenarios are shown in galactic coordinates. The \textit{astrophysicalR} (second row), \textit{primordial2R} (third row) and \textit{statistical} (fourth row) EGMF models are shown together with the ballistic case (first row). The left column of Fig.~\ref{sky_maps_den} corresponds to the source density $n_s=10^{-4}\,\text{Mpc}^{-3}$ and the right column to $n_s=10^{-2}\,\text{Mpc}^{-3}$.  We apply a Gaussian smoothing to the sky maps in Fig.~\ref{sky_maps_den} with a full width at half maximum (FWHM) of $45^\circ$ to remove small scale anisotropies while preserving the large scale structure. In the ballistic case, deviations from the isotropic expectation are observed due to the combination of the intrinsic structure of the source distribution and the interaction horizon. In particular, we see a hotspot of events at galactic coordinates $\sim(-60^\circ,30^\circ)$ with $\delta_\phi \gtrsim0.25$. As observed for the mean angular deflection, the position and structure of the hotspot is influenced by the magnetization level of the cosmic voids. Furthermore, the source density also affects the size of the UHECR excess flux in directions showing an excess. In the \textit{astrophysicalR} model the relative flux excess is reduced to $\delta_\phi\simeq0.15$ when $n_s=10^{-4}\,\text{Mpc}^{-3}$, while it is almost unchanged in the high density scenario. This is a direct consequence of the number of local sources, as shown in Fig.~\ref{source_distance}. The \textit{primordial2R} EGMF model introduces a general shift of the hotspot to the galactic coordinates $\sim(-90^\circ,45^\circ)$, independent of the source density. Finally, when the \textit{statistical} EGMF is considered, we can see that the hotspot of UHECRs is strongly scattered across the sky and is only visible in the high density case, with a relative excess of $\delta_\phi\simeq0.15-0.20$.
\begin{figure}[t]
\centering
\begin{minipage}{8.5cm}
\centering
\includegraphics[scale=0.5]{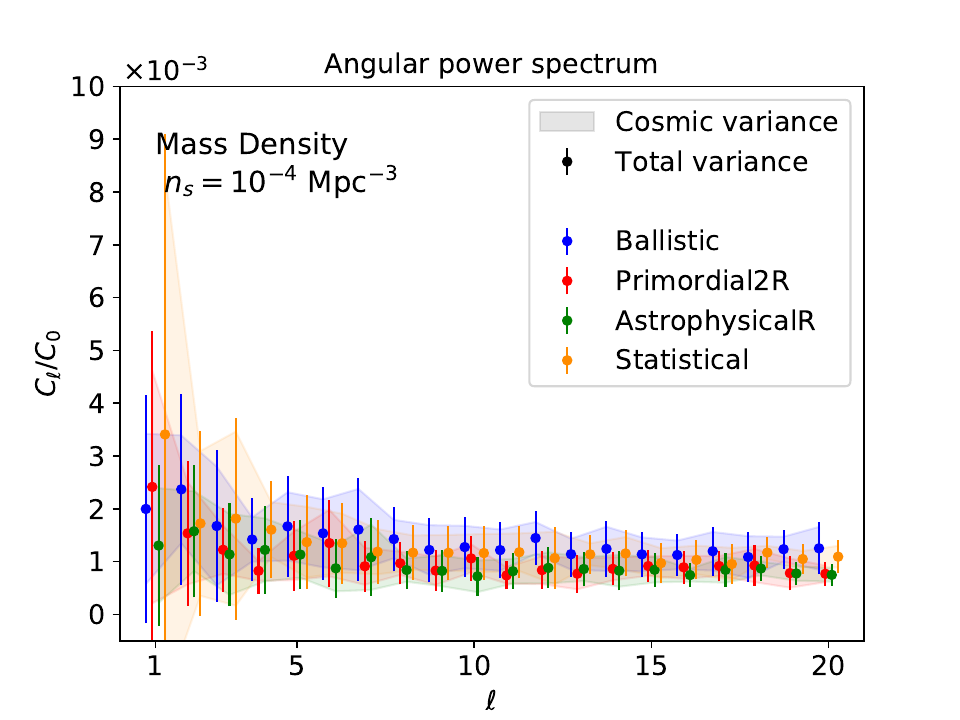}
\end{minipage}
\begin{minipage}{8.5cm}
\centering
\includegraphics[scale=0.5]{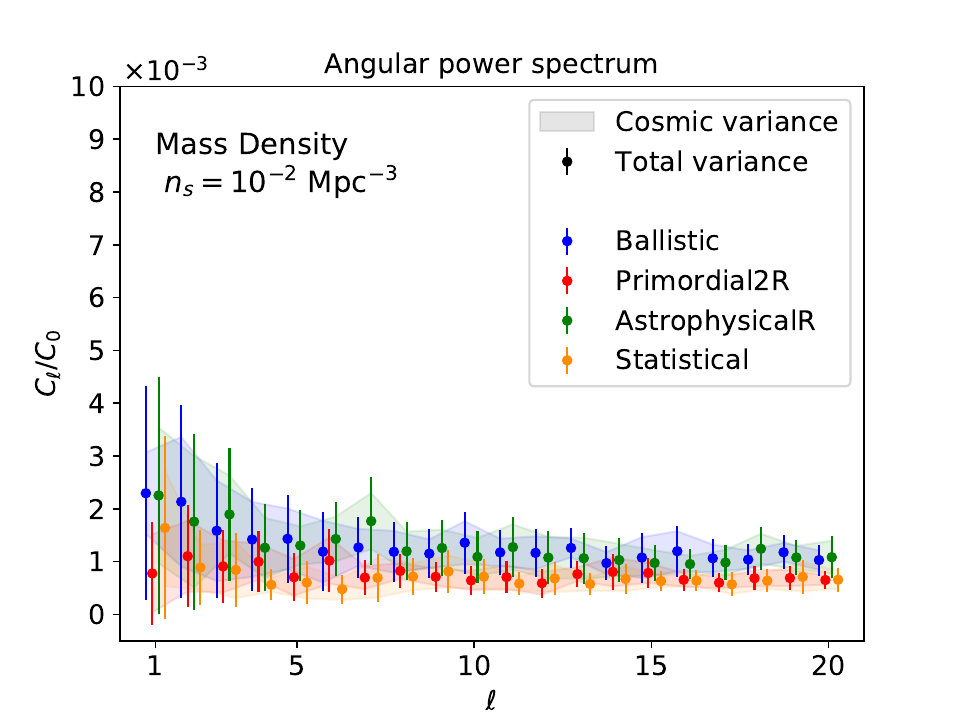}
\end{minipage}
\caption{Angular power spectra normalized to the monopole (i.e. $\mathcal{C}_l/\mathcal{C}_0$) of the UHECR arrival direction distributions above $8\,\text{EeV}$ for the \textit{mass density} source catalogues. The upper panel refers to the source density $n_s=10^{-4}\,\text{Mpc}^{-3}$ and the lower panel to $n_s=10^{-2}\,\text{Mpc}^{-3}$. Different EGMF models are shown in different colors. The shaded areas correspond to the cosmic variance and the error bar to the total variance (cosmic plus statistical variance from Eq.~\eqref{stat_var_C_l}).}
\label{angular_power_spectrum_den}
\end{figure}
\begin{figure*}[t]
\centering
\begin{minipage}{8.5cm}
\centering
\includegraphics[scale=0.2]{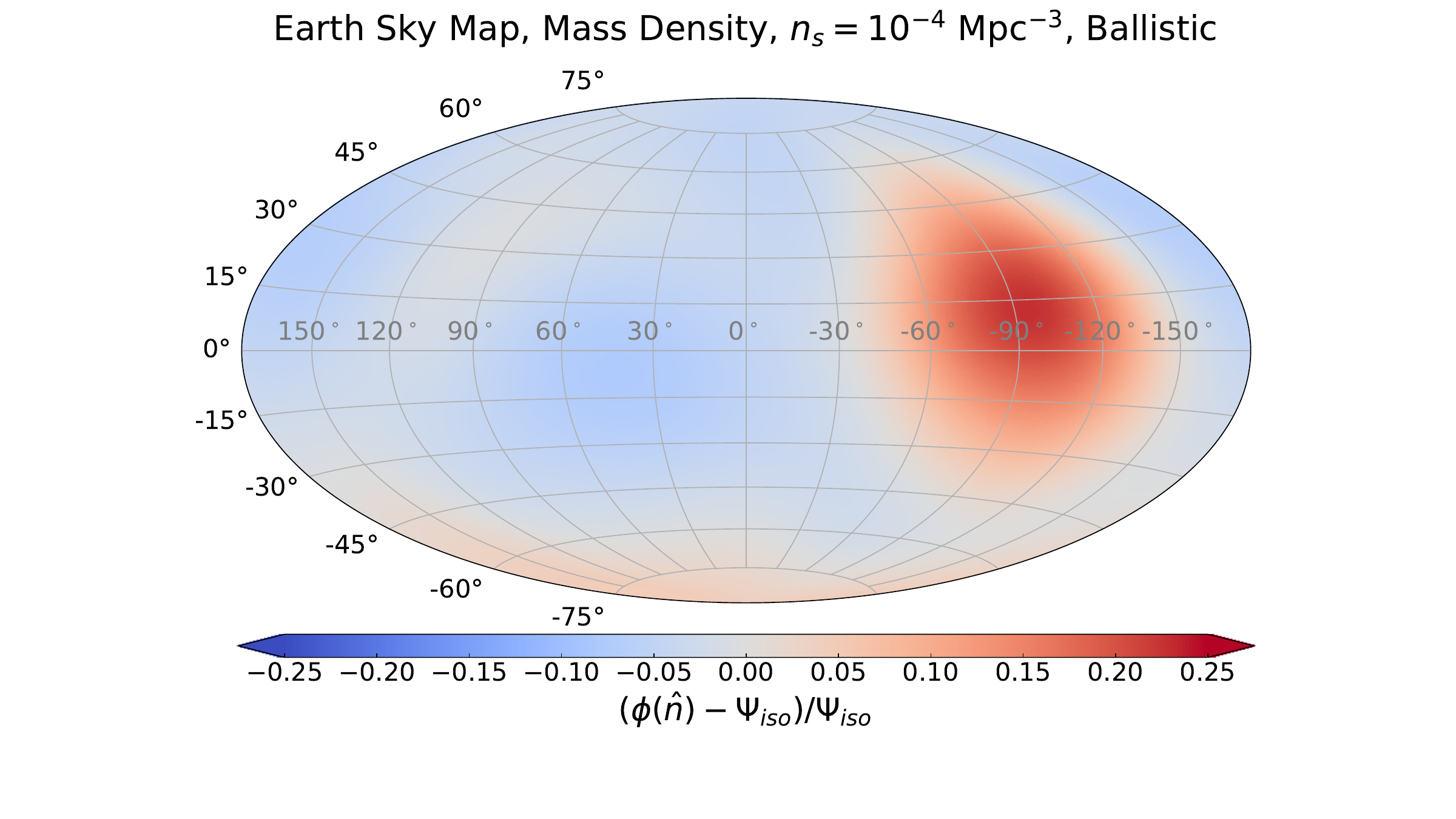}
\end{minipage}
\begin{minipage}{8.5cm}
\centering
\includegraphics[scale=0.2]{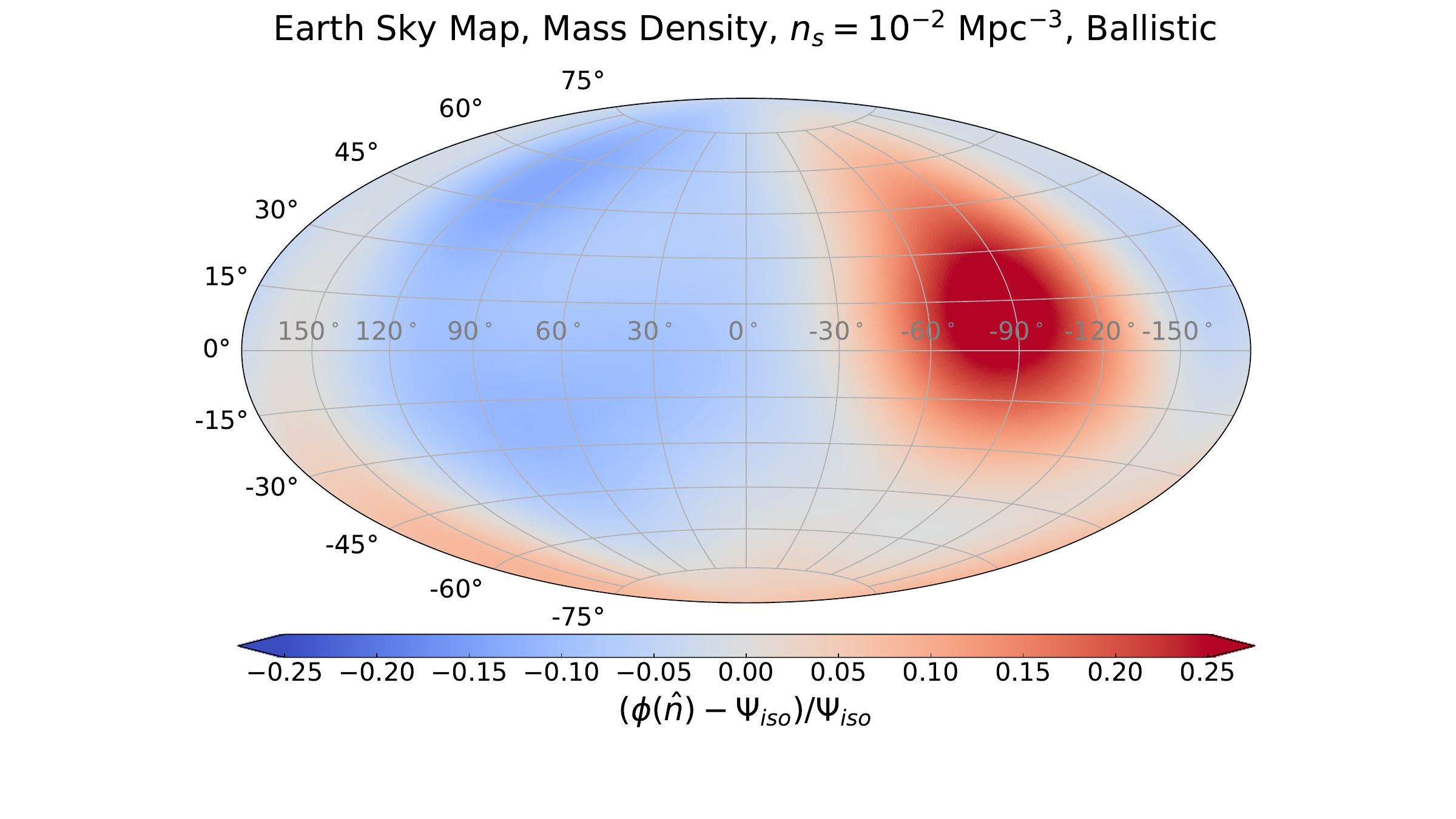}
\end{minipage}
\begin{minipage}{8.5cm}
\centering
\includegraphics[scale=0.2]{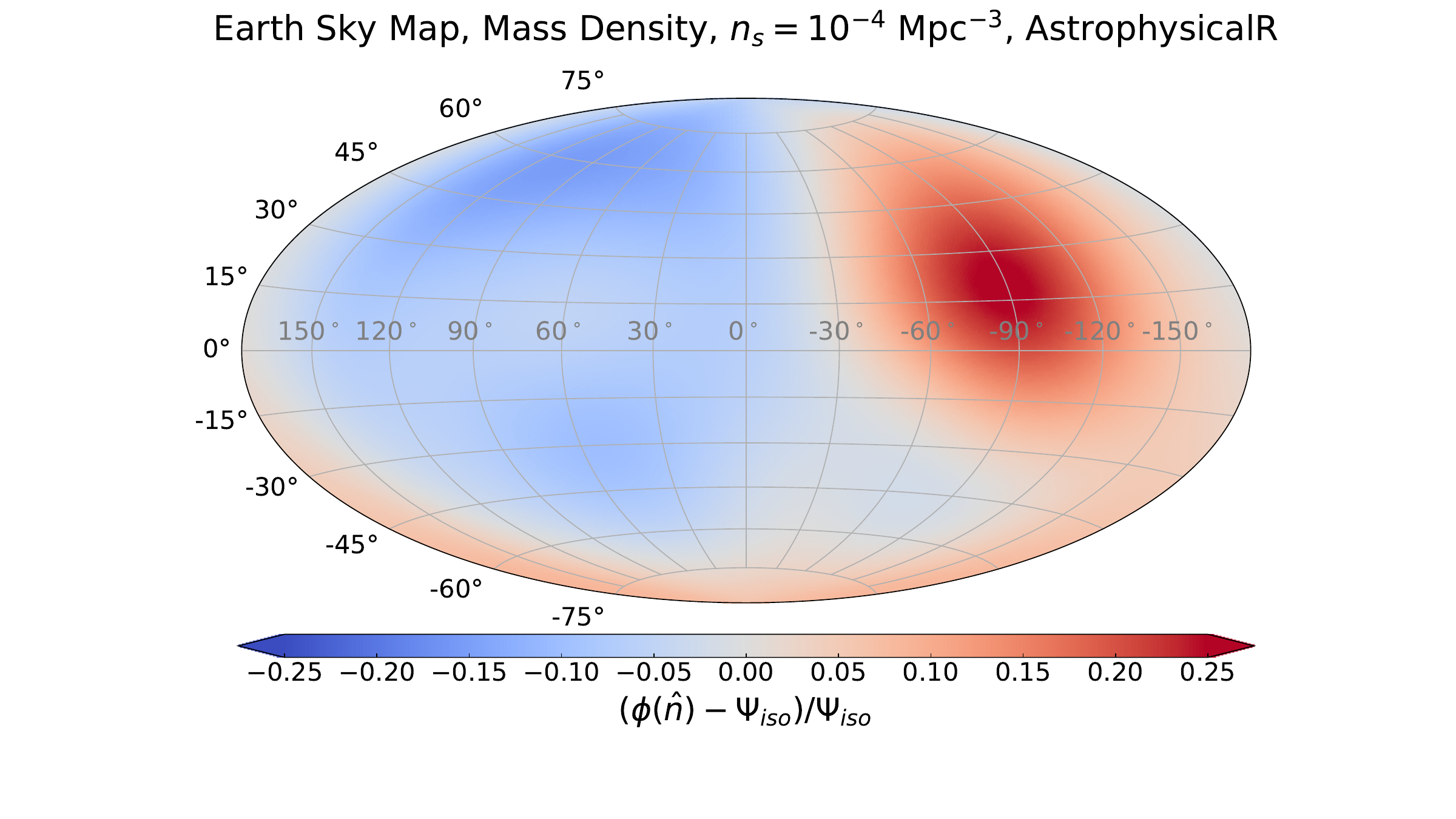}
\end{minipage}
\begin{minipage}{8.5cm}
\centering
\includegraphics[scale=0.2]{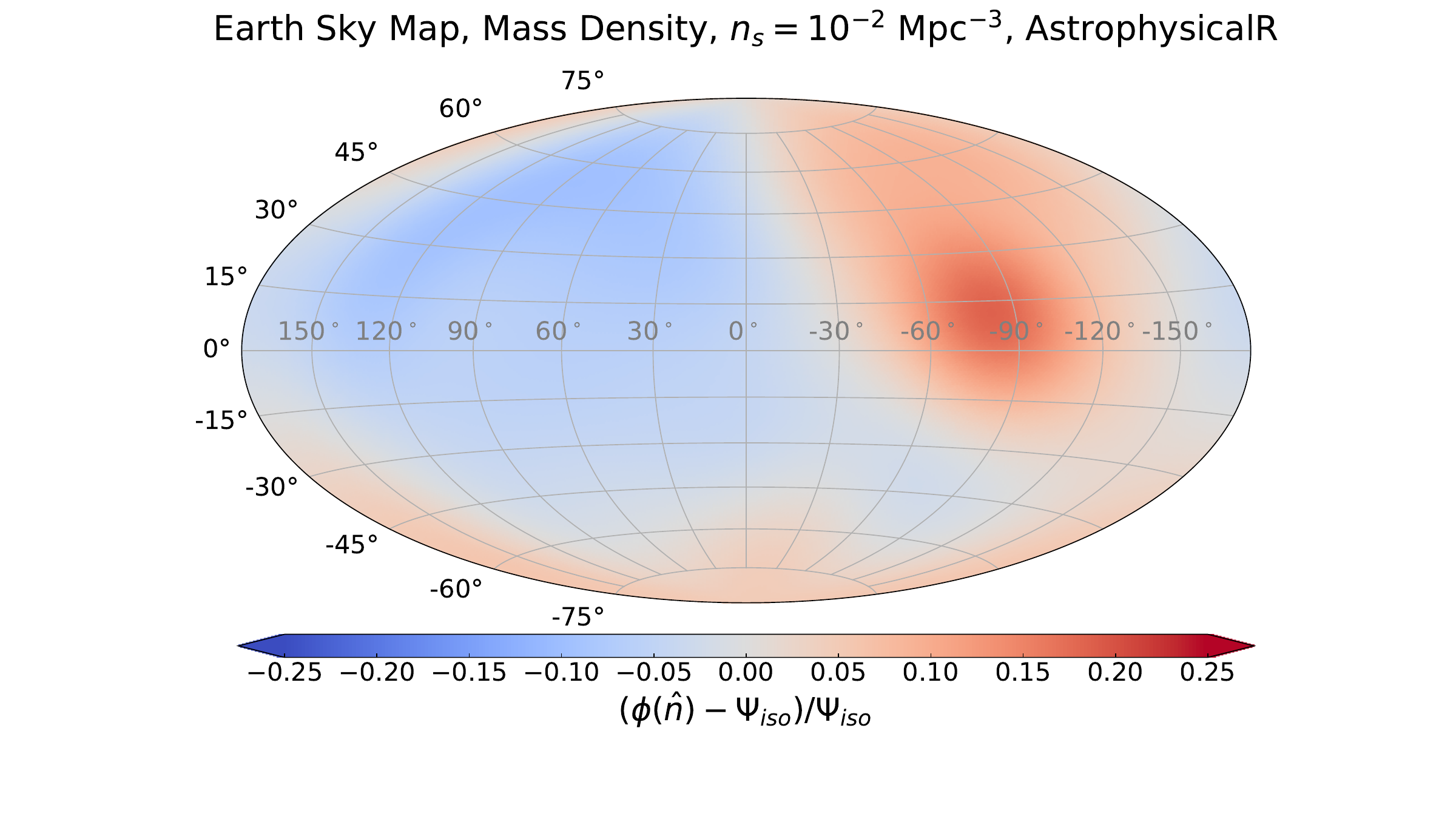}
\end{minipage}
\begin{minipage}{8.5cm}
\centering
\includegraphics[scale=0.2]{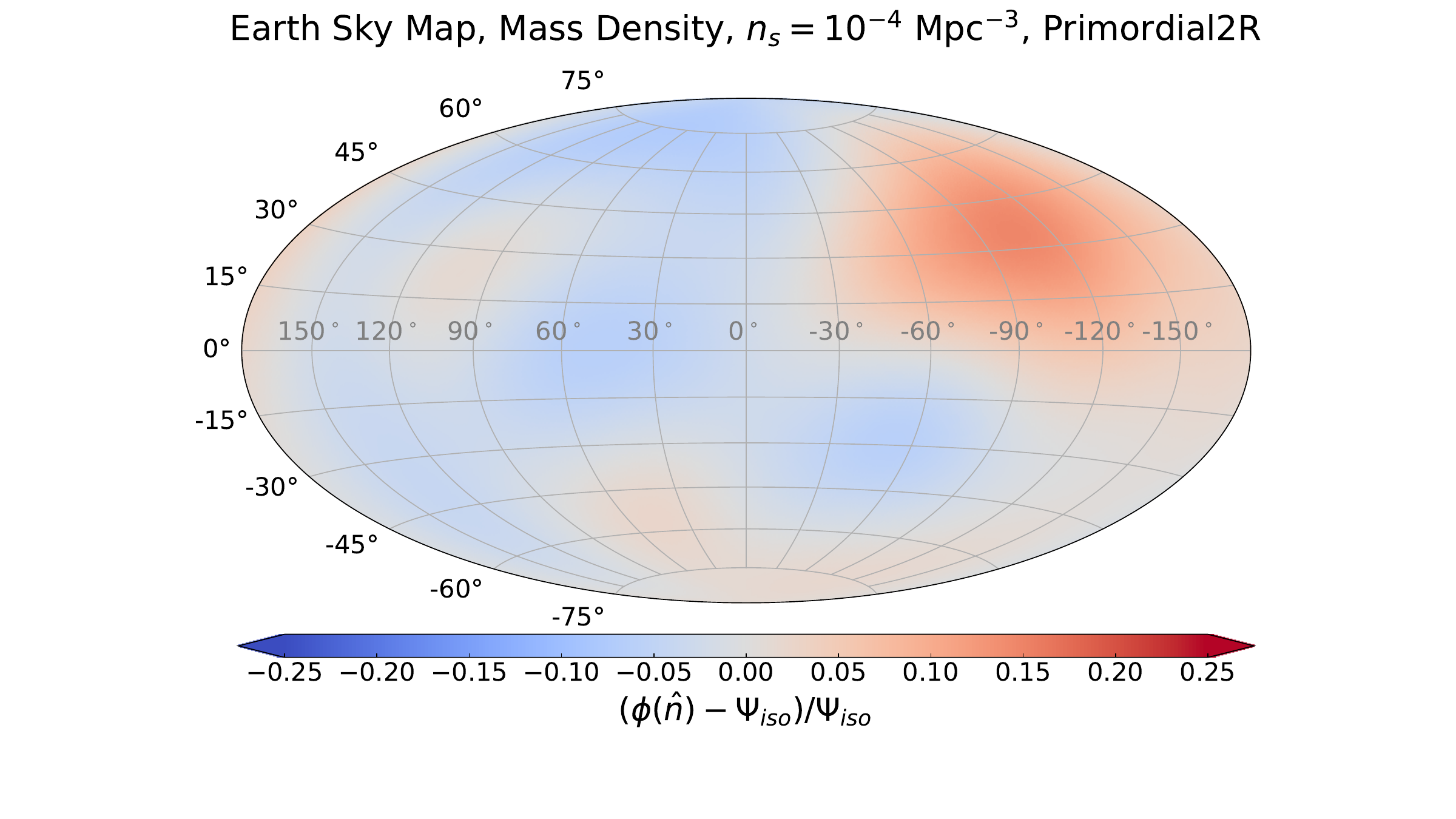}
\end{minipage}
\begin{minipage}{8.5cm}
\centering
\includegraphics[scale=0.2]{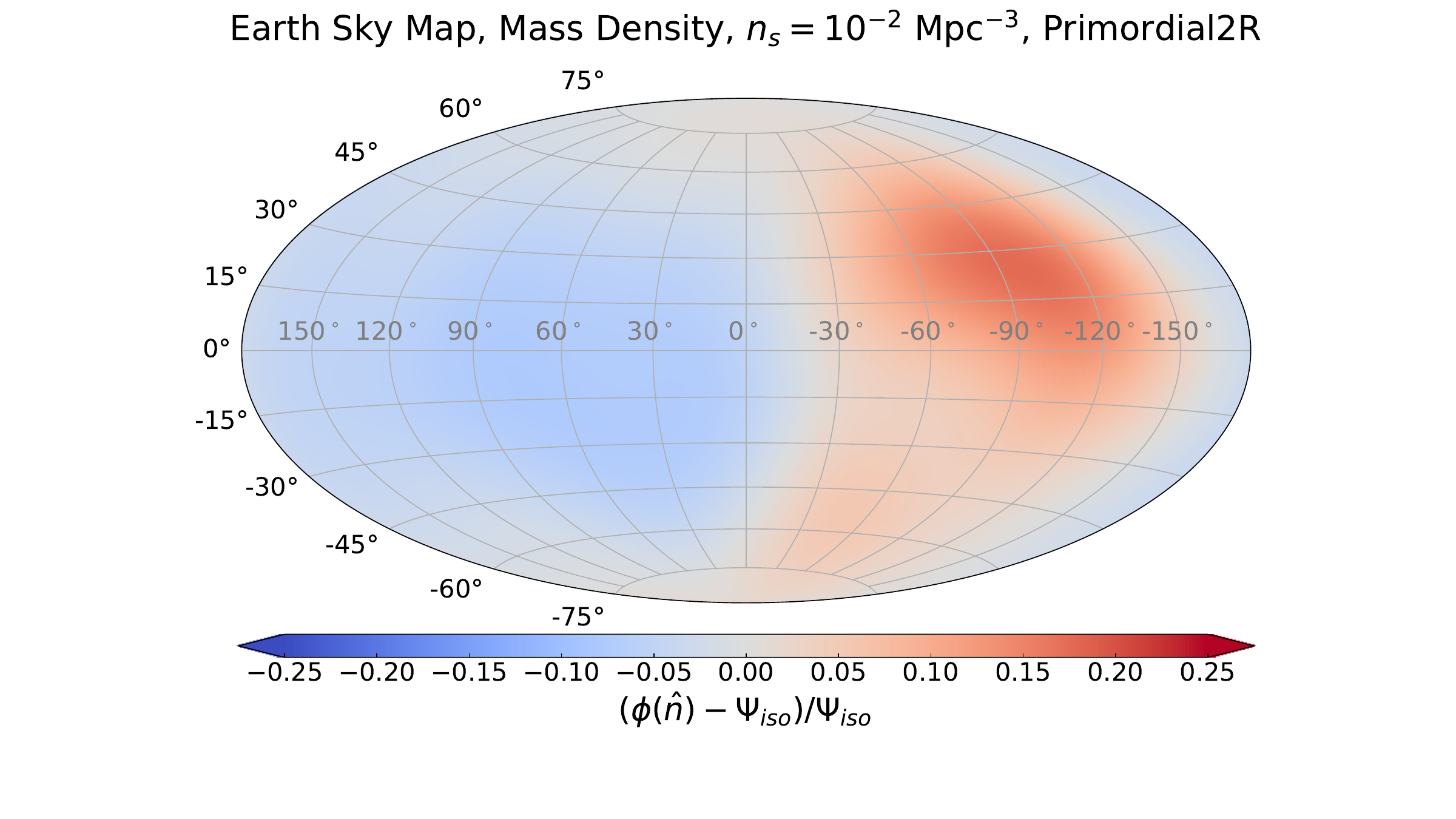}
\end{minipage}
\begin{minipage}{8.5cm}
\centering
\includegraphics[scale=0.2]{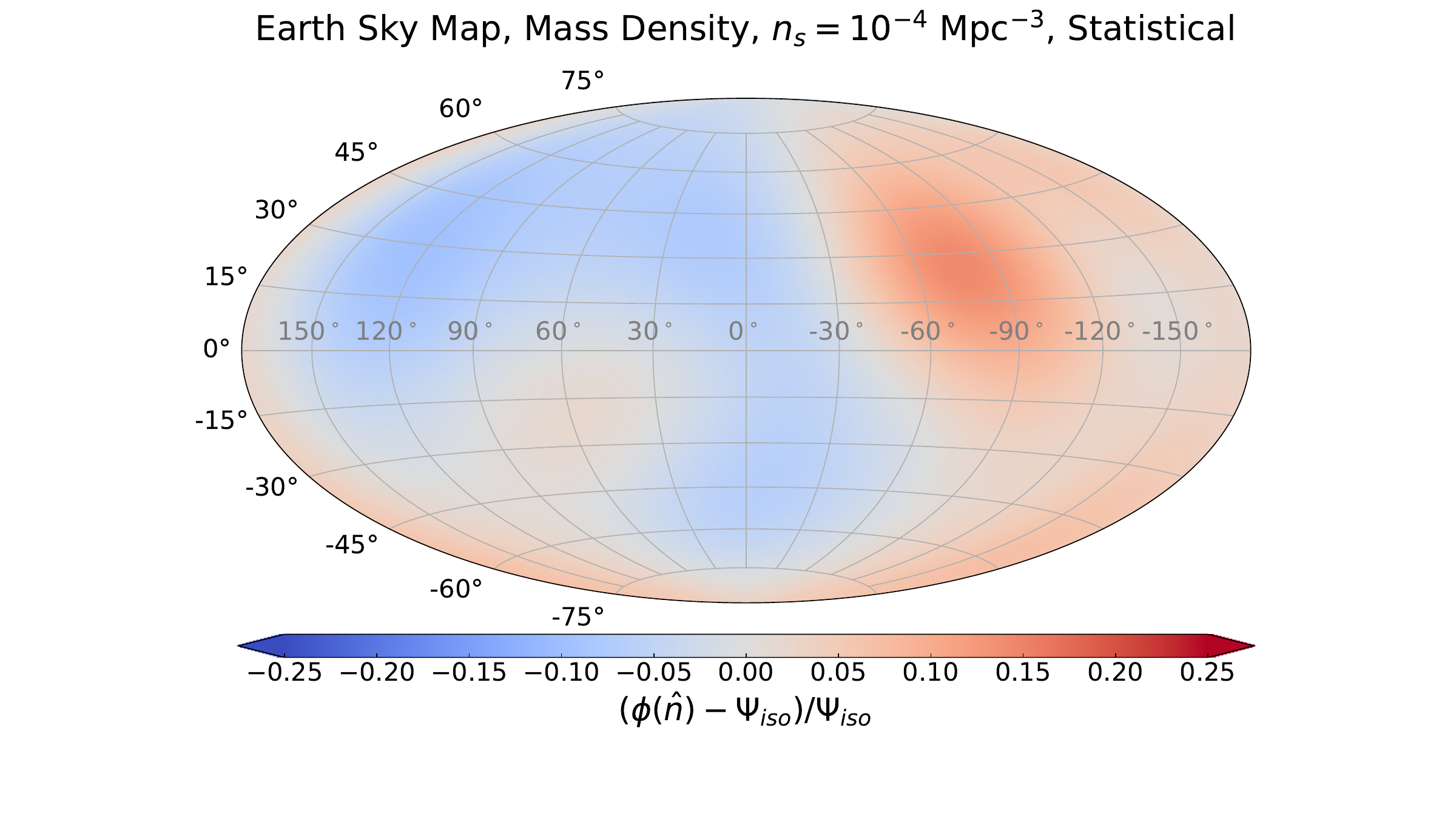}
\end{minipage}
\begin{minipage}{8.5cm}
\centering
\includegraphics[scale=0.2]{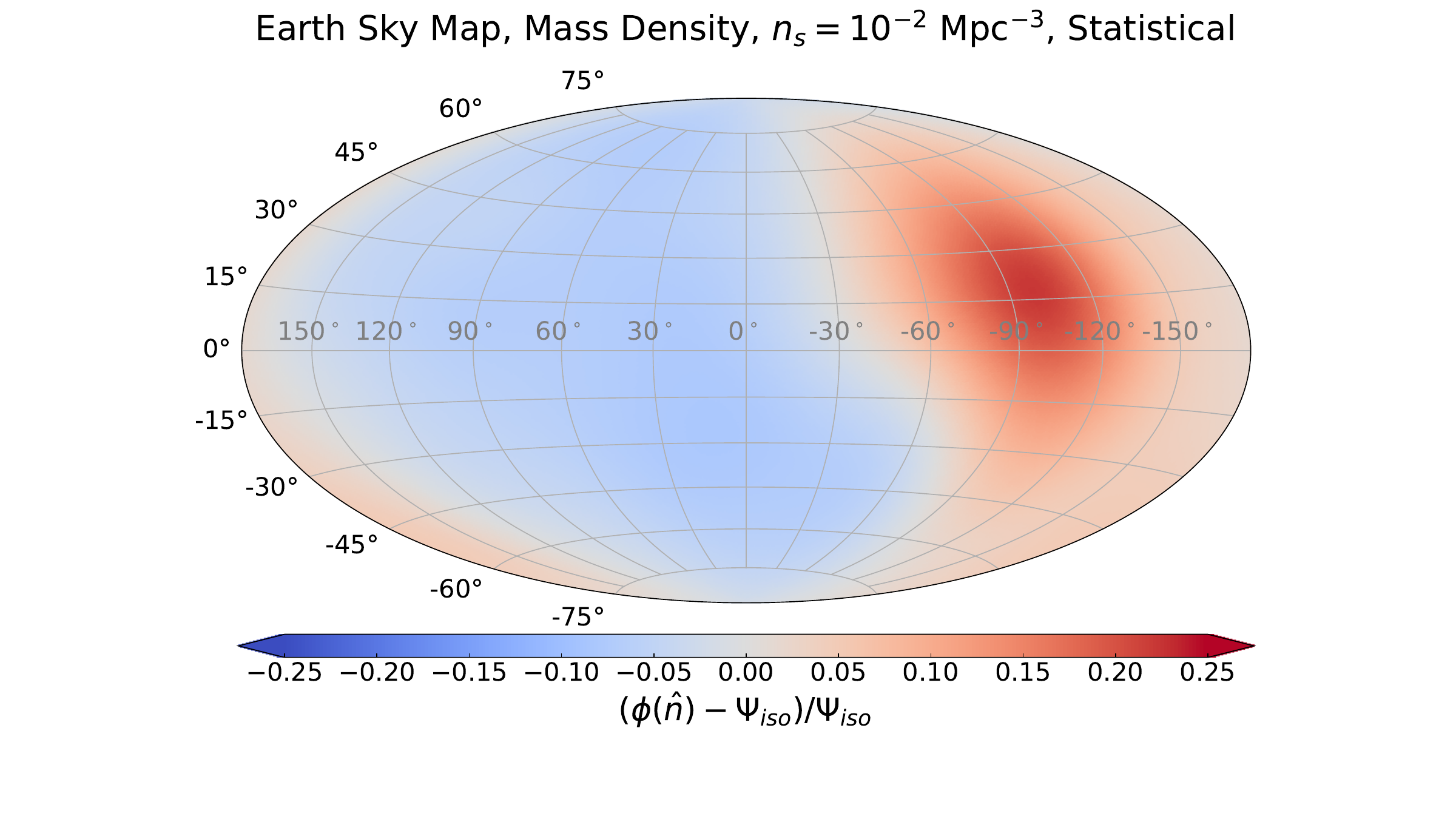}
\end{minipage}
\caption{Average maps of the fractional deviation from isotropy \eqref{frac_dev} of UHECRs above $8\,\text{EeV}$ obtained with the $10$ realizations of the \textit{mass density} source catalogues. Sky maps obtained at Earth after the propagation in the JF12 GMF model. Maps in galactic coordinates. The \textit{astrophysicalR} (second row), \textit{primordial2R} (third row) and \textit{statistical} (fourth row) EGMF models are shown together with the ballistic case (first row). The left column refers to the source density $n_s=10^{-4}\,\text{Mpc}^{-3}$ and the right column to $n_s=10^{-2}\,\text{Mpc}^{-3}$. We apply a Gaussian smoothing to the sky maps with a full width at half maximum (FWHM) of $45^\circ$. The color bar indicates the fractional deviation from isotropy from $-0.25$ to $0.25$.}
\label{sky_maps_den_lensed}
\end{figure*}
\begin{figure}[t]
\centering
\begin{minipage}{8.5cm}
\centering
\includegraphics[scale=0.5]{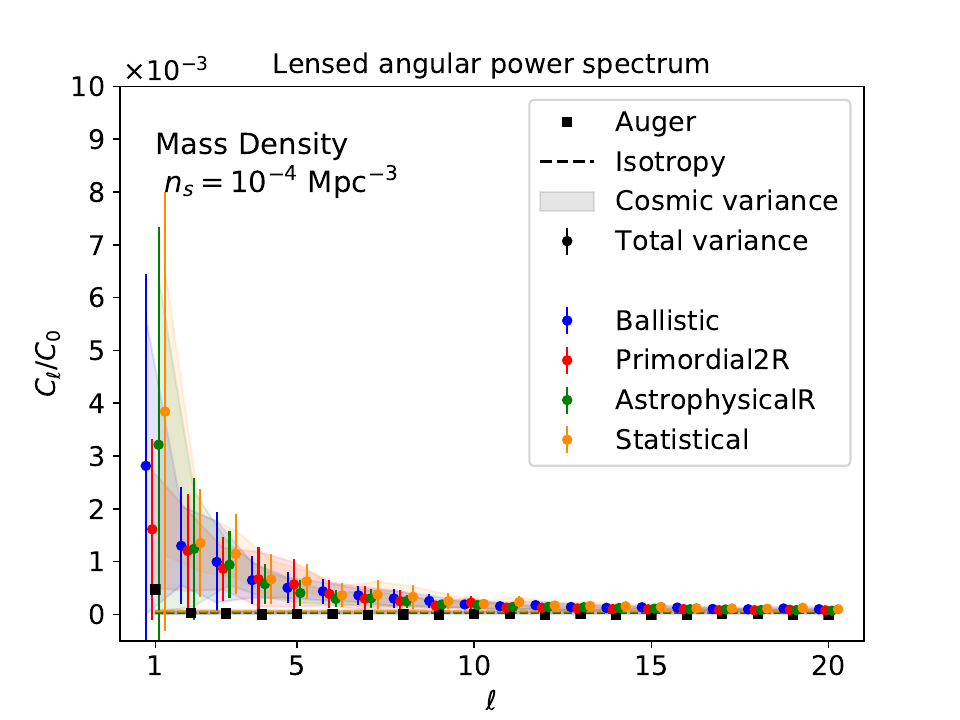}
\end{minipage}
\begin{minipage}{8.5cm}
\centering
\includegraphics[scale=0.5]{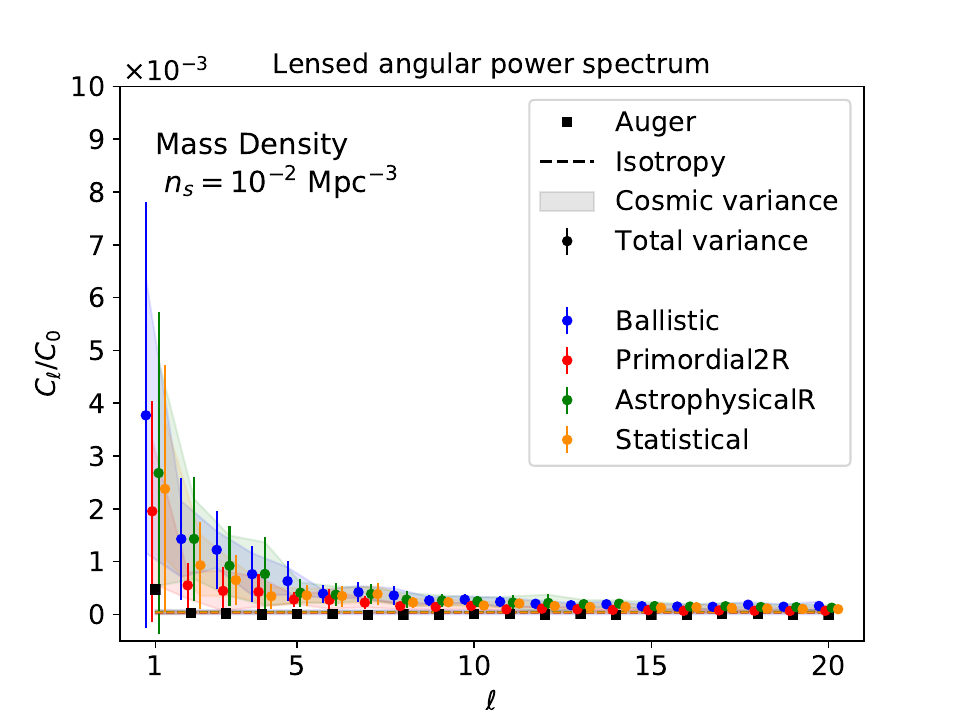}
\end{minipage}
\caption{Lensed angular power spectra normalized to the monopole (i.e. $\mathcal{C}_l/\mathcal{C}_0$) of the UHECR arrival direction distributions above $8\,\text{EeV}$ for the \textit{mass density} source catalogues. The upper panel refers to the source density $n_s=10^{-4}\,\text{Mpc}^{-3}$ and the lower panel to $n_s=10^{-2}\,\text{Mpc}^{-3}$. Different EGMF models are shown in different colors. The shaded areas correspond to the cosmic variance and the error bar to the total variance (cosmic plus statistical variance from Eq.~\eqref{stat_var_C_l}). The dashed lines correspond to the isotropic theoretical predictions from Eq.~\eqref{C_l_iso_text} of the scenarios considered. The black squares correspond to the observed angular power spectrum above $8\,\text{EeV}$ from \cite{PierreAuger:2024fgl}.}
\label{angular_power_spectrum_den_lensed}
\end{figure}
\begin{figure}[t]
\centering
\begin{minipage}{8.5cm}
\centering
\includegraphics[scale=0.5]{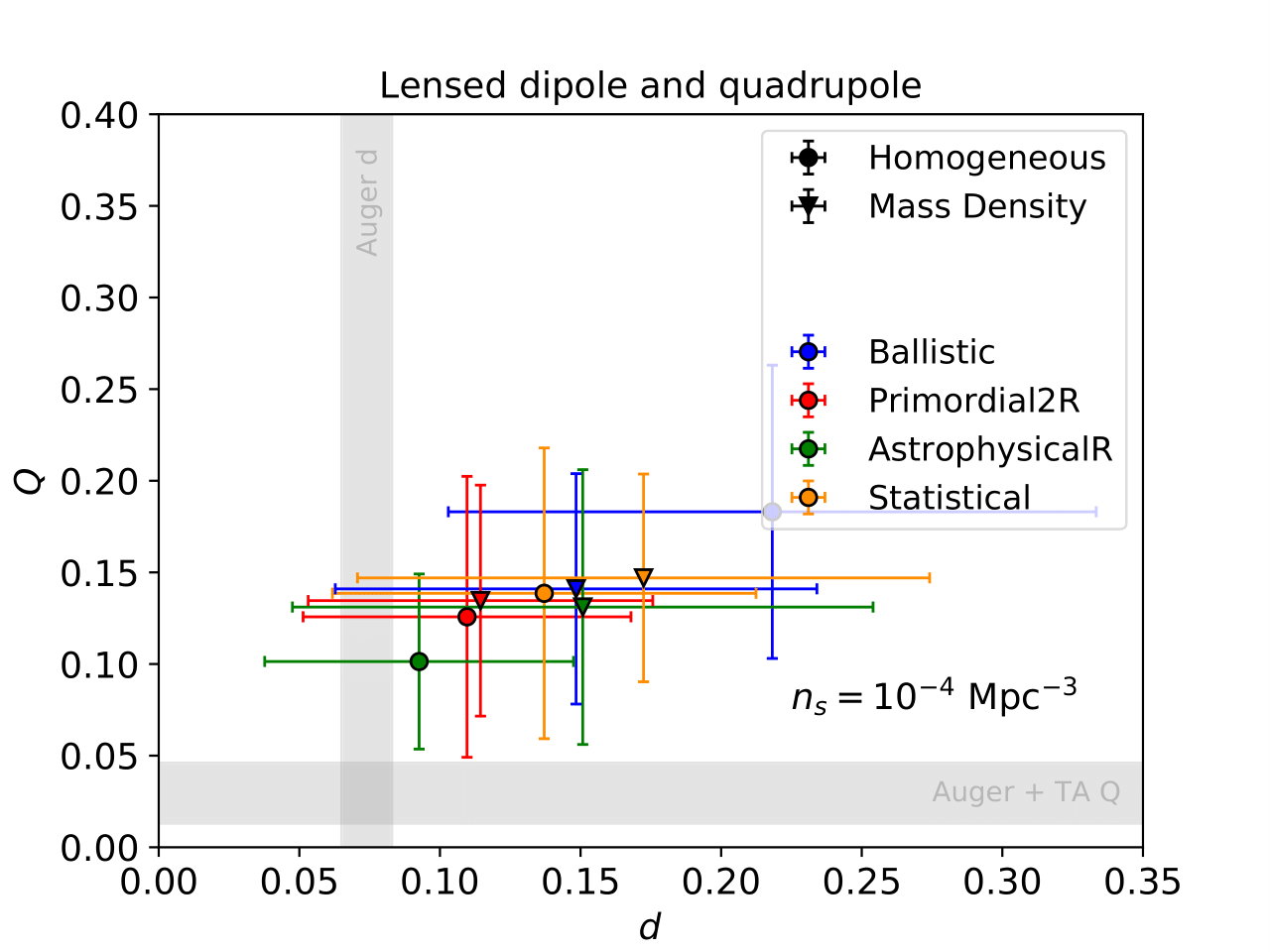}
\end{minipage}
\begin{minipage}{8.5cm}
\centering
\includegraphics[scale=0.5]{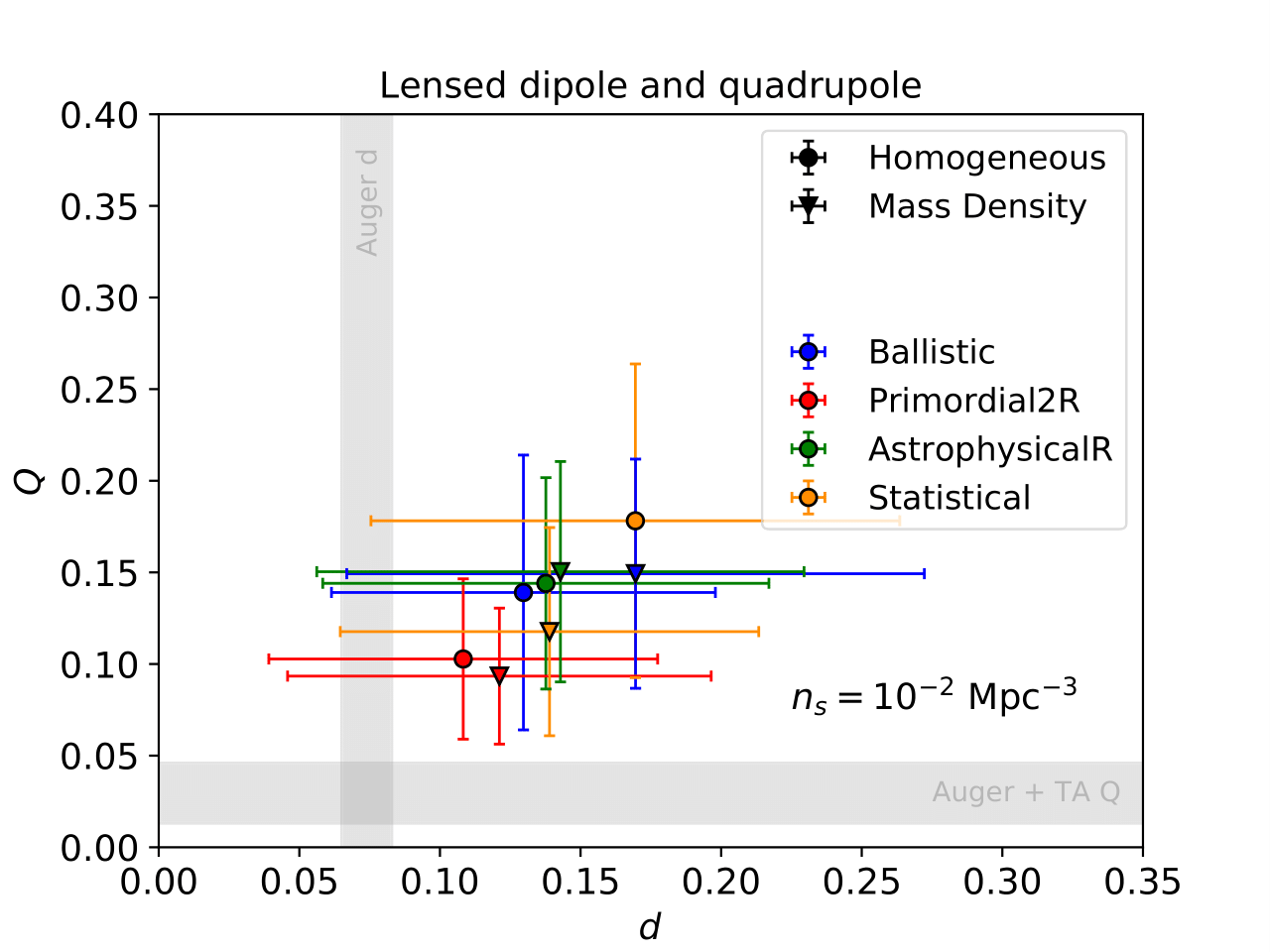}
\end{minipage}
\caption{Simulated lensed dipoles and quadrupoles above $8\,\text{EeV}$ obtained with Eqs.~\eqref{dip_amplitude} and~\eqref{quad_amplitude} shown as circles for the \textit{homogeneous} scenarios and triangles for the \textit{mass density}. The upper panel refers to the source density $n_s=10^{-4}\,\text{Mpc}^{-3}$ and the lower panel to $n_s=10^{-2}\,\text{Mpc}^{-3}$. Different EGMF models are shown in different colors. The error bars correspond to the total variance. The gray regions correspond to the observed dipole and quadrupole from \cite{PierreAuger:2023fcr} and \cite{PierreAuger:2023mvf}, respectively.}
\label{dipole_quadrupole_lensed}
\end{figure}
\par The level of anisotropy of the observed UHECR sky can be quantified by calculating the angular power spectrum $\mathcal{C}_l$ of the sky distributions $\phi(\hat{\bold{n}})$. To be independent of the normalization of $\phi(\hat{\bold{n}})$, we study the ratio $\mathcal{C}_l/\mathcal{C}_0$ (see Sec.~\ref{subsec_analysis} for more details). Fig.~\ref{angular_power_spectrum_den} shows the angular power spectra for the \textit{mass density} scenarios normalised to the monopole for the two source densities considered. We show the cosmic variance (shaded areas) and the total variance (error bars, cosmic plus statistical variance from Eq.~\eqref{stat_var_C_l}). It can be seen that the low multipole components of the angular power spectra are most affected by the value of the source density. As expected, a low source density corresponds to a slightly higher angular power spectrum and its cosmic variance, especially for $l\lesssim3$, due to the smaller number of nearby sources and their intrinsically structured distribution. Interestingly, we also observe this feature in the \textit{homogeneous} scenarios characterized by ballistic propagation (Fig.~\ref{angular_power_spectrum_homo}). As can be seen in Fig.~\ref{source_distance}, the \textit{homogeneous} catalogues contain more sources at small distances than the \textit{mass density} catalogues. Despite the structured distribution of the \textit{mass density} catalogues, the larger number of local sources in the \textit{homogeneous} catalogues has a direct impact on the UHECR multipole distribution. The fluctuation in the angular power spectra is due more to the cosmic variance of the source catalogue than to the statistical fluctuation due to the finite number of observed UHECRs. However, in all scenarios the angular power spectrum decreases at small angular sales (i.e. large $l$) because on small scales the average source distribution is rather smooth and only varies significantly on relatively larger scales. The same effect is also observed for the cosmic variance, due to the cosmological principle.  In fact, as we look at larger and larger volumes around the observer, the differences in multiple samples of the source catalogues become negligible. 
\par As previously discussed in Sec.~\ref{subsec_analysis}, we account for the influence of the galactic magnetic field by applying the magnetic lensing based on the JF12 GMF model to the UHECR sky at the edge of the galaxy. Fig.~\ref{sky_maps_den_lensed} shows the fractional deviation from isotropy of the lensed sky distributions. The sky maps are in galactic coordinates and organized as in Fig.~\ref{sky_maps_den}. We can see that the main hotspot present in the sky at the edge of the galaxy is still present after the propagation in the GMF, with basically the same relative excess $\delta_\phi$. We also observe a general shift of the hotspot towards lower galactic latitudes and a smearing of the hotspot angular size. Furthermore, we can clearly see that the GMF suppresses most of the small angular scale deviations present in Fig~\ref{sky_maps_den}. Only the large angular scale deviations are still present in the UHECR sky above $8\,\text{EeV}$. The effect of the galactic lensing is also visible in the angular power spectra of the lensed sky maps, shown in Fig.~\ref{angular_power_spectrum_den_lensed}. The angular power spectra are shown together with the theoretical prediction for an isotropic sky obtained with Eq.~\eqref{C_l_iso_text} (dashed lines) and the observed multipoles from \cite{PierreAuger:2024fgl} (black squares). The GMF introduces a strong suppression of multipoles with $l>5$, already visible in the lensed maps where small angular scale overdensities are suppressed by the GMF. The large multipole components of the lensed angular power spectra converge to the corresponding isotropic prediction as $l$ increases. However, it can be seen that the small multipole components of the lensed spectra (especially dipoles and quadrupoles) are larger than those observed. 
\par Fig.~\ref{dipole_quadrupole_lensed} shows the obtained dipole and quadrupole moments for the simulated scenarios after accounting for the GMF effect. The dipoles and quadrupoles are obtained from the lensed angular power spectra in Figs.~\ref{angular_power_spectrum_den_lensed} and~\ref{angular_power_spectrum_homo_lensed} with Eqs.~\eqref{dip_amplitude} and~\eqref{quad_amplitude}, respectively. The error bars indicate the total variance and the gray areas the observed dipole from \cite{PierreAuger:2023fcr} and quadrupole from \cite{PierreAuger:2023mvf}. It can be observed that in the \textit{mass density} scenarios we obtain on average larger dipole values than in the \textit{homogeneous} cases for $n_s=10^{-4}\,\text{Mpc}^{-3}$. This effect is less present in the high source density case due to the much larger number of local sources. This is again the result of the structured distribution of the local sources in our catalogues. We do not observe any significant influence of the EGMF model on the predicted dipole and quadrupole moments. It can be seen that most of the dipoles are compatible with observations only within the total variance, while the quadrupole moments are generally too large to be compatible with observations.

\section{Discussion and conclusions}
\label{sec_discussion_conclusions}
In this work, we have studied the propagation of UHECRs in constrained replicas of the local cosmic web by taking into account the effect of structured EGMF models from  \cite{Hackstein:2017pex}, as well as for a statistically uniform magnetic field with $B_\text{rms}=1\,\text{nG}$ and $\lambda_\text{c}=1\,\text{Mpc}$ and ballistic propagation. We have synthesised several cosmic replicas of source catalogues following homogeneous and structured distributions for low and high density scenarios. We have simulated the injection of five different nuclear species at the locations of sources with identical properties. We used the Monte Carlo code \texttt{CRPropa 3.2} to numerically integrate the equation of motion in a three-dimensional simulation volume. We considered periodic boundary conditions to account for the contribution of distant sources. Together with magnetic deflection, interactions with background photon fields, nuclear decay and adiabatic energy losses due to the expansion of the Universe were taken into account. We collect on the surface of our numerical observer (a sphere located at the position of the Milky Way with radius $R_\text{O}=1\,\text{Mpc}$) all the particles with energy still above $8\,\text{EeV}$. We combine the simulated species and re-weight the injection rate according to Eqs.~\eqref{best_fit_1} and~\eqref{best_fit_2} and we correct the observed arrival direction for the finite size of the observer (see Appendix.~\ref{appendix_ad_correction}). Finally, we account for the effect of the GMF by lensing the corrected arrival directions at the edge of the galaxy at Earth based on the GMF model JF12 \cite{Jansson:2012pc,Jansson:2012rt}.
\par To account for the cosmic variance in our results, we have considered $N_\text{R}=10$ realizations of the synthetic source catalogues. Although a larger number of realizations is possible in principle, three-dimensional simulations of UHECR propagation in structured magnetic fields are very time-consuming. As shown in Fig.~\ref{source_distance}, the cosmic variance of the number of sources within $\sim100\,\text{Mpc}$ from the observer is very small, and basically negligible for larger distances. Indeed, we find that the cosmic variance of the simulated UHECR observables does not affect the robustness of our results. The average number of sources within $100\,\text{Mpc}$ from the observer is $N_{\text{source,low}}\simeq4\cdot10^2$ for the low density catalogues and $N_{\text{source,high}}\simeq4\cdot10^4$ for the high density catalogues. Since the corresponding fluctuations in the number of sources within $100\,\text{Mpc}$ from the observer can be estimated as $\sqrt{N_{\text{source,low}}}\simeq20$ and $\sqrt{N_{\text{source,high}}}\simeq200$, we do not expect rare catalogue realizations characterized by large source overdensities to substantially change our conclusions.
\par In order to reproduce the observed UHECR energy spectrum and mass composition above $8\,\text{EeV}$, we found the spectral index $\gamma$, the rigidity cutoff $R_\text{cut}$ and the emitted energy fractions $I_A$ that minimize the variable $\chi^2=\chi^2_J+\chi^2_{\left< A\right>}$ defined in Eq.~\ref{chi_sq_def} for each combination of scenarios considered in this work. We found that a hard spectral index from $\gamma\simeq-1$ to $\gamma\simeq-3$ and a rigidity cutoff of $\log \left( R_\text{cut}/1\,\text{EV} \right)\simeq0.2$ are required to reproduce the observations. An injected composition dominated by nitrogen and silicon nuclei is also required. \textbf{However, the variance in the obtained injection parameters shows how different assumptions on the propagation scenario can strongly affect the predictions at the escape from the source environment.}
\par We have studied the total magnetic deflection due to the EGMF model by calculating the angle between the moment of the cosmic ray at the injection and at the observation at the surface of the observer. We have considered the distribution of the deflection angles as a function of the observed UHECR energy and source distance (see Figs.~\ref{deflection_den} and~\ref{deflection_homo}). Although deflections up to $\sim90^\circ$ are observed in all the EGMF scenarios, the level of magnetization of the cosmic voids clearly affects the distribution of the deflection angles in energy and distance. When the EGMF is mostly present in regions of high baryon density, large deflections are observed only for very low energy nuclei coming from distant sources. This is because cosmic rays are only deflected when they pass through magnetized regions, so large deflections correspond to large source distances. On the contrary, a high void magnetization corresponds to very large deflections even for cosmic rays emitted at $\sim100\,\text{Mpc}$ from the observer. Again, this is due to the fact that in these scenarios the UHECRs are deflected throughout their entire propagation.
\par The magnitude of the magnetization of the cosmic voids in the EGMF models also affects the maximum source distance of the observed UHECRs above $8\,\text{EeV}$. \textbf{We have found that the horizon resulting from the interaction with cosmic photon fields, obtained in the ballistic scenarios, is further reduced by the magnetic horizon, depending on the level of magnetisation of the cosmic voids.} In particular, the suppression is due to the large deflection experienced by the cosmic rays as they propagate through the magnetized voids. In the \textit{primordial2R} and \textit{statistical} cases, we find a suppression of the interaction horizon from $\sim500$ to $\sim100\,\text{Mpc}$. Correspondingly, the contribution of sources at distance $r\lesssim100\,\text{Mpc}$ increases with respect to the \textit{astrophysicalR} and ballistic cases. We have found that these results are not affected by the cosmic variance of the source catalogues. 
\par We have evaluated the probability distributions $\phi(\hat{\bold{n}})$ of the recorded UHECR arrival directions for our simulated scenarios. As expected, the intrinsic structure of the source distributions plays the major role in the arrival direction distributions. In the \textit{homogeneous} cases we do not observe any dominant hotspot or feature in the sky, while in the \textit{mass density} cases a significant excess in the sky was found with a fractional deviation from isotropy of $\delta_\phi\gtrsim0.1$. This effect was also observed in the angular power spectrum of the arrival direction distribution, calculated first at the edge of the galaxy. In Figs.~\ref{angular_power_spectrum_den} and~\ref{angular_power_spectrum_homo} it can be clearly seen how a structured source distribution corresponds to larger values of low multipoles ($l\lesssim3$) than in the \textit{homogeneous} scenarios. We have also confirmed that the main source of fluctuation in the multipole distribution is the cosmic variance of the source catalogues, which increases as the source density value decreases. No imprints of the considered EGMF model were found in the angular power spectrum distributions.
\par Finally, we repeated the analyses of the UHECR arrival directions, taking into account the deflection due to the GMF, considered with the magnetic lens of the JF12 model. The main effects of the GMF are the suppression of the small angular scale deviations observed at the edge of the galaxy, and a general shift of the main UHECR flux excess towards lower galactic latitudes. We found that all the angular power spectra of the considered scenarios converge to their corresponding isotropic prediction (obtained with Eq.~\eqref{C_l_iso_text}) for $l>5$. 
\par However, all the simulated scenarios show too large low multipoles compared to the Pierre Auger Observatory observations. In particular, it can be seen in Fig.~\ref{dipole_quadrupole_lensed} that \textbf{the dipole components are compatible with the observations only within the total variance, while the quadrupole components are generally too large}. This result is independent of the source density, distribution and EGMF model. This discrepancy between the simulated quadrupole amplitudes and the observations indicates the need to refine the assumptions of the model. In this study, we have considered only one GMF model and found that its main effect is to suppress the large multipole components of the angular power spectrum. The magnetic deflection angle of cosmic rays with energies above $8\,\text{EeV}$ in the GMF is $\theta\lesssim20^\circ$, corresponding to mutipoles of order $l\gtrsim10$. Therefore, we do not expect different GMF models to lead to a strong suppression of the quadrupole amplitude. Since the obtained dipole and quadrupole moments are independent of the EGMF model (even in the \textit{ballistic} case), one possible explanation for the observed UHECR anisotropies could be an intrinsic and almost purely dipolar distribution of their sources. This would put constraints on the general assumption that the UHECR sources follow the baryonic distribution in the local Universe.
\par The presence of cosmic magnetic fields and the distribution of UHECR sources are fundamental quantities for understanding the origin of these particles. Although we have not found explicit effects on the UHECR arrival directions, the details of their source distribution play a key role in the understanding of the observed anisotropies. In particular a source number density $>10^{-4}\,\text{Mpc}^{-3}$ seems to be preferred to explain both the observed dipole and the absence of higher multipoles. In the future, more efficient and high statistical three-dimensional simulations will be needed to include more cosmic replicas of the source catalogues and new EGMF models. In addition, future observations obtained with the upgrade of the Pierre Auger Observatory, AugerPrime \cite{Anastasi:2022vle}, and the Telescope Array, TAx4 \cite{TelescopeArray:2021dcx}, will provide more information to compare theoretical predictions with experimental results.

\begin{acknowledgments}
S.R. and G.S. acknowledge their participation to the Pierre Auger Collaboration. S.R. and G.S. acknowledge support by the Bundesministerium für Bildung und Forschung, under grants 05A20GU2 and 05A23GU3. This work is funded by the Deutsche Forschungsgemeinschaft (DFG, German Research Foundation) under Germany’s Excellence Strategy – EXC 2121 “Quantum Universe” – 390833306.
\end{acknowledgments}

\onecolumngrid

\appendix

\section{Energy loss lengths}
\label{appendix_ELL}
\begin{figure}[t]
\centering
\includegraphics[scale=0.7]{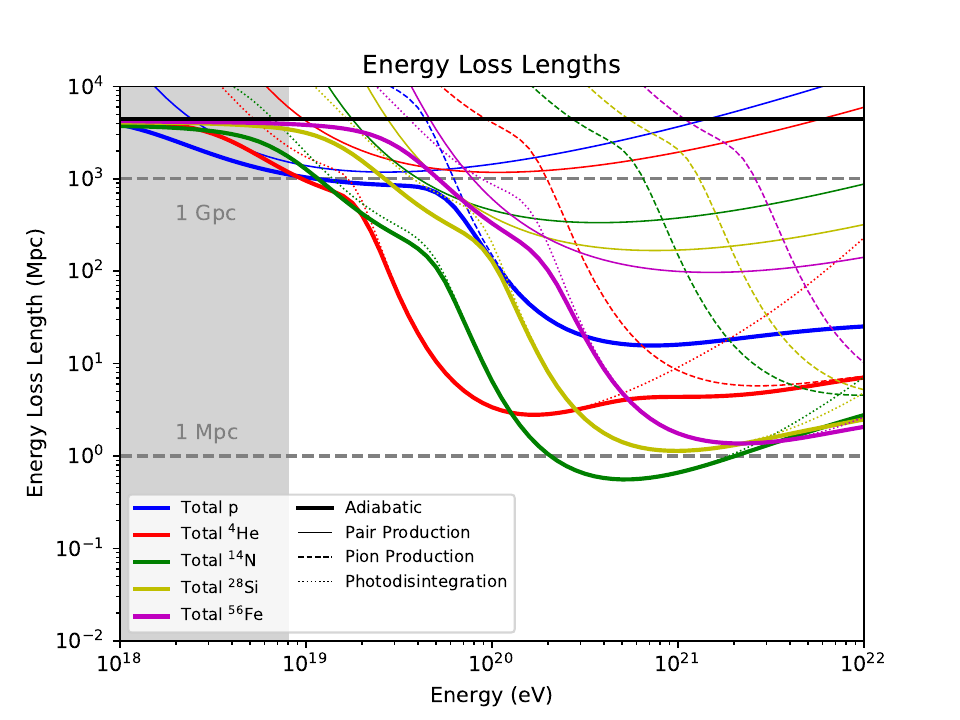}
\caption{Energy loss lengths corresponding to the interaction with both the CMB and the EBL \cite{Gilmore:2011ks} photon fields, extracted from \texttt{CRPropa 3.2} \cite{AlvesBatista:2022vem} for different nuclei: p in blue, $^4$He in red, $^{14}$N in green, $^{28}$Si in yellow and $^{56}$Fe in purple. The adiabatic energy loss is given by the black thick line. Different photohadronic interactions are shown with different line styles: thin continuous for pair production, dashed for pion production and dotted for photodisintegration. The total energy loss length of a given nucleus is given by the colored thick lines.}
\label{ELL_plot}
\end{figure}

In Fig.~\ref{ELL_plot} the energy loss lengths extracted from \texttt{CRPropa 3.2} \cite{AlvesBatista:2022vem} for different nuclei are shown in different colors (p in blue, $^4$He in red, $^{14}$N in green, $^{28}$Si in yellow and $^{56}$Fe in purple). Different interaction processes are shown with different line styles: for each process and nucleus, the energy loss length corresponding to the interaction with both CMB and EBL \cite{Gilmore:2011ks} photon fields is shown. The black thick line corresponds to the adiabatic energy loss due to the expansion of the Universe. The colored thick lines correspond to the total (adiabatic plus photohadronic) energy loss length of a given nucleus. 

\section{Weighting procedure}
\label{appendix_weight}
Here we discuss general calculations to reweight the simulation results for different injection scenarios. The simulated injection rate in the energy interval  between $E_\text{min}$ and $E_\text{max}$ is given by the sum of the injection rates of the simulated species 
\begin{equation}
Q_\text{sim}(E) = \sum_A Q_\text{sim}^A(E)\propto \sum_A b_A \left( \dfrac{E}{E_0}\right)^{-\alpha}\, ,
\end{equation}
where the coefficients $b_A$ characterize the simulated species abundances, such that $\sum_A b_A =1$, $E_0$ is a reference energy and $\alpha$ is the simulated spectral index. The corresponding injection probability of a cosmic ray with mass $A$ and energy $E$ is
\begin{equation}
\label{p_sim}
p_\text{sim}^A (E) = \dfrac{b_A}{N_\text{sim}}\left( \dfrac{E}{E_0}\right)^{-\alpha} \,  ,
\end{equation}
where the normalization factor $N_\text{sim}$ is given by
\begin{equation}
N_\text{sim} = \sum_A b_A \int_{E_\text{min}}^{E_\text{max}} dE \, \left(\dfrac{E}{E_0}\right)^{-\alpha} = E_0^\alpha \mathcal{I}_\text{sim}\, ,
\end{equation}
and the quantity $\mathcal{I}_\text{sim}$ is defined as
\begin{equation}
\mathcal{I}_\text{sim} = \int_{E_\text{min}}^{E_\text{max}} dE \, E^{-\alpha} \, .
\end{equation}
We define the injection rate that we want to reproduce as
\begin{equation}
Q_\text{rep} (E) =\sum_A Q_\text{rep}^A (E)\propto \sum_A a_A \left(\dfrac{E}{E_0}\right)^{-\gamma} f_\text{cut} \left(\dfrac{E}{Z_A R_\text{cut}}\right) \, ,
\end{equation}
where the coefficients $a_A$ represent the new abundances, such that $\sum_A a_A =1$  and $\gamma$ is the new spectral index. The cutoff function $f_\text{cut}$ now depends on the charge of the injected particle $Z_A$ and a given rigidity cutoff value $R_\text{cut}$, so that the corresponding cutoff energy of the species $A$ is given by $E_{\text{cut},A}=Z_A R_\text{cut}$. The corresponding injection probability is given by 
\begin{equation}
\label{p_rep}
p_\text{rep}^A(E) = \dfrac{a_A}{N_\text{rep}}\left(\dfrac{E}{E_0}\right)^{-\gamma} f_\text{cut} \left(\dfrac{E}{Z_A R_\text{cut}}\right) \, ,
\end{equation}
with normalization factor  
\begin{equation}
N_\text{rep} = \sum_A a_A \int_{E_\text{min}}^{E_\text{max}} dE \, \left(\dfrac{E}{E_0} \right)^{-\gamma} f_\text{cut} \left(\dfrac{E}{Z_A R_\text{cut}}\right) = E_0^\gamma R_\text{cut}^{1-\gamma} \sum_A a_A Z_A^{1-\gamma} \mathcal{I}_\text{rep}^A \, ,
\end{equation}
where we have defined the integrals 
\begin{equation}
\mathcal{I}_\text{rep}^A = \int_{x_{\text{min},A}}^{x_{\text{max},A}} dx_A \, x_A^{-\gamma} f_\text{cut} (x_A) \, ,
\end{equation}
over the variable $x_A=E/(Z_A R_\text{cut})$. The weight to be assigned to an observed event at the end of the simulation with injected energy $E_i$ and mass $A_i$ is given by the ratio of the probabilities \eqref{p_rep} and \eqref{p_sim}
\begin{equation}
\omega_{A_i} (E_i) = \dfrac{p_\text{rep}^{A_i}(E_i)}{p_\text{sim}^{A_i}(E_i)} =  \dfrac{\mathcal{I}_\text{sim}}{\sum_{A'} a_{A'} E_{\text{cut},A'}^{1-\gamma} \mathcal{I}_\text{rep}^{A'}} \dfrac{a_{A_i}}{b_{A_i}} E_i^{-\gamma+\alpha}f_\text{cut}\left(\dfrac{E_i}{Z_{A_i} R_\text{cut}}\right) \, .
\end{equation}
\par Previous calculations have assumed that all cosmic ray sources inject particles at the same rate. In general, we can define the relative injection rate factor of the source as $\tilde\omega(s_i)$, where $s_i$ denotes the source in the catalogue from which the cosmic ray recorded at the end of the simulation was emitted with injection energy $E_i$ and mass $A_i$. The final weight factor of the particle $i$ will be given by $\omega_i = \tilde\omega(s_i)\cdot \omega_{A_i}(E_i)$.

\section{Sky distribution analysis}
\label{appendix_spherical}
This appendix provides mathematical details and definitions on the arrival direction analysis used in this article.

\subsection{Arrival directions correction}
\label{appendix_ad_correction}
\par As described in Sec.~\ref{subsec_crpropa}, the simulated UHECRs are recorded by a spherical observer with radius $R_\text{O}=1\,\text{Mpc}$, while the radius of the Milky Way is $R_\text{MW}\simeq20\,\text{kpc}$. The use of a large observer is almost mandatory in order to collect enough statistics, but it introduces numerical effects on the arrival direction distribution. In this work, we consider the following geometric procedure to correct for the finite size of the observer (for a discussion of the role of a finite observer size in 3D UHECR simulations, see also \cite{Hackstein:2016pwa}). 
\par
\begin{figure*}[t]
\centering
\includegraphics[scale=0.4]{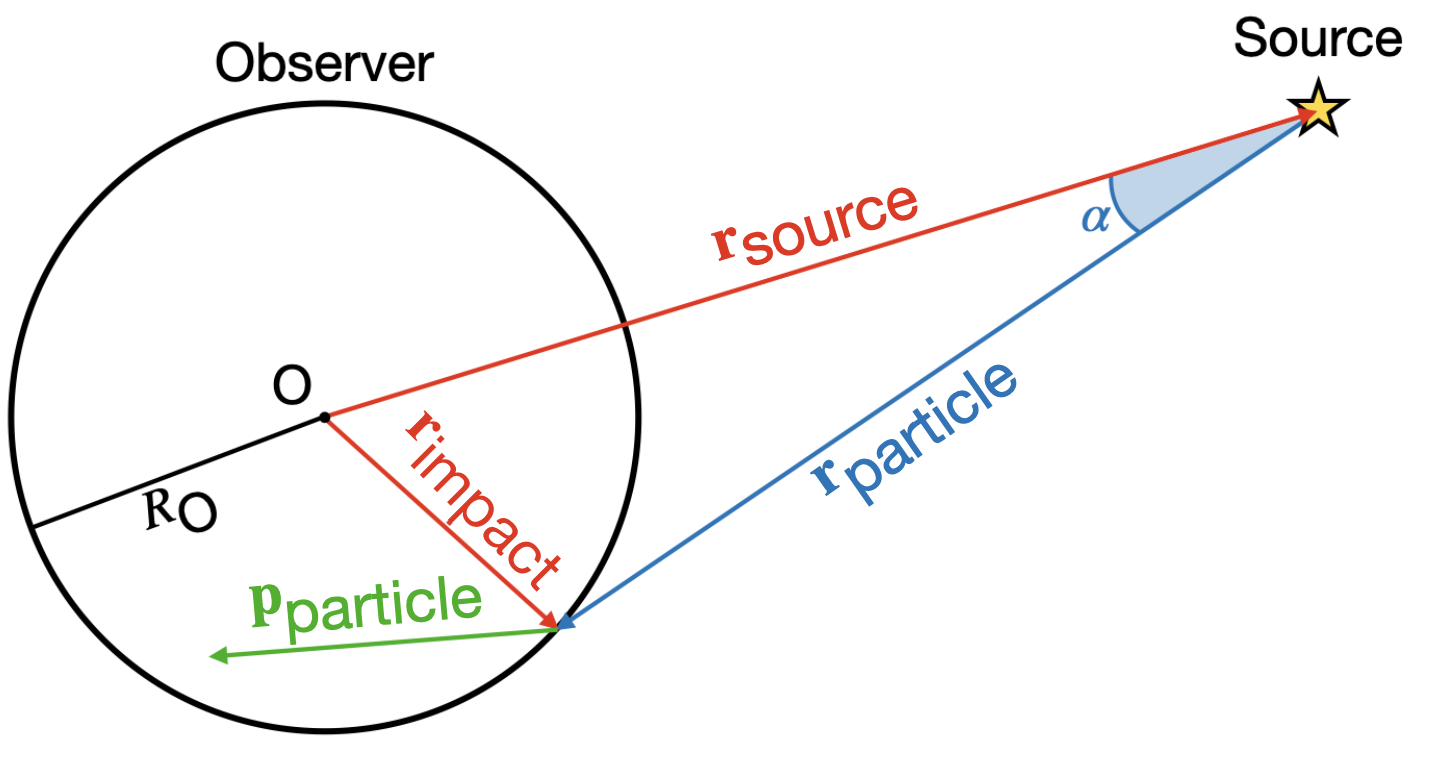}
\caption{Schematic representation for the correction of UHECR arrival directions on an observer with finite size. Vectors shown: position of the cosmic ray source relative to the center of the observer $\bold{r}_\text{source}$ (red), position of the cosmic ray impact point on the surface of the observer relative to the center of the observer $\bold{r}_\text{impact}$ (red), position of the cosmic ray impact point on the surface of the observer relative to the source $\bold{r}_\text{particle}=\bold{r}_\text{impact}-\bold{r}_\text{source}$ (blue) and recorded momentum vector at the end of the simulation $\bold{p}_\text{particle}$ (green). The normalized rotation axis is given by $\hat{\bold{n}}\propto\bold{r}_\text{particle}\times\bold{r}_\text{source}$. The rotation angle $\alpha$ is also shown in blue.}
\label{correction_plot}
\end{figure*}

 Given the position of the cosmic ray source relative to the center of the observer $\bold{r}_\text{source}$ and the position of the cosmic ray impact point on the surface of the observer $\bold{r}_\text{impact}$, we define the normalized rotation vector $\hat{\bold{n}}\propto\bold{r}_\text{source}\times\bold{r}_\text{particle}$, where $\bold{r}_\text{particle}=\bold{r}_\text{impact}-\bold{r}_\text{source}$. The corresponding rotation angle is given by $\sin \alpha=\left| \bold{r}_\text{source}\times\bold{r}_\text{particle}\right|/(|\bold{r}_\text{source}||\bold{r}_\text{particle}|)$ (see Fig.~\ref{correction_plot} for a schematic representation of the correction procedure). Given a rotation axis $\hat{\bold{n}}=(n_x,n_y,n_z)$, such that $n_x^2+n_y^2+n_z^2=1$, and a rotation angle $\alpha$, the rotation matrix $R(\hat{\bold{n}},\alpha)$ is given by 
\begin{equation}
R(\hat{\bold{n}},\alpha) =
\begin{pmatrix}
n_x^2(1-\cos\alpha) + \cos\alpha & n_x n_y (1-\cos\alpha) -n_z\sin\alpha & n_x n_z (1-\cos\alpha) +n_y\sin\alpha \\
n_x n_y (1-\cos\alpha) + n_z\sin\alpha & n_y^2(1-\cos\alpha) + \cos\alpha & n_y n_z (1-\cos\alpha) -n_x\sin\alpha \\
n_x n_z (1-\cos\alpha) -n_y\sin\alpha & n_y n_z (1-\cos\alpha) +n_x\sin\alpha & n_z^2(1-\cos\alpha) + \cos\alpha \\
\end{pmatrix}
\,.
\end{equation}
Given the rotation matrix $R(\hat{\bold{n}},\alpha)$, we compute the corrected momentum vector as $\bold{p}_\text{corrected}=R(\hat{\bold{n}},\alpha)\cdot\bold{p}_\text{particle}$, where $\bold{p}_\text{particle}$ is the momentum vector at the end of the simulation. 

\subsection{Number of pixels}
\label{appendix_pixels}
The required angular resolution on the sky is related to the maximum spherical coefficient considered in the analysis. In particular, each spherical parameter $l$ is related to the angular scale $\theta\sim\pi/l$. The corresponding solid angle is given by 
\begin{equation}
\Delta\Omega = 4\pi \sin^2 \left(\dfrac{\theta}{2}\right) \simeq \pi \,\theta^2 \, ,
\end{equation}
where the last equality holds for the approximation of small angles. Therefore, for the solid angle we obtain that
\begin{equation}
\label{solid_resolution}
\Delta\Omega \sim \pi^3/l^2 \, .
\end{equation}
\par In the HEALPix scheme, the spherical surface is divided into quadrilaterals of equal area and different shapes. The fundamental resolution of the spherical grid in the HEALPix scheme uses $12$ pixels. The parameter $N_\text{side}$, usually defined as a power of $2$, gives the number of divisions along the side of a fundamental quadrilateral\footnote{More details can be found in \href{https://healpix.jpl.nasa.gov/pdf/intro.pdf}{https://healpix.jpl.nasa.gov/pdf/intro.pdf}}. This means that the final number of pixels is equal to 
\begin{equation}
\label{N_pixels}
N_\text{pix} = 12\cdot N_\text{side}^2,
\end{equation}
and then, the corresponding angular size of a single pixel is 
\begin{equation}
\label{pixel_size}
\Delta\Omega_\text{pix} = \dfrac{4\pi}{N_\text{pix}}=  \dfrac{\pi}{3\cdot N_\text{side}^2}.
\end{equation}
\par In this work we consider a maximum spherical parameter of $l_\text{max}=20$, which corresponds to an angular resolution of $\Delta\Omega \sim 7.8\cdot 10^{-2}\,\text{sr}$. So the minimum number of pixels is given by the inverse of the relation \eqref{pixel_size}, where $\Delta\Omega_\text{pix}$ is given by \eqref{solid_resolution}. This corresponds to a minimum number of pixels of $N_\text{pix}=4\pi/\Delta\Omega_{l=20} \simeq 160$. We use a division parameter $N_\text{side}=64$, which corresponds to $N_\text{pix}=49152$. The corresponding angular resolution is $\Delta\Omega_\text{pix}\sim 2.6\cdot10^{-4}\,\text{sr}$, or $\theta\sim 0.5^\circ$.
\par The angular size of the pixels is used to obtain the discretized version of the arrival direction probability density function $\phi(\hat{\bold{n}})$. If the index $k$ indicates the direction of the pixel, then $\mathcal{N}_k$ is defined as the sum of the particle weights $\omega_i$ within the pixel $k$. The discretized arrival direction map is then given by
\begin{equation}
\phi_k = \dfrac{\mathcal{N}_k}{\mathcal{N}\Delta\Omega_\text{pix}},
\end{equation}
where $\mathcal{N}=\sum_{k=1}^{N_\text{pix}}\mathcal{N}_k$. The normalization of $\phi_k$ is given by $\sum_{k=1}^{N_\text{pix}}\phi_k\Delta\Omega_\text{pix}=1$, where the sum is over all the pixels of the sphere.

\subsection{Spherical harmonics}
\label{appendix_spherical_harmonics}
Given a complex representation of the spherical harmonic functions $Y_{lm}(\theta,\varphi)$, where $l$ and $m$ are the spherical indices $l=0,\, 1,\,2,\,... $ and $m=-l,\,...,\,l$ and the angles $\theta$ and $\varphi$ are the spherical coordinates, such that $0\leq\theta\leq\pi$ and $0\leq\varphi\leq 2\pi$, the condition of orthonormality is given by 
\begin{equation}
\int d\Omega\,Y_{lm}(\Omega)Y_{l'm'}^*(\Omega) = \mathcal{A}\,\delta_{ll'}\delta_{mm'}\, ,
\end{equation}
where $\Omega$ is the solid angle corresponding to the spherical angles $\theta$ and $\varphi$, $\delta_{xx'}$ is the Kronecker delta symbol, $Y_{l'm'}^*$ is the complex conjugate of the spherical harmonic and $\mathcal{A}$ is the normalization factor. In this work we consider the complex representation of the spherical harmonics such that
\begin{equation}
\label{spherical_harmonics_def}
Y_{lm}(\theta,\varphi) = (-1)^m \sqrt{\dfrac{2l+1}{4\pi}\dfrac{(l-m)!}{(l+m)!}} P_{lm}(\cos \theta)e^{im\varphi},
\end{equation}
where $P_{lm}(x)$ are the suitably normalized associate Legendre polynomials. Therefore, the first harmonic is
\begin{equation}
\label{first_harmonic}
Y_{00} = \dfrac{1}{\sqrt{4\pi}},
\end{equation}
and the corresponding normalization factor is $\mathcal{A}=1$. An important property of spherical harmonic functions, that will be used afterwards, is
\begin{equation}
\label{spherical_harm_prop}
Y_{l-m}(\theta,\varphi)=(-1)^m Y_{lm}^*(\theta,\varphi).
\end{equation}
If we assume a spherical decomposition as in \eqref{spherical_decomp}, it can be shown that
\begin{equation}
\label{spherical_coeff_prop}
a_{l-m}=(-1)^m a_{lm}^*,
\end{equation}
where we have assumed that the function $\phi(\hat{\bold{n}})$ in \eqref{spherical_decomp} is a real valued function.

\subsection{Angular power spectrum expectation}
\label{appendix_aps_estimation}
In this appendix, we recall general concepts for theoretical calculations of the angular power spectrum from binned data. More detailed studies can be found in \cite{Sommers:2000us,Anchordoqui:2003bx,Aublin:2005nv}. We also compute the expected angular power spectrum under the assumption of a purely isotropic sky.
\par We consider a finite number of pixels in the sky centered in the direction $\hat{\bold{p}}_i$, where $i=1,...,N_p$, and $N_p$ is the total number of pixels. Given the unit vector $\hat{\bold{n}}$ and a sequence of $N$ events with arrival directions $\left( \hat{\bold{n}}_1,\,\hat{\bold{n}}_2,\,...,\,\hat{\bold{n}}_N\right)$, we can define the arrival direction distribution $\phi(\hat{\bold{n}})$ as 
\begin{equation}
\label{signal_function}
\phi (\hat{\bold{n}}) = \dfrac{1}{\mathcal{N}} \sum_{i=1}^{N_p} \mathcal{N}_i \,\Delta (\hat{\bold{n}}-\hat{\bold{p}}_i) \, ,
\end{equation}
where $\mathcal{N}= \sum_{i=1}^{N_p}\mathcal{N}_i$ and $\mathcal{N}_i$ is defined as the sum of the particle weights $\omega_i$ (which are independent of the arrival direction) within the pixel $i$. The function $\Delta (\hat{\bold{n}}-\hat{\bold{p}}_i)$ is such that 
\begin{equation}
\Delta (\hat{\bold{n}}-\hat{\bold{p}}_i)= \left\{ \begin{aligned} 
  & \dfrac{1}{\Delta \Omega_i} \,\,\,\,\,\,\,\,\,\,\,\,\,\,\,\,\,\,\,\,\,\,\, \text{for $\hat{\bold{n}}$ in the pixel $\hat{\bold{p}}_i$}; \\
  & 0 \,\,\,\,\,\,\,\,\,\,\,\,\,\,\,\,\,\,\,\,\,\,\,\,\,\,\,\,\,\,\,\,  \text{otherwise}, 
\end{aligned} \right.
\end{equation}
where $\Delta \Omega_i$ is the angular size of the pixel $\hat{\bold{p}}_i$. It can be seen that $\phi(\hat{\bold{n}})$ is normalised to 1. The quantity $\mathcal{N}_i$ can then be written as
\begin{equation}
\label{N_i_def}
\mathcal{N}_i = \Delta\Omega_i\sum_{j=1}^N \omega_j \, \Delta (\hat{\bold{n}}_j-\hat{\bold{p}}_i) \, .
\end{equation}
The spherical decomposition of \eqref{signal_function} is given by the spherical coefficients $a_{lm}$, which can be calculated as 
\begin{equation}
a_{lm} = \int d\hat{\bold{n}}\, \phi(\hat{\bold{n}})Y_{lm}^*(\hat{\bold{n}}) = \dfrac{1}{\mathcal{N}}\sum_{i=1}^{N_p} \mathcal{N}_i \int d\hat{\bold{n}} \, \Delta (\hat{\bold{n}}-\hat{\bold{p}}_i) Y_{lm}^*(\hat{\bold{n}}) \, ,
\end{equation}
where $Y_{lm}^*(\hat{\bold{n}})$ are the spherical harmonic functions (see Appendix \ref{appendix_spherical_harmonics} for the spherical harmonic normalisation) and $d\hat{\bold{n}}$ is the differential of the solid angle. The coefficients $f^*_{lm,i}$ can be defined as
\begin{equation}
\label{def_f_lmi}
f^*_{lm,i} =\dfrac{1}{\Delta\Omega_i} \int_{\Delta\Omega_i} d\hat{\bold{n}}\, Y_{lm}^*(\hat{\bold{n}}) \, ,
\end{equation}
hence 
\begin{equation}
\label{a_lm-signal}
a_{lm} = \dfrac{1}{\mathcal{N}}\sum_{i=1}^{N_p} \mathcal{N}_i f^*_{lm,i} \, .
\end{equation}
From Eq.~\eqref{def_f_lmi} we can see that the coefficient $f_{lm,i}$ corresponds to the average value of the spherical harmonic $Y_{lm}$ within the pixel in the direction $\hat{p}_i$. Considering the spherical harmonic normalization used in this work, we have that $f_{00,i}^* =  (4\pi)^{-1/2}$, and then $a_{00} = (4\pi)^{-1/2}$, independent of the number of pixels. The angular power spectrum $\mathcal{C}_l$ is defined as the average of the spherical coefficients with the same $l$ value
\begin{equation}
\label{APS-def}
\mathcal{C}_l = \dfrac{1}{2l+1} \sum_{m=-l}^l \left| a_{lm}\right|^2 \, ,
\end{equation}
where $\left| a_{lm}\right|^2=a_{lm}a_{lm}^*$. The angular power spectrum of the function \eqref{signal_function} is then given by
\begin{equation}
\label{C_l_pix}
\mathcal{C}_l = \dfrac{1}{2l+1}\sum_{m=-l}^l \left| a_{lm}\right|^2 = \dfrac{1}{2l+1}\dfrac{1}{\mathcal{N}^2}\sum_{m=-l}^l \left|\sum_{i=1}^{N_p} \mathcal{N}_i f^*_{lm,i} \right|^2 \, ,
\end{equation}
where $\mathcal{C}_0=(4\pi)^{-1}$, again independent of the number of pixels.
\par Given a probability distribution function of arrival directions in the sky $\Psi(\hat{\bold{n}})$, we can compute the expected value of the spherical decomposition \eqref{a_lm-signal} assuming the distribution $\Psi(\hat{\bold{n}})$. The expected value is 
\begin{equation}
\left< a_{lm}\right>_{\Psi} = \dfrac{1}{\mathcal{N}} \sum_{i=1}^{N_p}\left< \mathcal{N}_i\right>_{\Psi}f_{lm,i}^* \, ,
\end{equation}
where $\left< ...\right>_{\Psi}$ is the average with respect to the distribution $\Psi$. The quantity $\left< \mathcal{N}_i\right>_{\Psi}$ is given by 
\begin{equation}
\left< \mathcal{N}_i\right>_{\Psi} = \Delta\Omega_i \sum_{j=1}^N \omega_j\,\int d\hat{\bold{n}}_j \, \Delta (\hat{\bold{n}}_j-\hat{\bold{p}}_i) \Psi (\hat{\bold{n}}_j)  = \mathcal{N} \Delta\Omega_i \overline{\Psi}(\hat{\bold{p}}_i)
\end{equation}
where $\overline{\Psi}(\hat{\bold{p}}_i)$ is the average value of $\Psi(\hat{\bold{n}})$ in the pixel $i$. The expected value of the spherical coefficients is 
\begin{equation}
\left< a_{lm} \right>_{\Psi} =\sum_{i=1}^{N_p} \Delta\Omega_i \overline{\Psi}(\hat{\bold{p}}_i) f_{lm,i}^* \, .
\end{equation}
In the case of an isotropic distribution $\Psi_\text{iso}=(4\pi)^{-1}$, we have that $\overline{\Psi}_\text{iso}(\hat{\bold{p}}_i)=(4\pi)^{-1}$. We can then calculate the expected value of the spherical coefficients as \footnotetext{In the calculation of Eq~\eqref{prediction_iso_a_lm} we have used the normalization condition described in Appendix~\ref{appendix_spherical_harmonics} for the product of the spherical harmonics $Y_{lm}(\hat{\bold{n}})$ and $Y_{00}^*(\hat{\bold{n}})$. The result is
\begin{equation}
\int d\hat{\bold{n}}\,Y_{lm}(\hat{\bold{n}})Y_{00}^*(\hat{\bold{n}}) = \dfrac{1}{\sqrt{4\pi}}\int d\hat{\bold{n}}\,Y_{lm}(\hat{\bold{n}}) = \,\delta_{l0}\delta_{m0}\, .
\end{equation}
}
\begin{equation}
\label{prediction_iso_a_lm}
\left< a_{lm} \right>_{\Psi_\text{iso}} = \dfrac{1}{4\pi} \sum_{i=1}^{N_p} \Delta\Omega_i f_{lm,i}^* = \dfrac{1}{4\pi} \sum_{i=1}^{N_p} \int_{\Delta\Omega_i} d\hat{\bold{n}}\, Y_{lm}^*(\hat{\bold{n}}) = \dfrac{1}{4\pi}  \int 
d\hat{\bold{n}}\, Y_{lm}^*(\hat{\bold{n}}) = \dfrac{\delta_{l0}\delta_{m0}}{\sqrt{4\pi}} \, .
\end{equation}
As expected, the only spherical coefficient different from zero under the assumption of an isotropic sky is $a_{00}=(4\pi)^{-1/2}$. 
\par The expectation value of the angular power spectrum, given a distribution $\Psi(\hat{\bold{n}})$, can be obtained by estimating the modulus square of the angular coefficients. The modulus square of the spherical coefficients is given by 
\begin{align}
\left| a_{lm}\right|^2 & = \dfrac{1}{\mathcal{N}^2} \sum_{i,j=1}^{N_p} \mathcal{N}_i \mathcal{N}_j f_{lm,i}^* f_{lm,j}  \nonumber \\
& = \dfrac{1}{\mathcal{N}^2} \left[ \sum_{i=1}^{N_p} \mathcal{N}_i^2 \left| f_{lm,i} \right|^2 + \sum_{i\neq j}\mathcal{N}_i \mathcal{N}_j f_{lm,i}^* f_{lm,j} \right] \, .
\end{align}
We can write the factors $\mathcal{N}_i^2$ and $\mathcal{N}_i \mathcal{N}_j $ using the definition \eqref{N_i_def}. The general result is
\begin{align}
\mathcal{N}_i \mathcal{N}_j & =  \Delta\Omega_i\Delta\Omega_j\sum_{s,t=1}^N \omega_s\,\omega_t\,\Delta (\hat{\bold{n}}_s-\hat{\bold{p}}_i) \Delta (\hat{\bold{n}}_t-\hat{\bold{p}}_j)  \nonumber \\
& = \delta_{ij}\Delta\Omega_j\sum_{s=1}^N \omega_s^2\,\Delta(\hat{\bold{n}}_s-\hat{\bold{p}}_j) + \Delta\Omega_i\Delta\Omega_j\sum_{s\neq t} \omega_s\,\omega_t\,\Delta (\hat{\bold{n}}_s-\hat{\bold{p}}_i)\Delta (\hat{\bold{n}}_t-\hat{\bold{p}}_j) \, .\,
\end{align}
where the value of $\mathcal{N}_i^2$ corresponds to the case $i=j$. The expected values of $\mathcal{N}_i^2$ and $\mathcal{N}_i \mathcal{N}_j$ with the distribution $\Psi(\hat{\bold{n}})$ are given by 
\begin{align}
\left<N_i N_j \right>_{\Psi} & = \delta_{ij}\Delta\Omega_j\sum_{s=1}^N \omega_s^2\int d\hat{\bold{n}}_s  \,\Psi(\hat{\bold{n}}_s) \Delta(\hat{\bold{n}}_s-\hat{\bold{p}}_j) \nonumber \\
& \,\,\,\,\,\,+ \Delta\Omega_i\Delta\Omega_j\sum_{s\neq t} \omega_s\,\omega_t\int d\hat{\bold{n}}_s  \,\Psi(\hat{\bold{n}}_s)\,\Delta (\hat{\bold{n}}_s-\hat{\bold{p}}_i)\Delta (\hat{\bold{n}}_t-\hat{\bold{p}}_j) \nonumber \\
& = \delta_{ij}\Delta\Omega_j\sum_{s=1}^N \omega_s^2\, \overline{\Psi}(\hat{\bold{p}}_i) +\Delta\Omega_i\Delta\Omega_j \sum_{s\neq t} \omega_s\,\omega_t\, \overline{\Psi}(\hat{\bold{p}}_i)\overline{\Psi}(\hat{\bold{p}}_j) \, ,
\end{align}
We then obtain the expected values of the spherical coefficients 
\begin{equation}
\left< \left| a_{lm}\right|^2\right>_{\Psi} =  \dfrac{\sum_{j=1}^N \omega_j^2 }{\mathcal{N}^2} \sum_{i=1}^{N_p} \Delta\Omega_i \overline{\Psi}(\hat{\bold{p}}_i) \left|f_{lm,i} \right|^2+\dfrac{\sum_{j\neq k}^N \omega_j \omega_k }{\mathcal{N}^2} \left| \sum_{i=1}^{N_p} \Delta\Omega_i \overline{\Psi}(\hat{\bold{p}}_i)  f_{lm,i}^* \right|^2 \, .
\end{equation}
If we again consider the prediction under an isotropic sky distribution, and that the sky is divided into pixels with the same angular size $\Delta\Omega$, the expected values of the spherical coefficients become
\begin{equation}
\left< \left| a_{lm}\right|^2\right>_{\Psi_\text{iso}} =  \dfrac{\Delta\Omega}{4\pi}\dfrac{\sum_{j=1}^N \omega_j^2 }{\mathcal{N}^2} \sum_{i=1}^{N_p}\left|f_{lm,i} \right|^2+ \dfrac{1}{4\pi}\dfrac{\sum_{j\neq k} \omega_j \omega_k }{\mathcal{N}^2} \delta_{l0}\delta_{m0} \, ,
\end{equation}
and then the angular power spectrum
\begin{equation}
\label{C_iso_pixel}
 \left< \mathcal{C}_l \right>_{\Psi_\text{iso}} = \left\{ \begin{aligned} 
  & \dfrac{1}{4\pi} \,\,\,\,\,\,\,\,\,\,\,\,\,\,\,\,\,\,\,\,\,\,\,\,\,\,\,\,\,\,\,\,\,\,\,\,\,\,\,\,\,\,\,\,\,\,\,\,\,\,\,\,\,\,\,\,\,\,\,\,\,\,\,\,\,\,\,\,\,\,\,\,\,\,\,\,\,\,\,\,\,\,\,\,\,\,\,\,\,\,\,\,\,\,\,\,\,\,\,\,\,\,\,\,\,\,\,\,   l=0; \\
  &\dfrac{1}{2l+1}\dfrac{\Delta\Omega}{4\pi}\dfrac{\sum_{j=1}^N \omega_j^2 }{\mathcal{N}^2}\sum_{m=-l}^l\sum_{i=1}^{N_p} \left| f_{lm,i} \right|^2 \,\,\,\,\,\,\,\,\,\,\,\,\,\,\,\,\,\,\,\,\,\, l\neq0. 
\end{aligned} \right.
\end{equation}
As expected, the monopole component of \eqref{C_iso_pixel} does not depend on the number of pixels or on the number of arrival directions. The components for $l\neq0$ go to zero for $N$ going to infinity and depend on the pixelation used. If we assume $\omega_i=1$ for all the particles, the prediction of the angular power spectrum becomes 
\begin{equation}
\label{C_iso_pixel_bis}
 \left< \mathcal{C}_l \right>_{\Psi_\text{iso}} = \left\{ \begin{aligned} 
  & \dfrac{1}{4\pi} \,\,\,\,\,\,\,\,\,\,\,\,\,\,\,\,\,\,\,\,\,\,\,\,\,\,\,\,\,\,\,\,\,\,\,\,\,\,\,\,\,\,\,\,\,\,\,\,\,\,\,\,\,\,\,\,\,\,\,\,\,\,\,\,\,\,\,\,\,\,\,\,\,\,\,\,\,\,\,\,\,\,\,\,\,\,\,   l=0; \\
  &\dfrac{1}{2l+1}\dfrac{\Delta\Omega}{4\pi N}\sum_{m=-l}^l\sum_{i=1}^{N_p} \left| f_{lm,i} \right|^2 \,\,\,\,\,\,\,\,\,\,\,\,\,\,\,\,\,\,\,\,\,\, l\neq0. 
\end{aligned} \right.
\end{equation}
For an infinite number of pixels (i.e. infinitesimally small $\Delta\Omega$) the function in \eqref{signal_function} is $\phi(\hat{\bold{n}})=\mathcal{N}^{-1}\sum_{i=1}^N \omega_i \, \delta (\hat{\bold{n}}-\hat{\bold{n}}_i)$, where $\delta(x)$ is the Dirac delta function. The angular power spectrum is then given by
\begin{equation}
\label{C_l_no_pix}
\mathcal{C}_l = \dfrac{1}{2l+1}\sum_{m=-l}^l \left| a_{lm}\right|^2 = \dfrac{1}{2l+1}\dfrac{1}{\mathcal{N}^2}\sum_{m=-l}^l \left|\sum_{i=1}^{N} \omega_i \, Y^*_{lm}(\hat{\bold{n}}_i) \right|^2 \, .
\end{equation}
As the number of pixels goes to infinity, the sum of the coefficients $f_{lm,i}$ is such that $\sum_{i=1}^{N_p}\left| f_{lm,i} \right|^2 \rightarrow\Delta\Omega$, with our normalization for the spherical harmonics. Therefore, the predictions \eqref{C_iso_pixel} and \eqref{C_iso_pixel_bis} converge to 
\begin{equation}
\label{C_l-iso}
 \left< \mathcal{C}_l \right>_{\Psi_\text{iso}} = \left\{ \begin{aligned} 
  & \dfrac{1}{4\pi} \,\,\,\,\,\,\,\,\,\,\,\,\,\,\,\,\,\,\,\,\,\,\,\,\,\,\,\,\,\,\,\,\,\,\,\,\,\,\,\,\, l=0; \\
  & \dfrac{1}{4\pi \mathcal{N}^2}\sum_{i=1}^N \omega_i^2 \,\,\,\,\,\,\,\,\,\,\,\,\,\,\,\, l\neq0; 
\end{aligned} \right. 
\end{equation}
and 
\begin{equation}
\label{C_l-iso-bis}
 \left< \mathcal{C}_l \right>_{\Psi_\text{iso}} = \left\{ \begin{aligned} 
  & \dfrac{1}{4\pi} \,\,\,\,\,\,\,\,\,\,\,\,\,\,\,\,\,\,\,\,\,\,\,\,\,\,\,\,\,\,\,\,\,\,\,\,\,\,\,\,\, l=0; \\
  & \dfrac{1}{4\pi N}\,\,\,\,\,\,\,\,\,\,\,\,\,\,\,\,\,\,\,\,\,\,\,\,\,\,\,\,\,\,\,\,\,\,\,\, l\neq0;
\end{aligned} \right.
\end{equation}
respectively. 
\begin{figure}[t]
\centering
\begin{minipage}{8.5cm}
\centering
\includegraphics[scale=0.5]{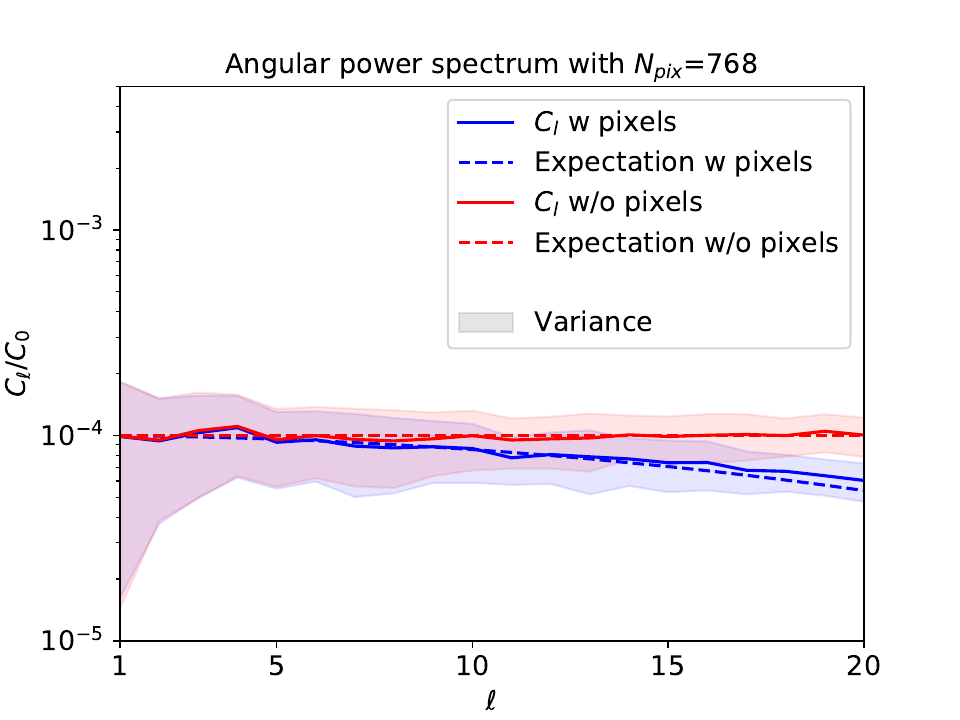}
\end{minipage}
\begin{minipage}{8.5cm}
\centering
\includegraphics[scale=0.5]{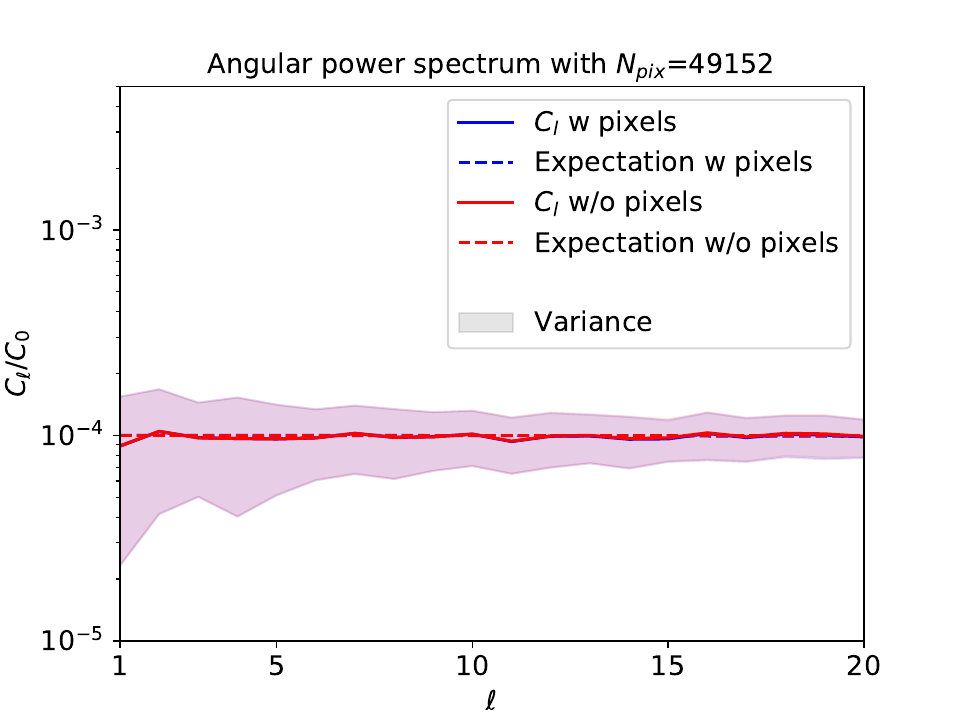}
\end{minipage}
\caption{Mean angular power spectra normalized to the monopole (i.e. $\mathcal{C}_l/\mathcal{C}_0$) of $100$ isotropic skies with $15,000$ events each. The angular power spectra of \eqref{C_l_pix} (blue line) and \eqref{C_l_no_pix} (red line) are shown assuming all weighting factors $\omega_i=1$, along with the isotropic predictions \eqref{C_iso_pixel_bis} (blue dashed line) and \eqref{C_l-iso-bis} (red dashed line). Different numbers of pixels are considered: $N_\text{pix}=768$ (left panel) and $N_\text{pix}=49.152$ (right panel). The shaded areas indicate the estimated variance of the angular power spectra.}
\label{aps_prediction}
\end{figure}
\par We note that the introduction of a finite sky resolution implies a maximum resolvable angular scale in the sky. In particular, the study of the angular power spectrum is limited by the maximum spherical parameter $l_\text{max} \sim \sqrt{\pi^3/\Delta\Omega}$ (see Appendix \ref{appendix_pixels}), where $ \Delta\Omega_\text{pix}$ is the angular size of the pixel, given by $\Delta\Omega_\text{pix} = 4\pi/N_{p}$. In Fig.~\ref{aps_prediction} the mean angular power spectrum of $100$ isotropic skies with $15,000$ events each is shown. The angular power spectra of \eqref{C_l_pix} (blue line) and \eqref{C_l_no_pix} (red line) are shown assuming all weight factors $\omega_i=1$, along with the isotropic predictions \eqref{C_iso_pixel_bis} (blue dashed line) and \eqref{C_l-iso-bis} (red dashed line), respectively. The left panel shows the case of $N_\text{pix}=768$. The maximum spherical parameter corresponds to $l_\text{max} \sim 13$. It can be seen that both the angular power spectrum and the isotropic prediction deviate from the results obtained with infinite resolution. In particular, for $l\gtrsim15$ the deviation becomes larger than $1\sigma$. This is due to the fact that the coefficients $f_{lm,i}$ correspond to the average value of the spherical harmonic $Y_{lm}(\hat{\bold{n}})$ in the pixel. Therefore, if the angular size of the pixel is larger than the solid angle associated with the spherical parameter $l$, several oscillations of the spherical harmonic are contained within the pixel, leading to a decrease in the mean value of the spherical harmonic $Y_{lm}(\hat{\bold{n}})$. In the right panel we consider $N_\text{pix}=49.152$, corresponding to $l_\text{max}\sim348$. We can see that all the predictions for $l\leq20$ have the same value for the angular power spectrum.

\begin{table*}
\centering
\bgroup
\def\arraystretch{1.25}
\begin{tabular}{c|c|c|c|c|c|c|c|c}
& \textit{ballistic} & \textit{astrophysicalR} & \textit{primordial2R} & \textit{statistical} & \textit{ballistic} & \textit{astrophysicalR} & \textit{primordial2R} & \textit{statistical} \\
\hline
\hline
$\gamma$ & $-3.06$ & $-0.66$ & $-2.13$ & $-2.25$ & $-1.49$ &$-0.93$ &$-1.52$ & $-2.01$\\
$\log (R_\text{cut}/1\,\text{EV})$ & $0.09$ & $0.28$ &$0.13$ & $0.17$ & $0.28$&$0.45$ & $0.25$& $0.24$\\
$I_{p}$ & $4.2\cdot 10^{-2}$ & $4.2\cdot 10^{-2}$ & $8.2\cdot 10^{-2}$& $1.6\cdot 10^{-1}$ & $1.4\cdot 10^{-1}$& $0.0$& $6.6\cdot 10^{-2}$& $7.3\cdot 10^{-3}$\\
$I_{He}$ & $0.0$ & $2.0\cdot 10^{-1}$& $1.4\cdot 10^{-1}$& $1.8\cdot 10^{-2}$ & $0.0$& $0.0$& $1.1\cdot 10^{-1}$&$0.0$ \\
$I_{N}$ & $7.1\cdot 10^{-1}$& $4.6\cdot 10^{-1}$& $4.2\cdot 10^{-1}$ & $5.5\cdot 10^{-1}$ & $5.4\cdot 10^{-1}$ & $6.3\cdot 10^{-1}$& $5.3\cdot 10^{-1}$& $8.0\cdot 10^{-1}$\\
$I_{Si}$ & $2.1\cdot 10^{-1}$& $1.5\cdot 10^{-1}$&$3.6\cdot 10^{-1}$ & $2.4\cdot 10^{-1}$ &$3.2\cdot 10^{-1}$ &$3.7\cdot 10^{-1}$ &$2.2\cdot 10^{-1}$& $1.3\cdot 10^{-1}$\\
$I_{Fe}$ & $4.2\cdot 10^{-2}$ &$1.5\cdot 10^{-1}$ &$3.0\cdot 10^{-3}$ & $3.9\cdot 10^{-2}$  & $6.5\cdot 10^{-3}$&$0.0$ & $7.4\cdot 10^{-2}$& $5.7\cdot 10^{-2}$\\
\hline
& \multicolumn{4}{c|}{$n_s=10^{-4}\,\text{Mpc}^{-3}$} & \multicolumn{4}{c}{$n_s=10^{-2}\,\text{Mpc}^{-3}$} \\
\end{tabular}
\egroup
\caption{Minimization results of Eq.~\eqref{chi_sq_def} for the \textit{homogeneous} source scenarios. Different columns correspond to different EGMF models. The source number density of a given model is indicated at the bottom of the table.}
\label{table_homo}
\end{table*}
\begin{figure}[h]
\centering
\begin{minipage}{8.5cm}
\centering
\includegraphics[scale=0.475]{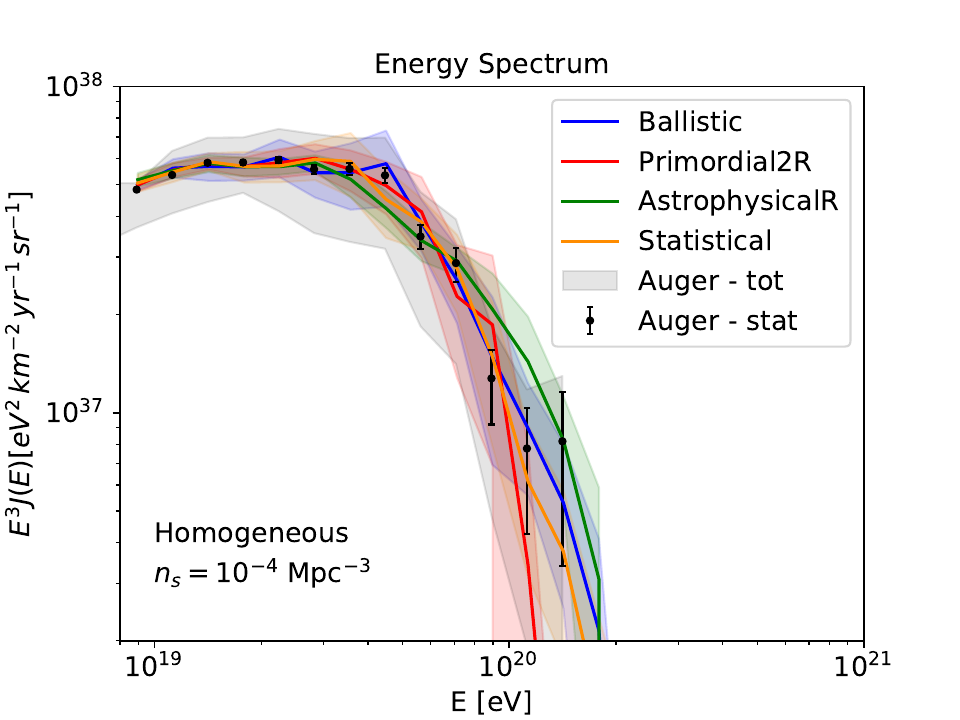}
\end{minipage}
\begin{minipage}{8.5cm}
\centering
\includegraphics[scale=0.475]{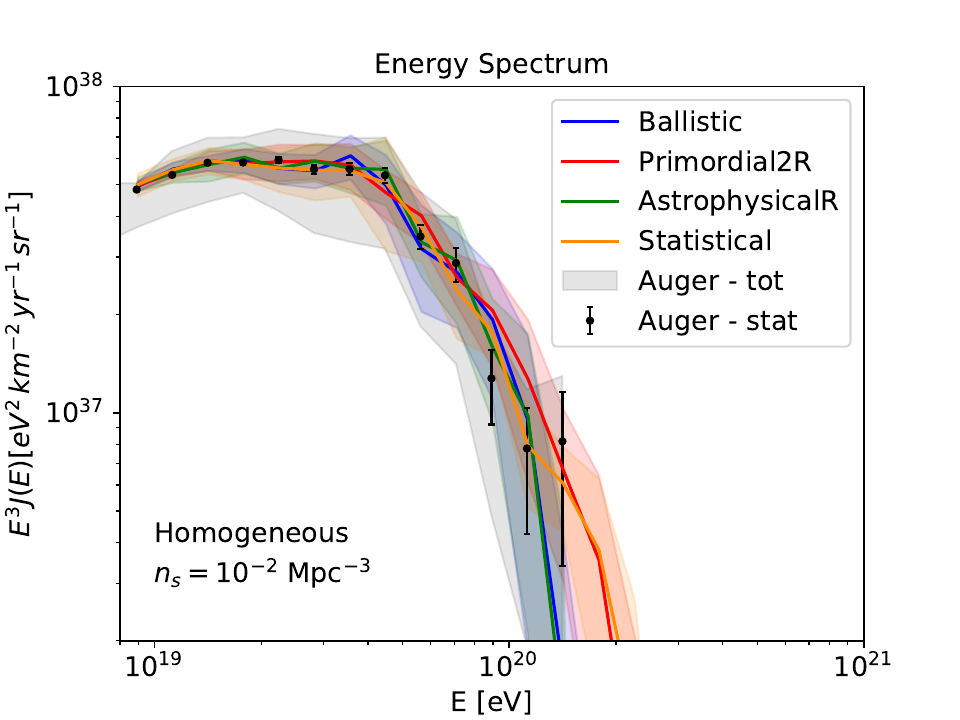}
\end{minipage}
\begin{minipage}{8.5cm}
\centering
\includegraphics[scale=0.475]{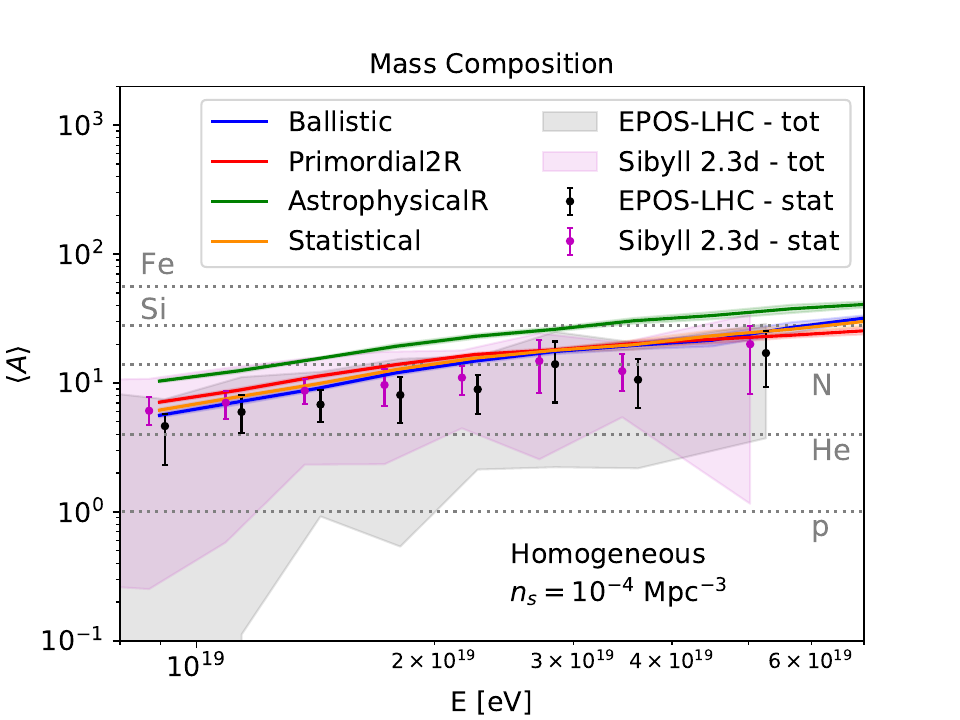}
\end{minipage}
\begin{minipage}{8.5cm}
\centering
\includegraphics[scale=0.475]{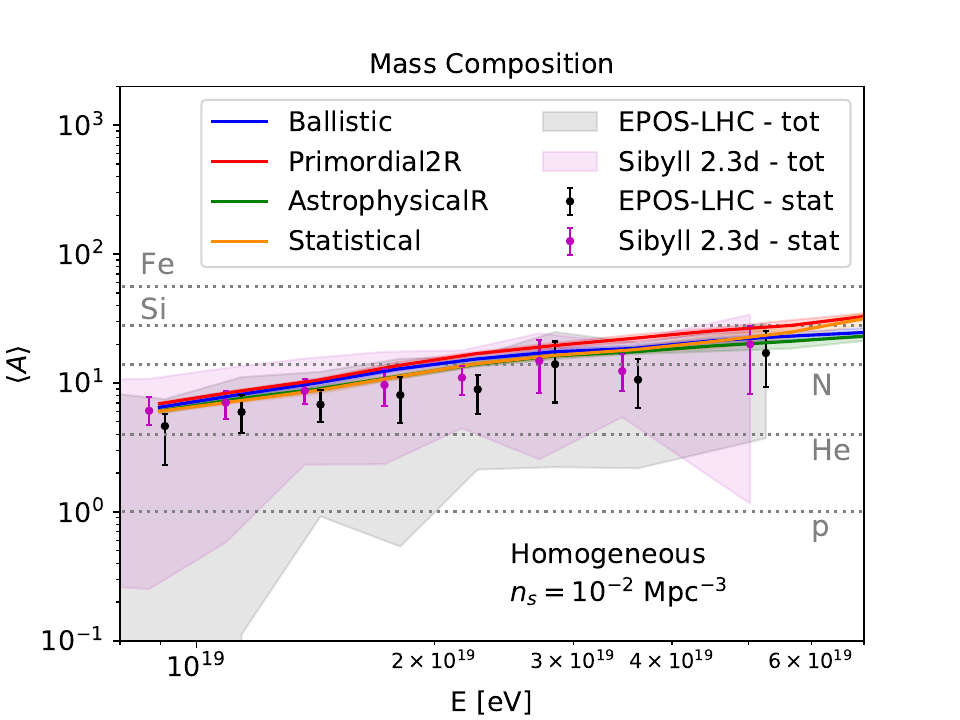}
\end{minipage}
\caption{Same as Fig.~\ref{spectrum_composition_den}, but for the \textit{homogeneous} scenarios.}
\label{spectrum_composition_homo}
\end{figure}
\begin{figure}[th!]
\centering
\begin{minipage}{8.5cm}
\centering
\includegraphics[scale=0.5]{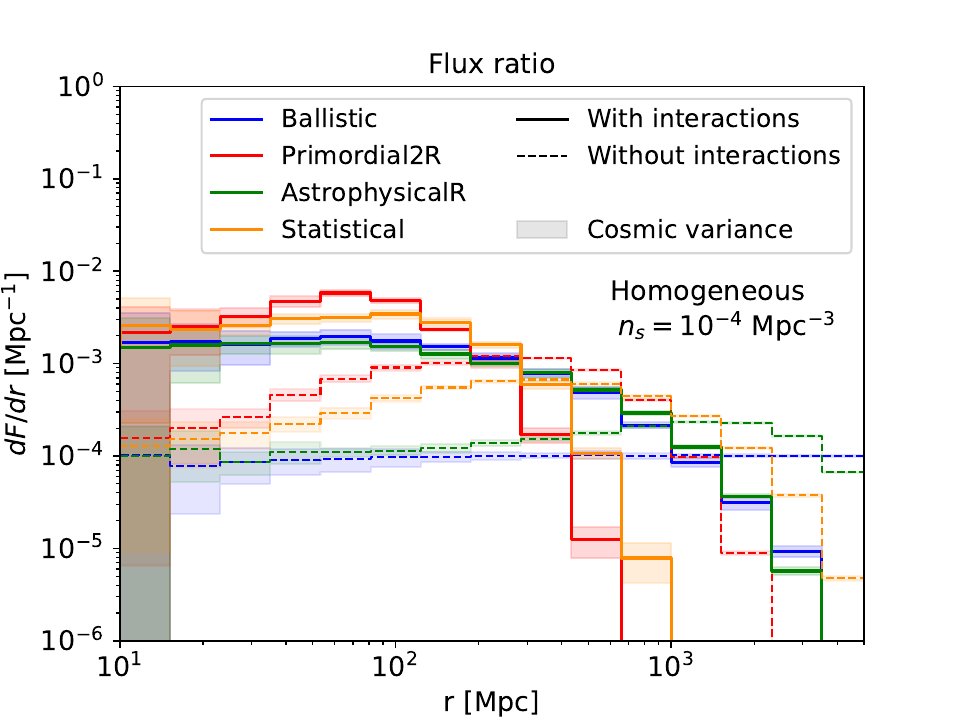}
\end{minipage}
\begin{minipage}{8.5cm}
\centering
\includegraphics[scale=0.5]{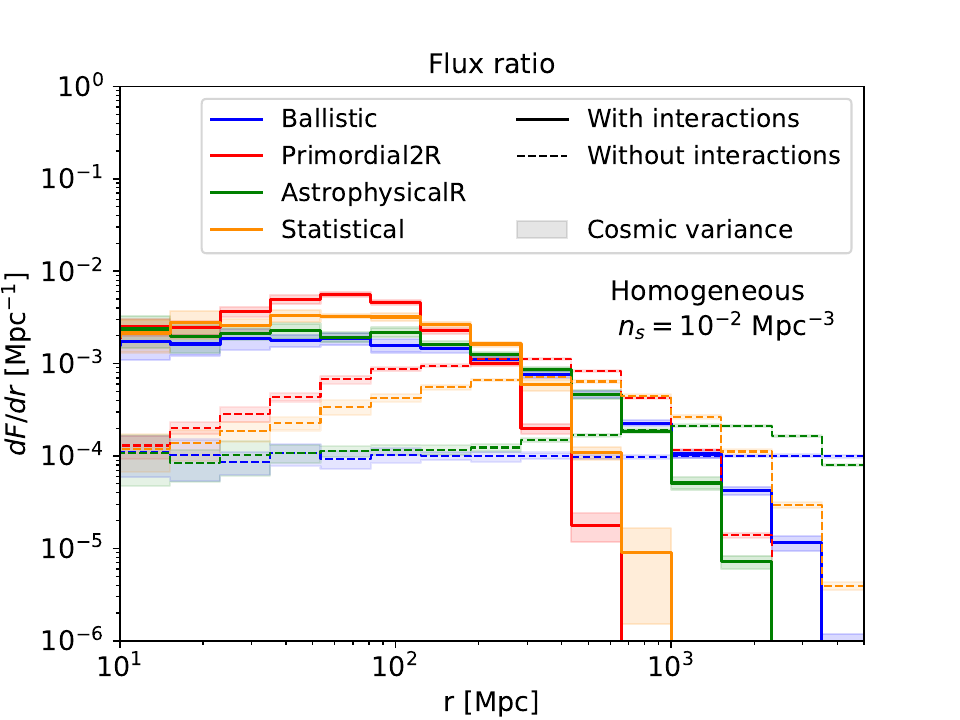}
\end{minipage}
\caption{Same as Fig.~\ref{number_sources_den}, but for the \textit{homogeneous} scenarios.}
\label{number_sources_homo}
\end{figure}
\begin{figure}[t]
\centering
\begin{minipage}{7.5cm}
\centering
\includegraphics[scale=0.5]{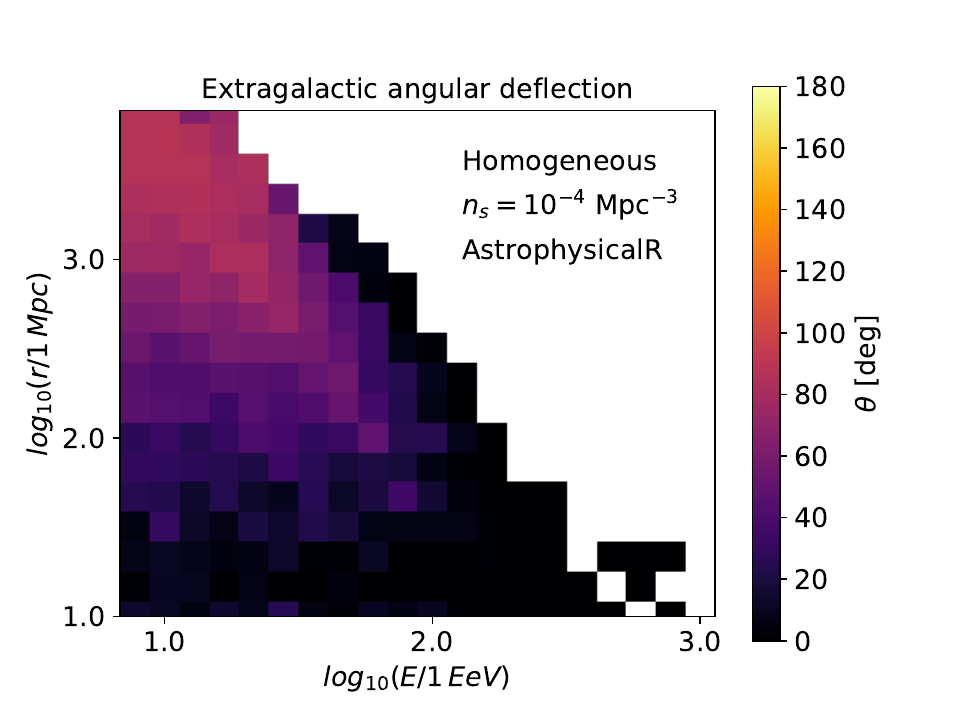}
\end{minipage}
\begin{minipage}{7.5cm}
\centering
\includegraphics[scale=0.5]{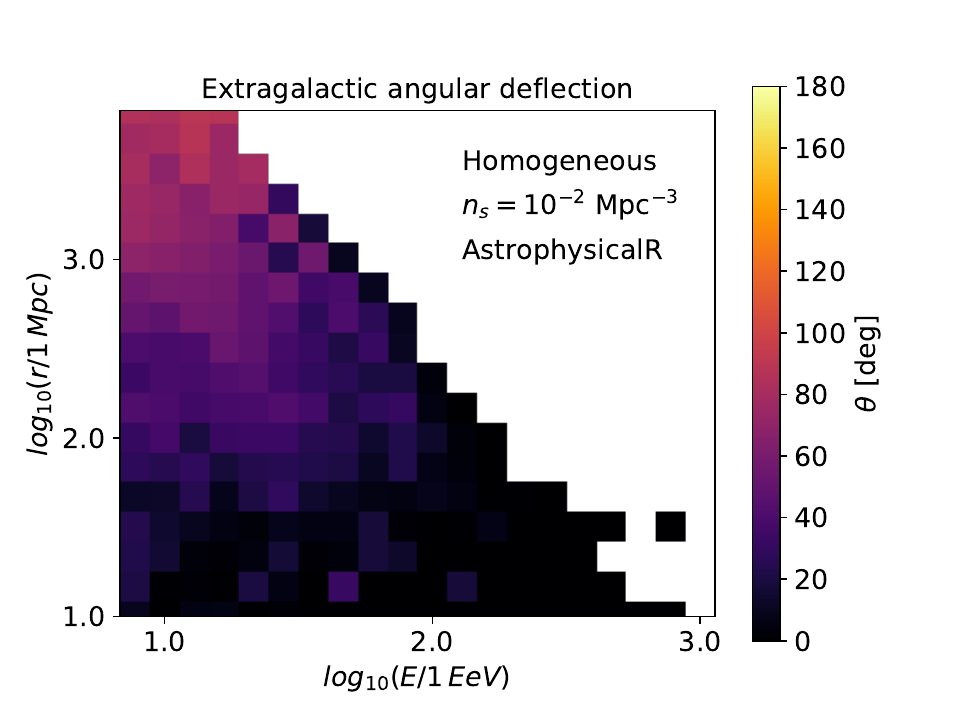}
\end{minipage}
\begin{minipage}{7.5cm}
\centering
\includegraphics[scale=0.5]{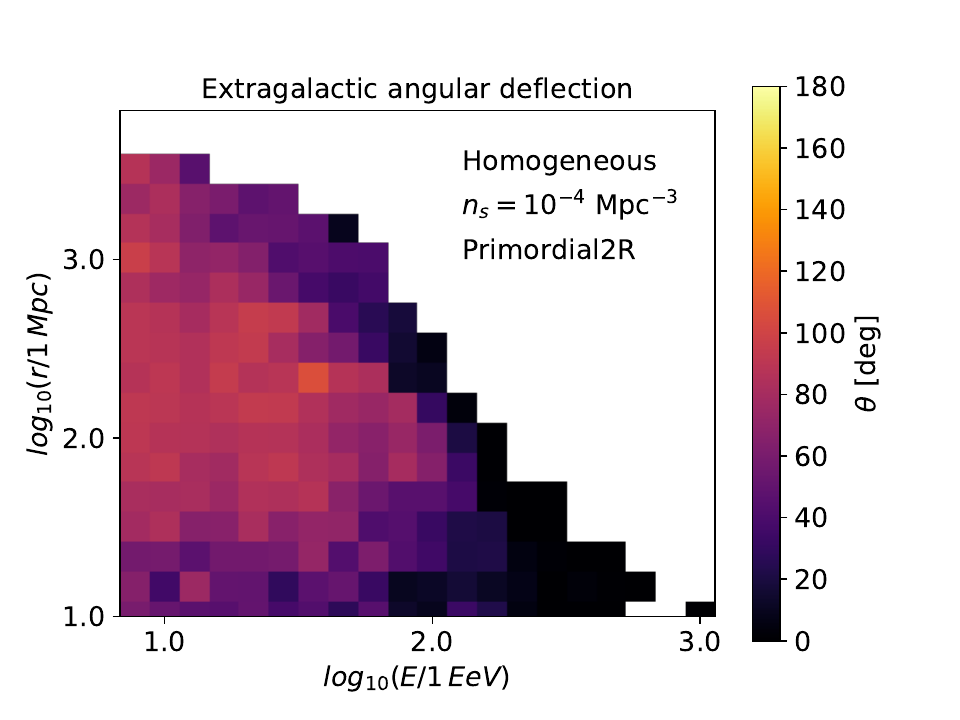}
\end{minipage}
\begin{minipage}{7.5cm}
\centering
\includegraphics[scale=0.5]{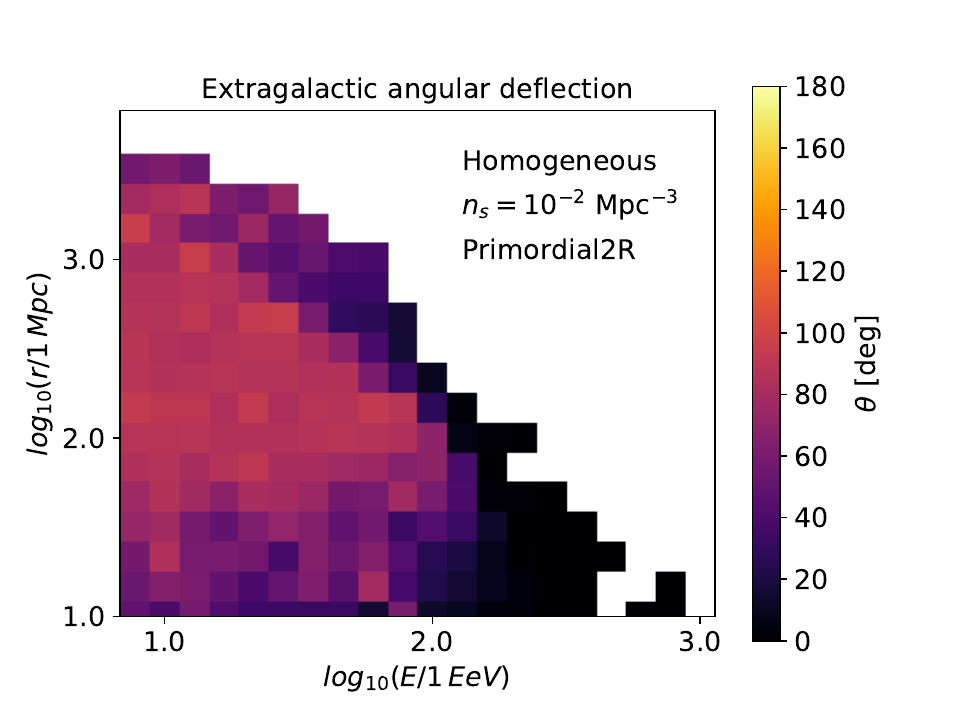}
\end{minipage}
\begin{minipage}{7.5cm}
\centering
\includegraphics[scale=0.5]{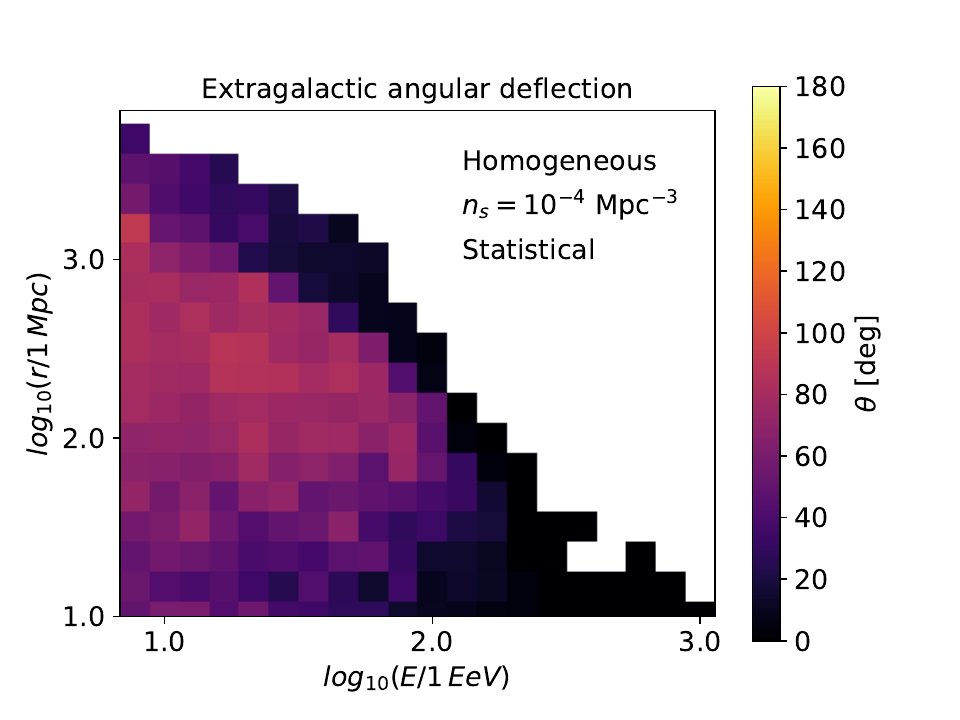}
\end{minipage}
\begin{minipage}{7.5cm}
\centering
\includegraphics[scale=0.5]{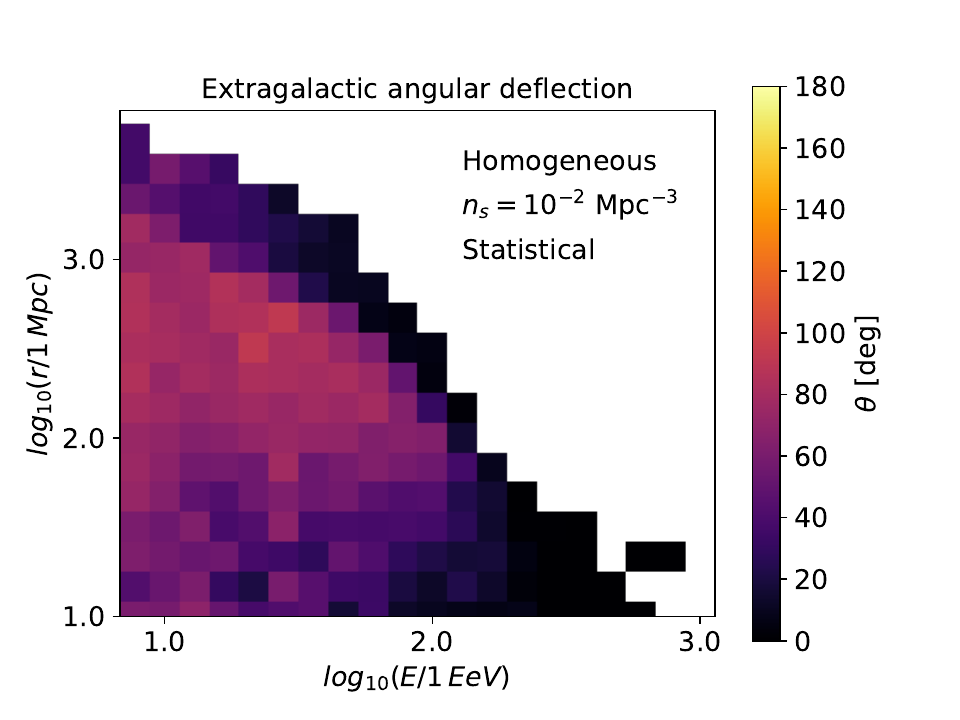}
\end{minipage}
\caption{Same as Fig.~\ref{deflection_den}, but for the \textit{homogeneous} scenarios.}
\label{deflection_homo}
\end{figure}
\section{Homogeneous source scenario}
\label{appendix_homo_source}
\begin{figure}[t]
\centering
\begin{minipage}{8.5cm}
\centering
\includegraphics[scale=0.2]{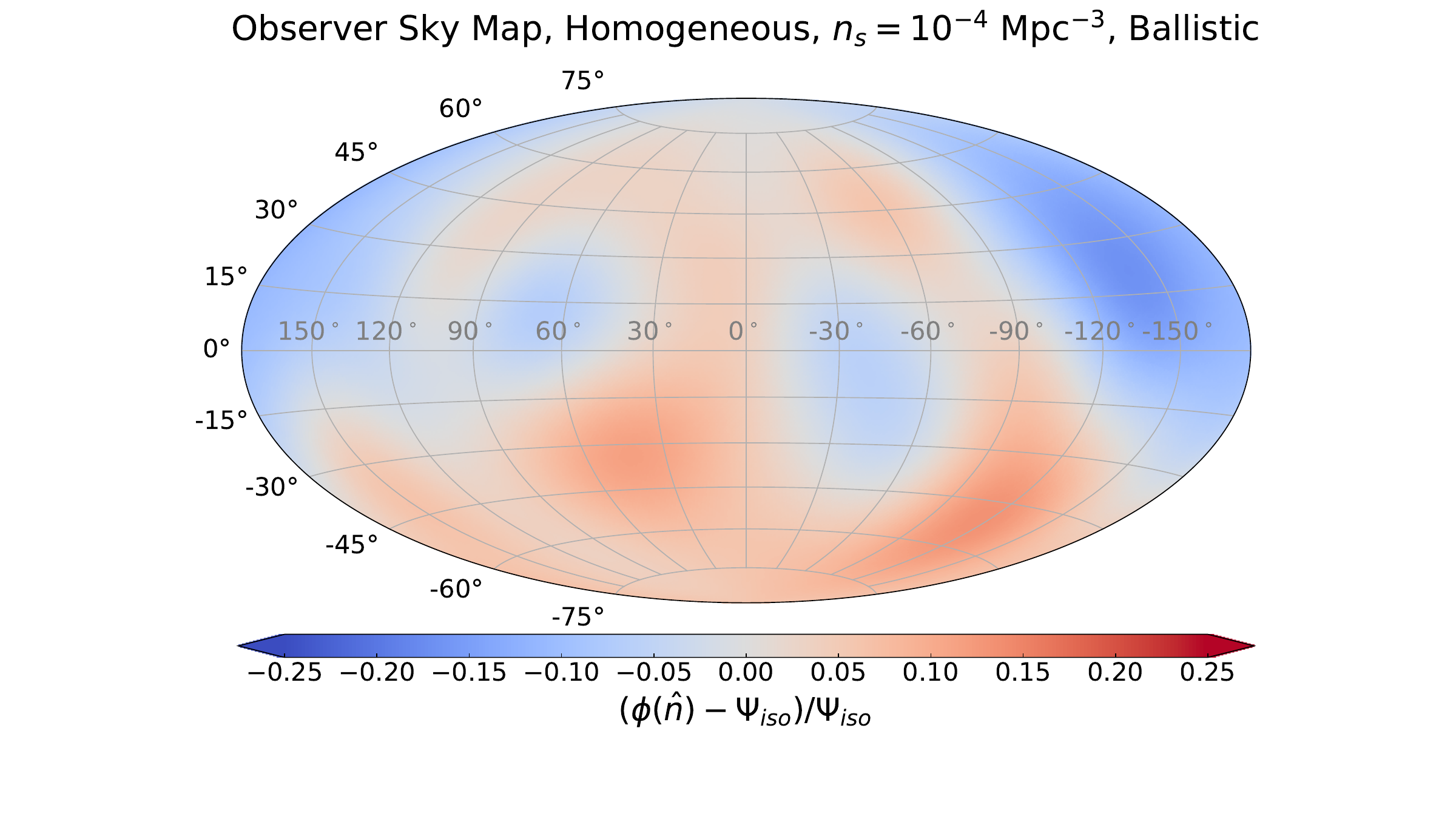}
\end{minipage}
\begin{minipage}{8.5cm}
\centering
\includegraphics[scale=0.2]{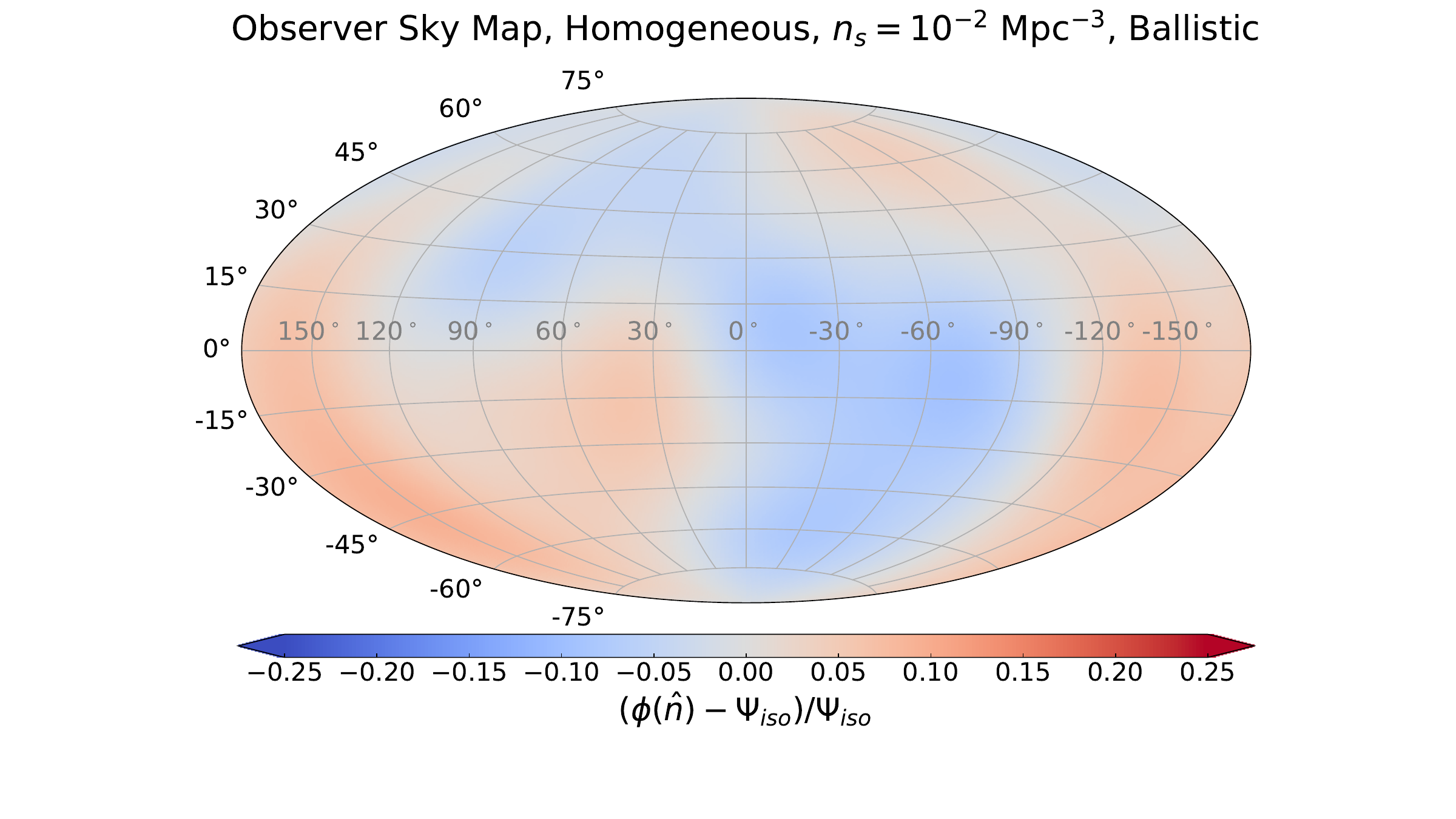}
\end{minipage}
\begin{minipage}{8.5cm}
\centering
\includegraphics[scale=0.2]{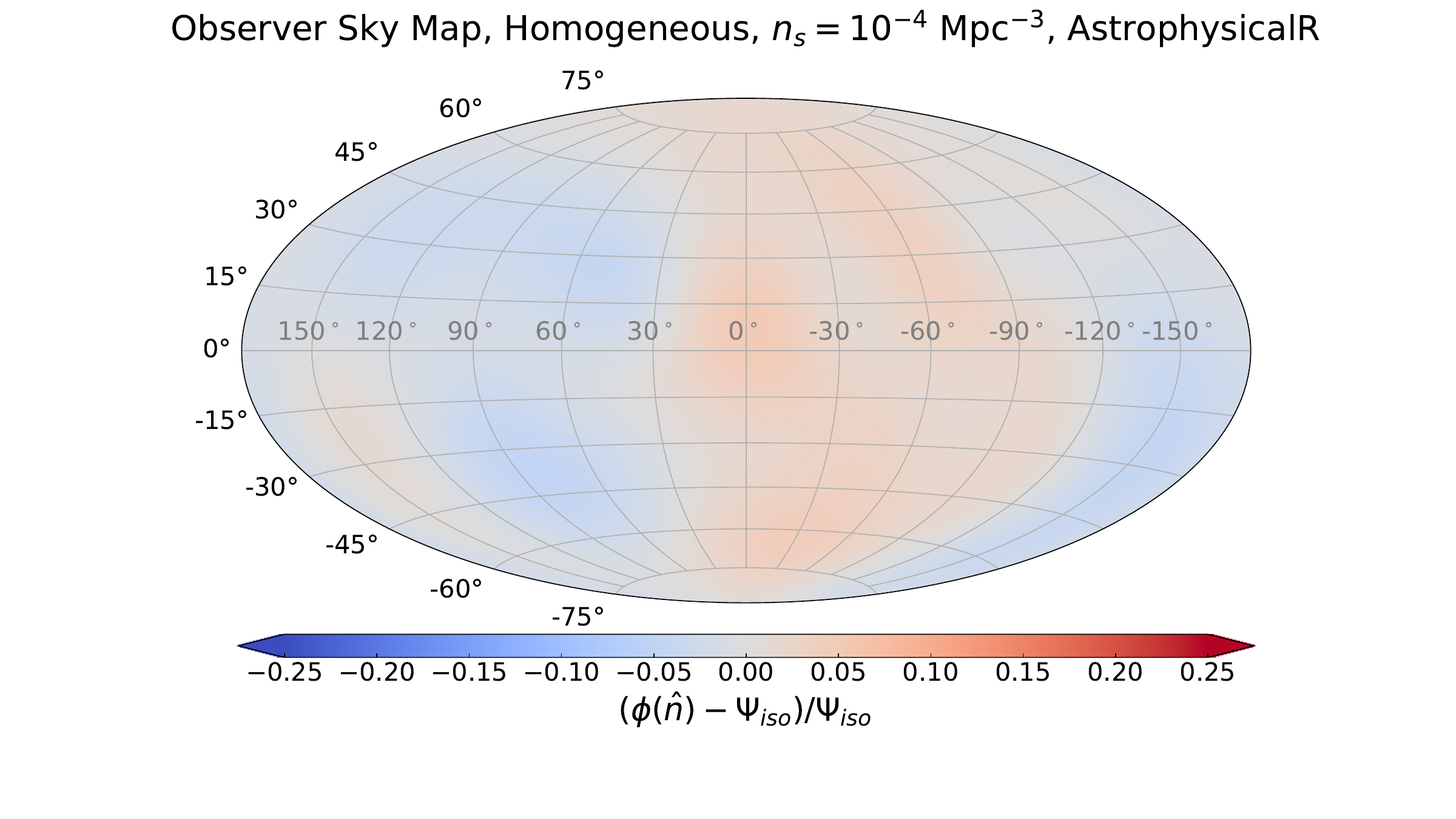}
\end{minipage}
\begin{minipage}{8.5cm}
\centering
\includegraphics[scale=0.2]{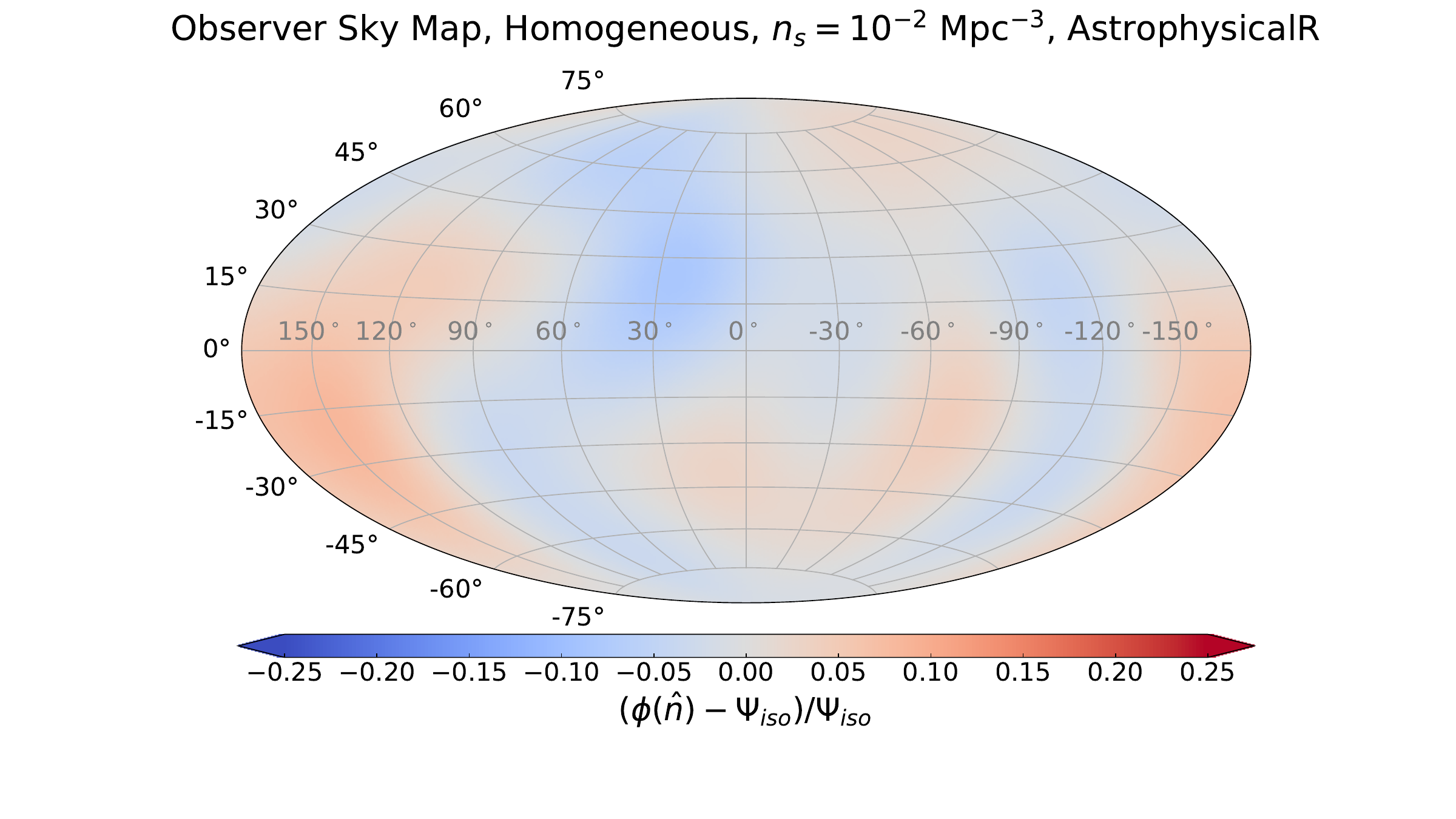}
\end{minipage}
\begin{minipage}{8.5cm}
\centering
\includegraphics[scale=0.2]{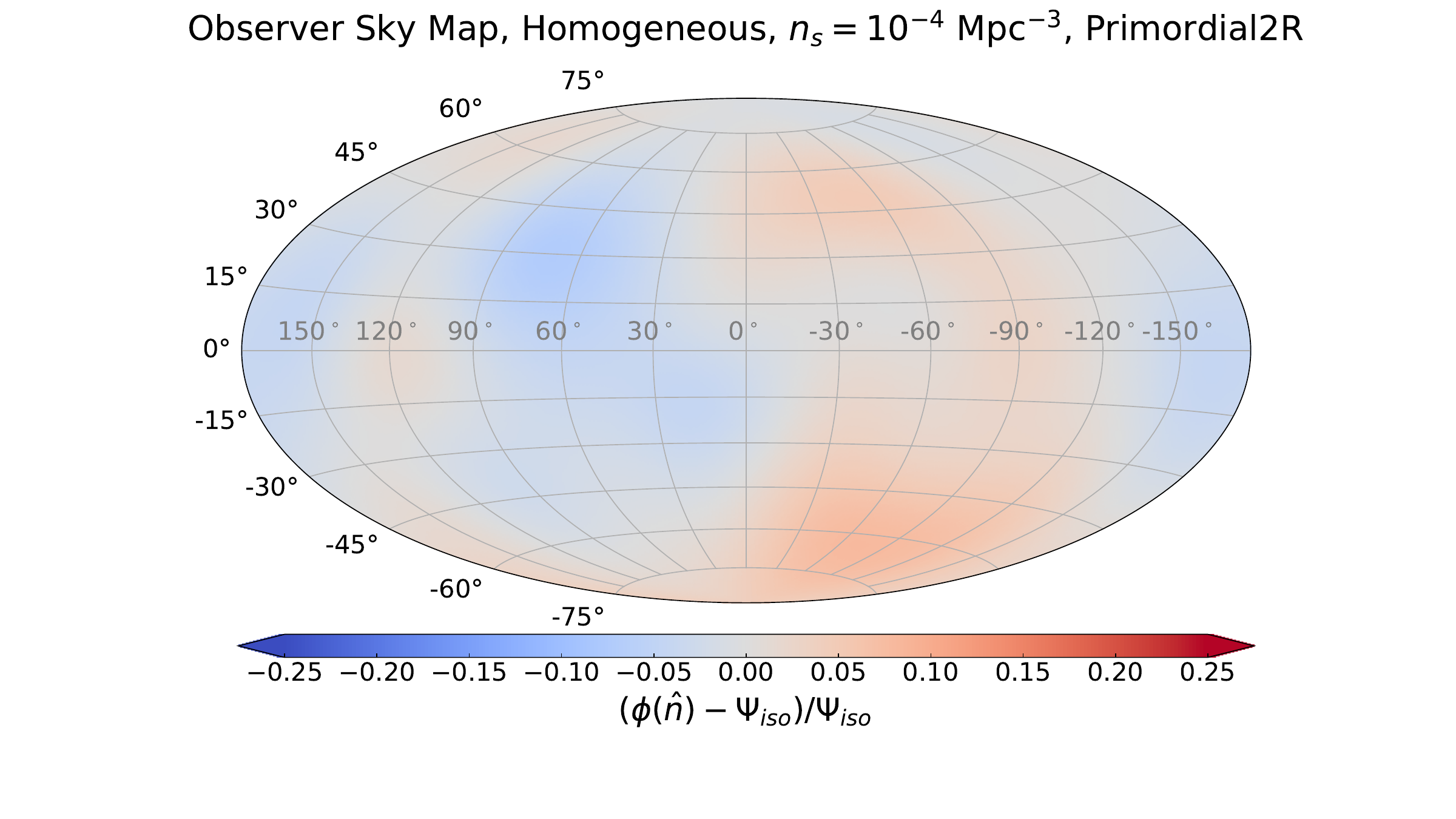}
\end{minipage}
\begin{minipage}{8.5cm}
\centering
\includegraphics[scale=0.2]{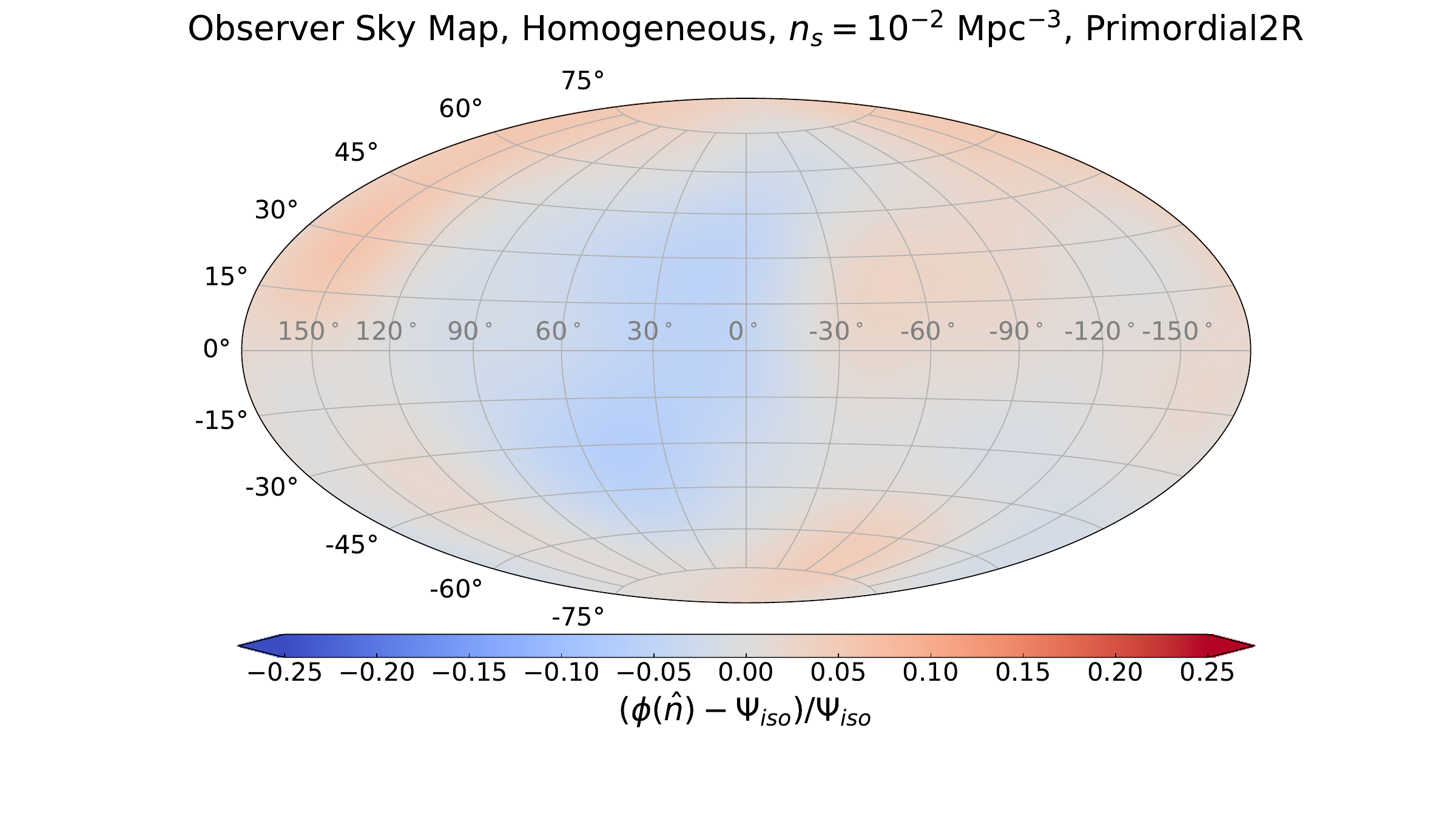}
\end{minipage}
\begin{minipage}{8.5cm}
\centering
\includegraphics[scale=0.2]{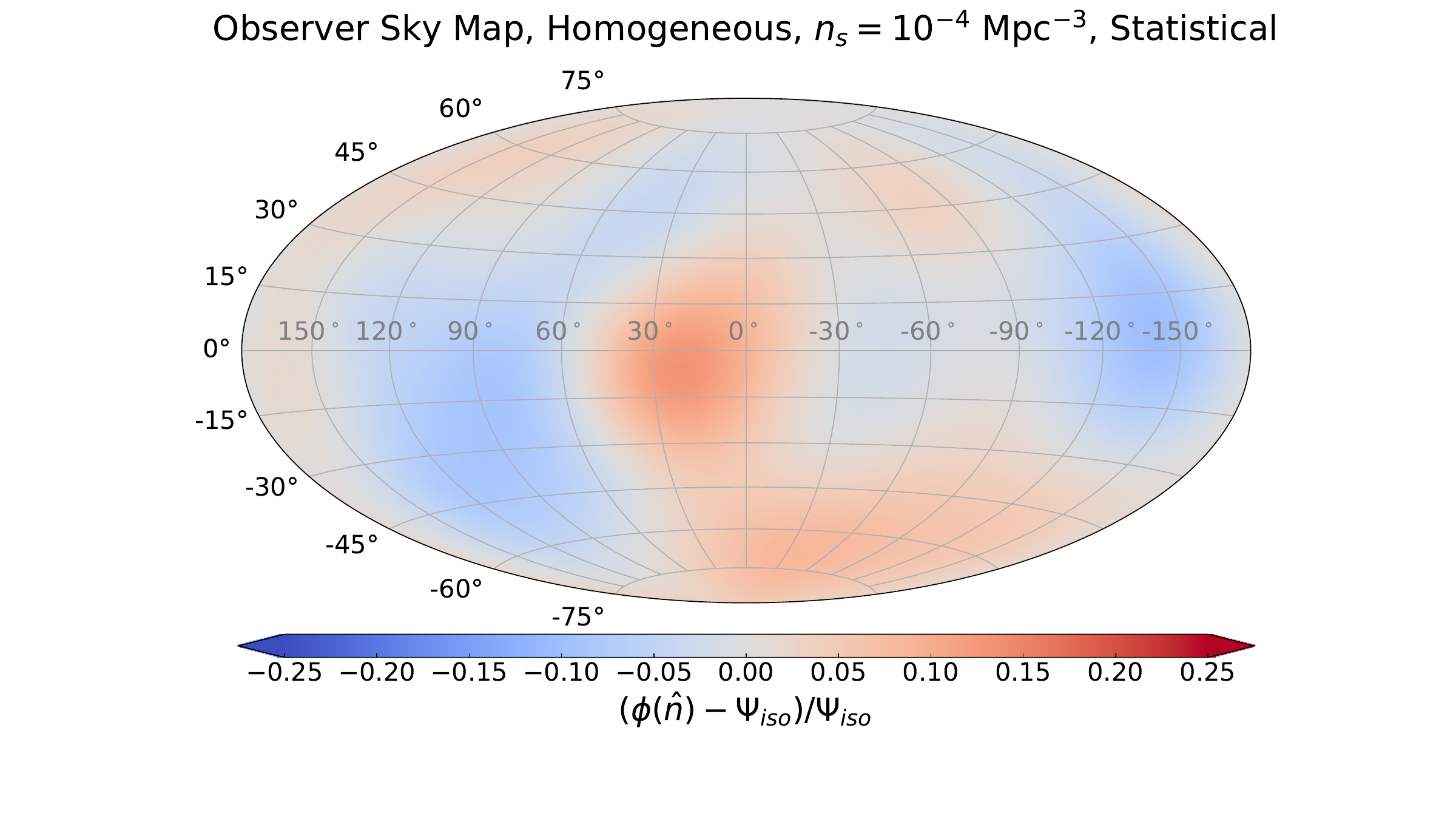}
\end{minipage}
\begin{minipage}{8.5cm}
\centering
\includegraphics[scale=0.2]{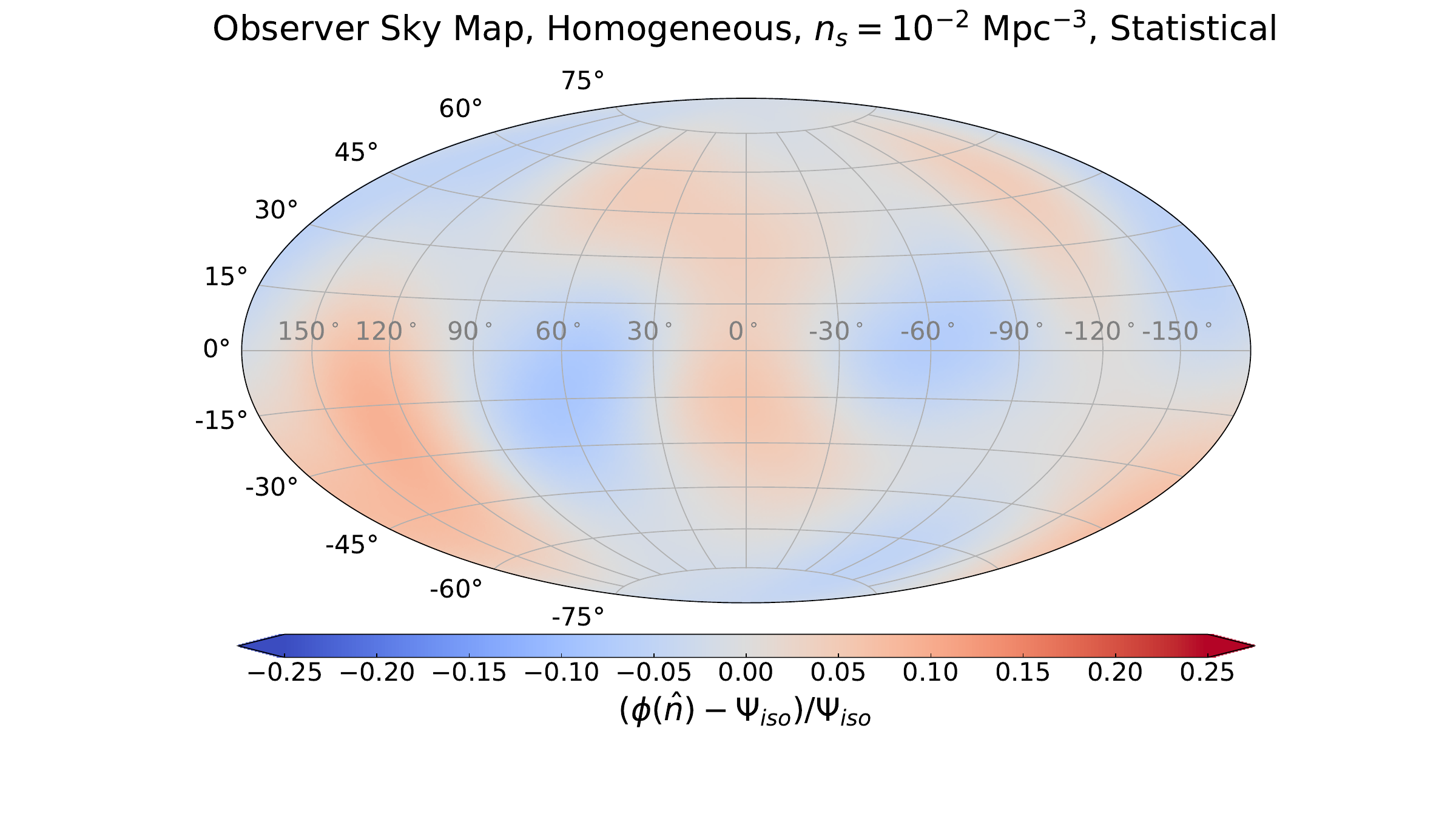}
\end{minipage}
\caption{Same as Fig.~\ref{sky_maps_den}, but for the \textit{homogeneous} scenarios.}
\label{sky_maps_homo}
\end{figure}

\begin{figure}[t]
\centering
\begin{minipage}{8.5cm}
\centering
\includegraphics[scale=0.2]{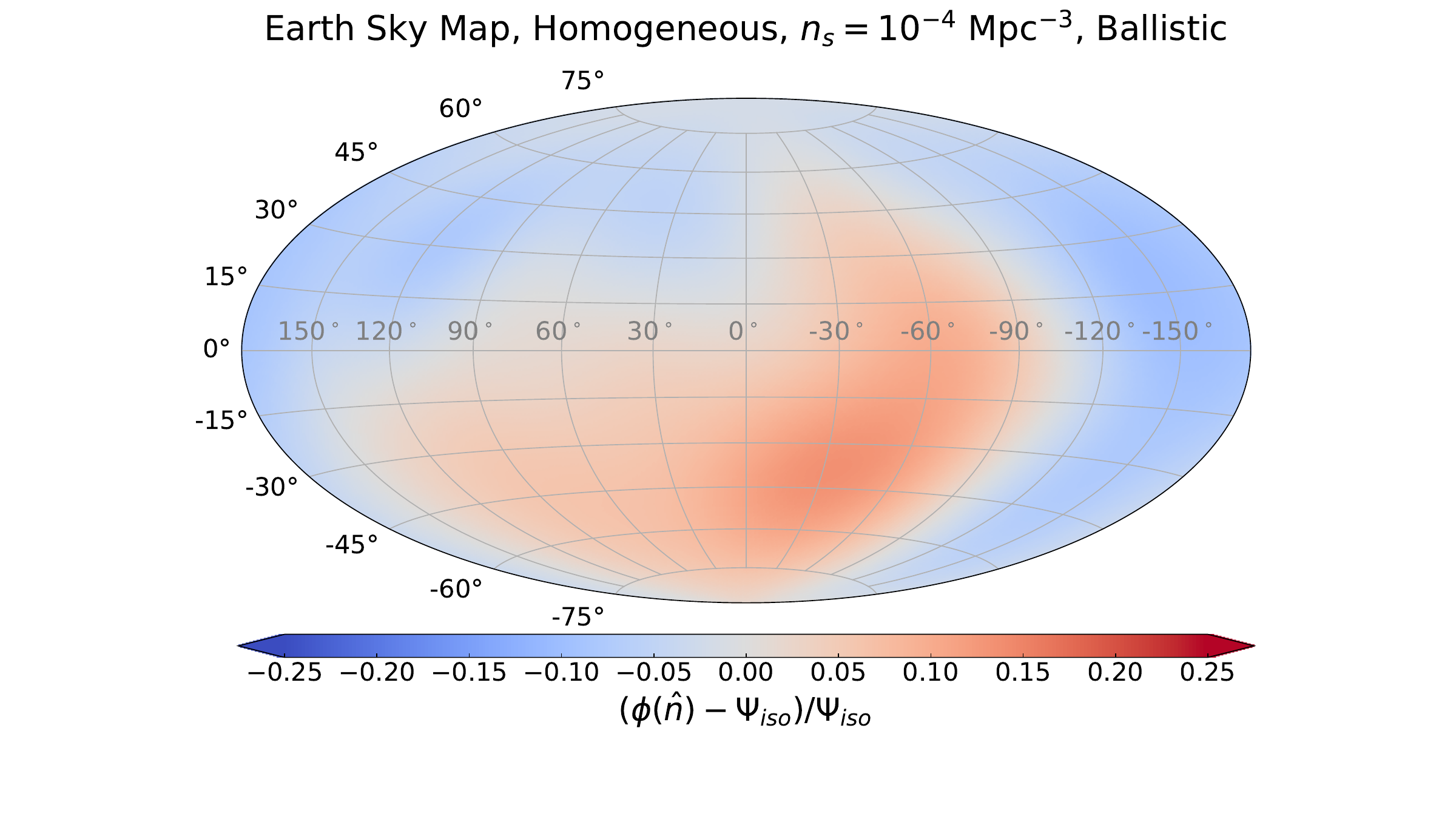}
\end{minipage}
\begin{minipage}{8.5cm}
\centering
\includegraphics[scale=0.2]{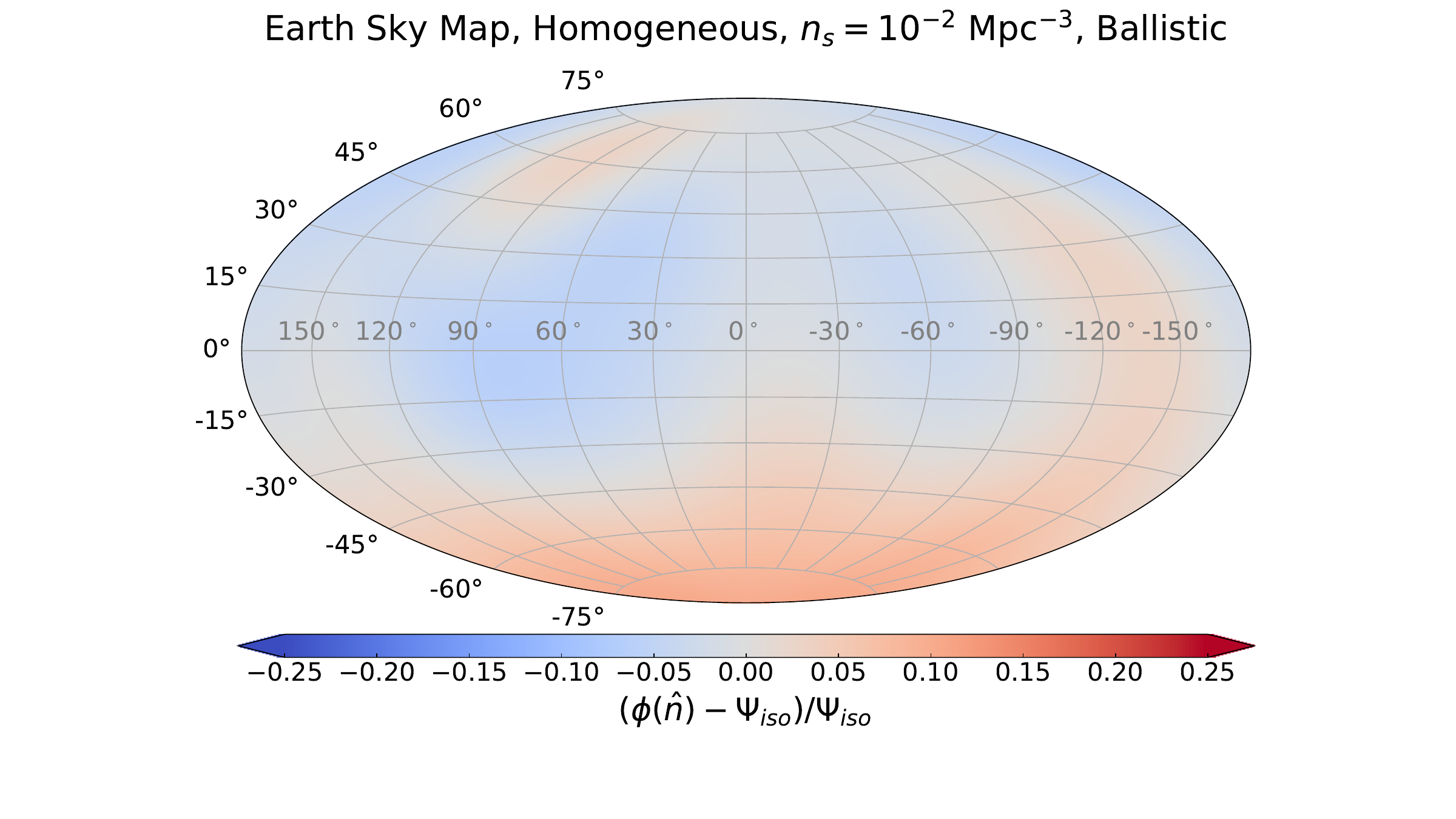}
\end{minipage}
\begin{minipage}{8.5cm}
\centering
\includegraphics[scale=0.2]{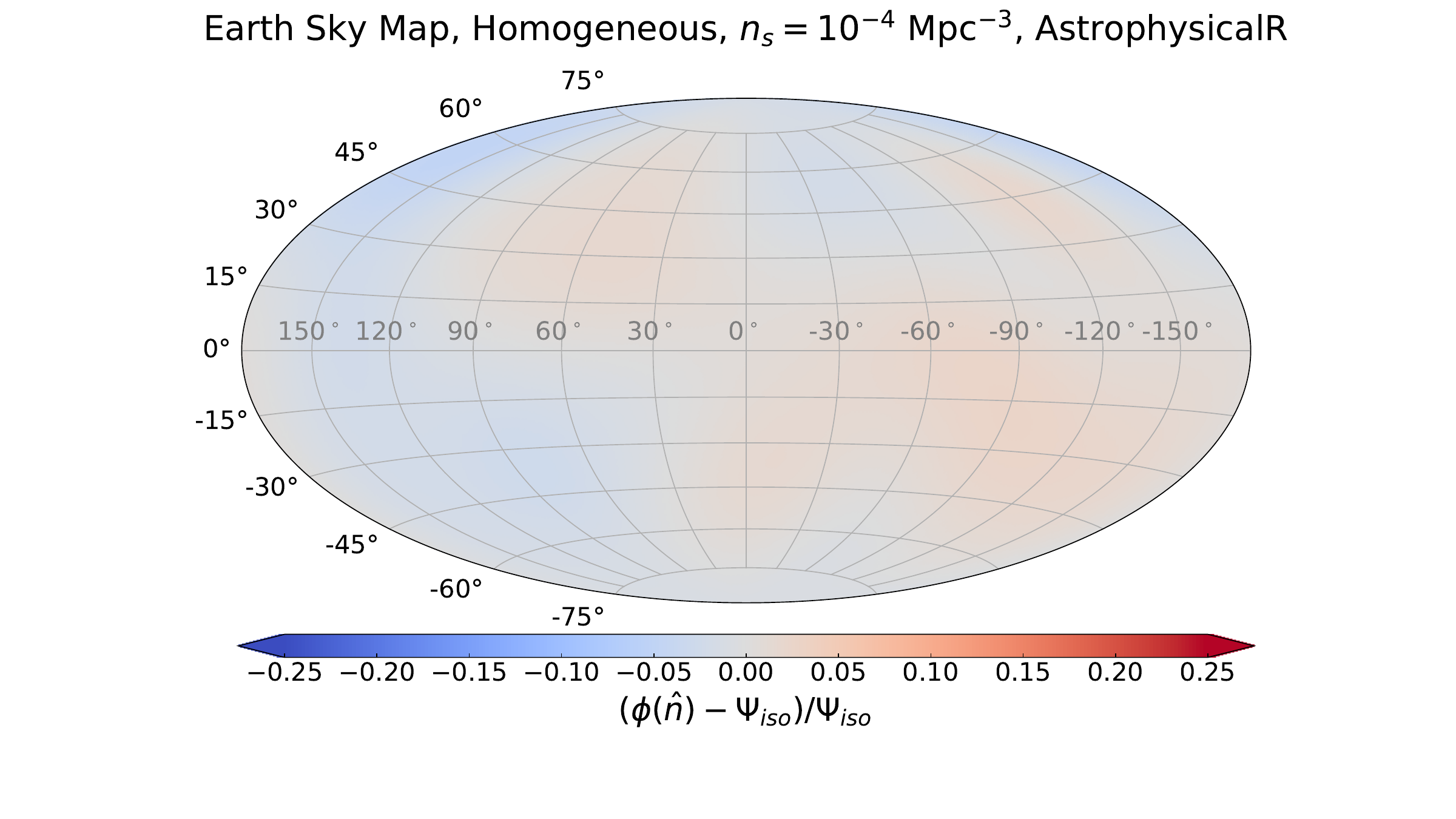}
\end{minipage}
\begin{minipage}{8.5cm}
\centering
\includegraphics[scale=0.2]{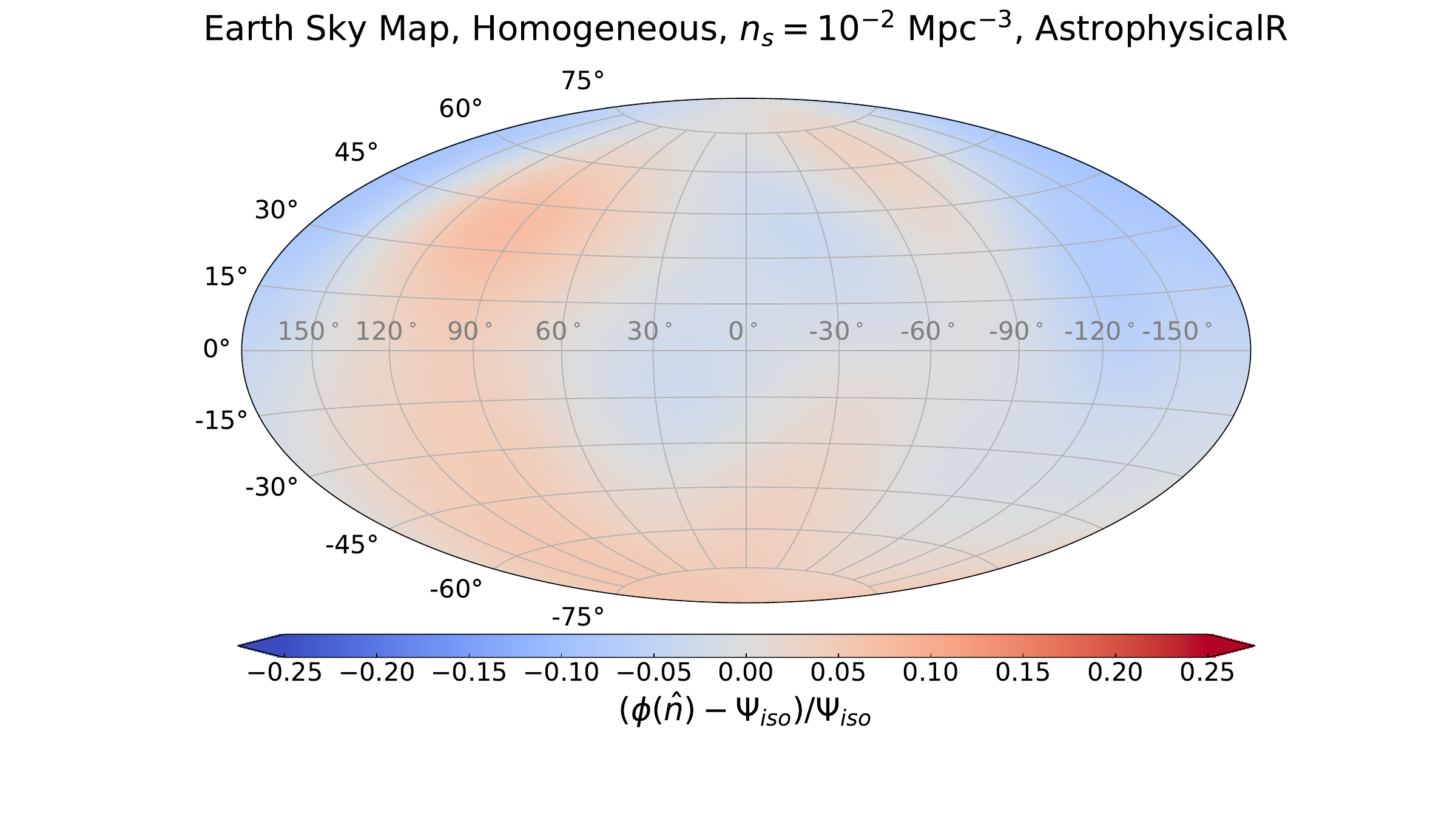}
\end{minipage}
\begin{minipage}{8.5cm}
\centering
\includegraphics[scale=0.2]{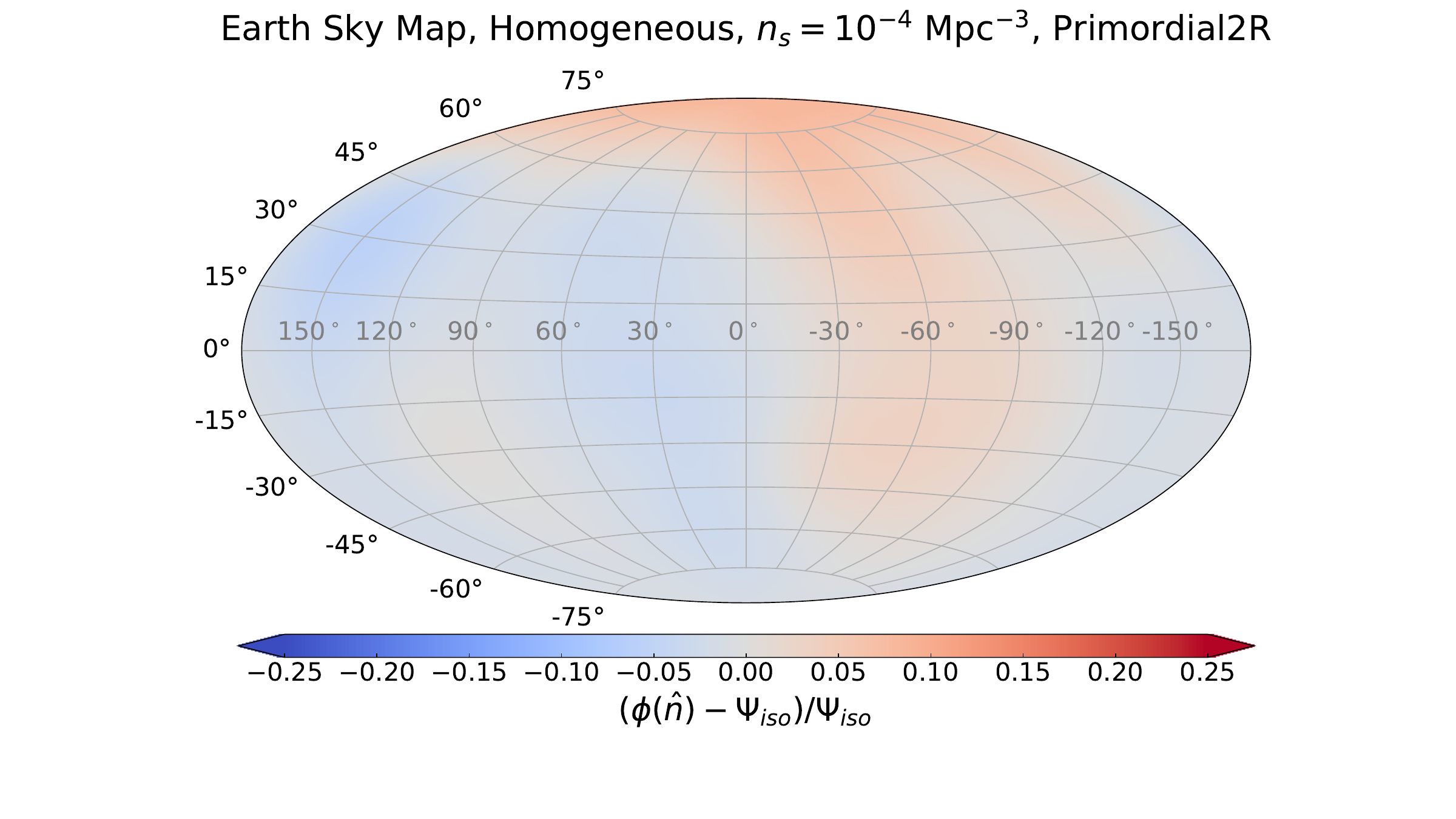}
\end{minipage}
\begin{minipage}{8.5cm}
\centering
\includegraphics[scale=0.2]{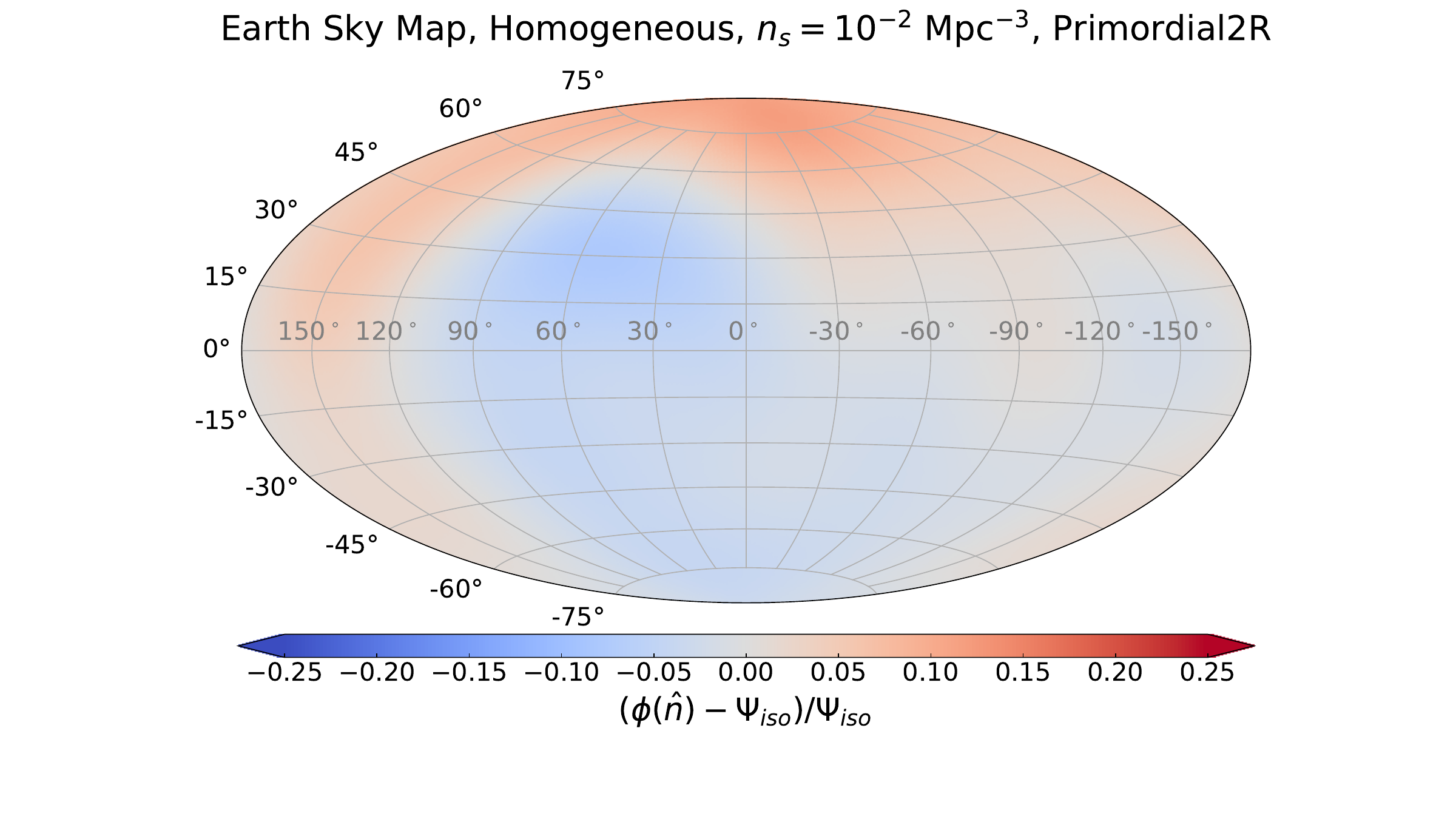}
\end{minipage}
\begin{minipage}{8.5cm}
\centering
\includegraphics[scale=0.2]{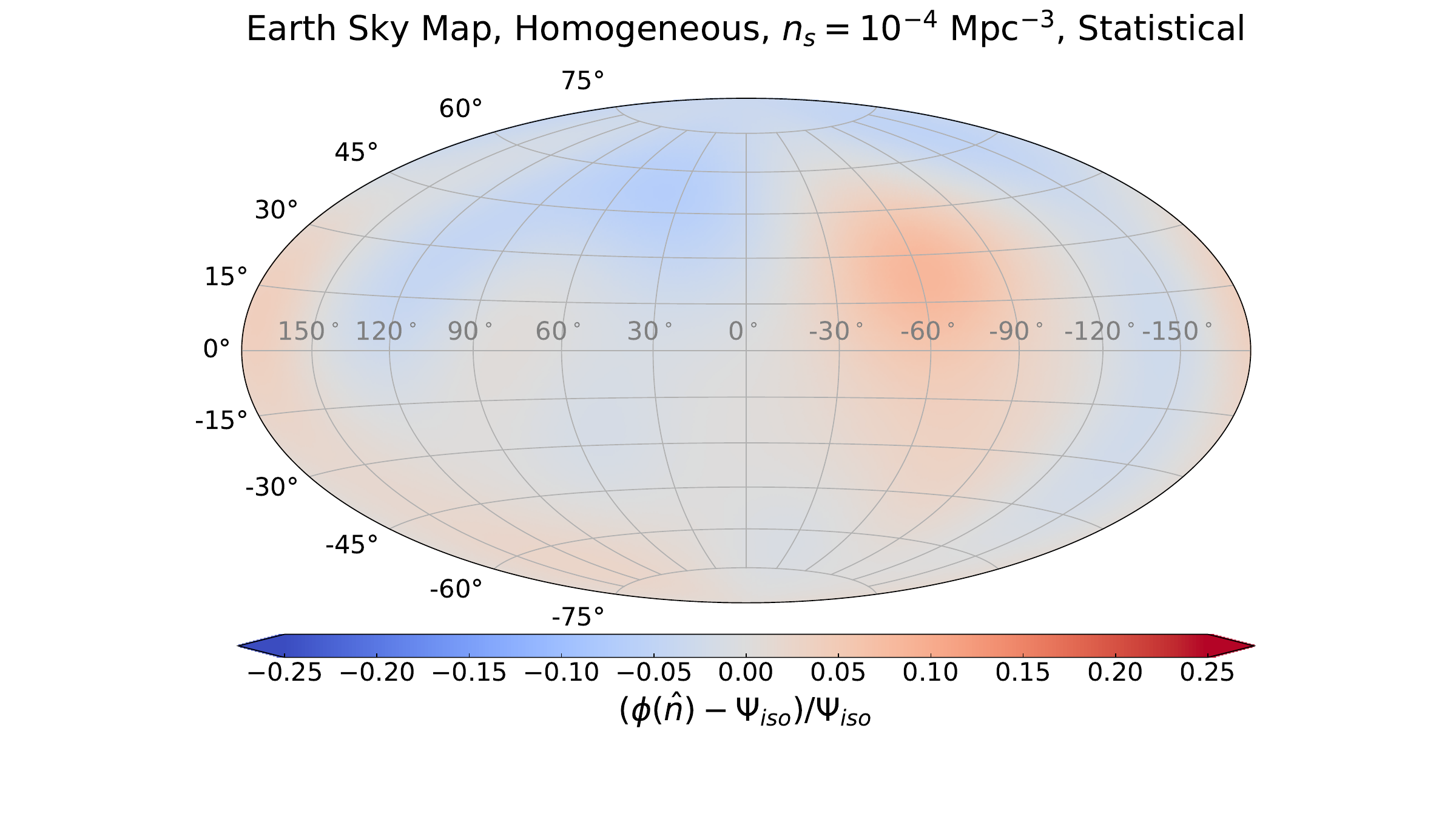}
\end{minipage}
\begin{minipage}{8.5cm}
\centering
\includegraphics[scale=0.2]{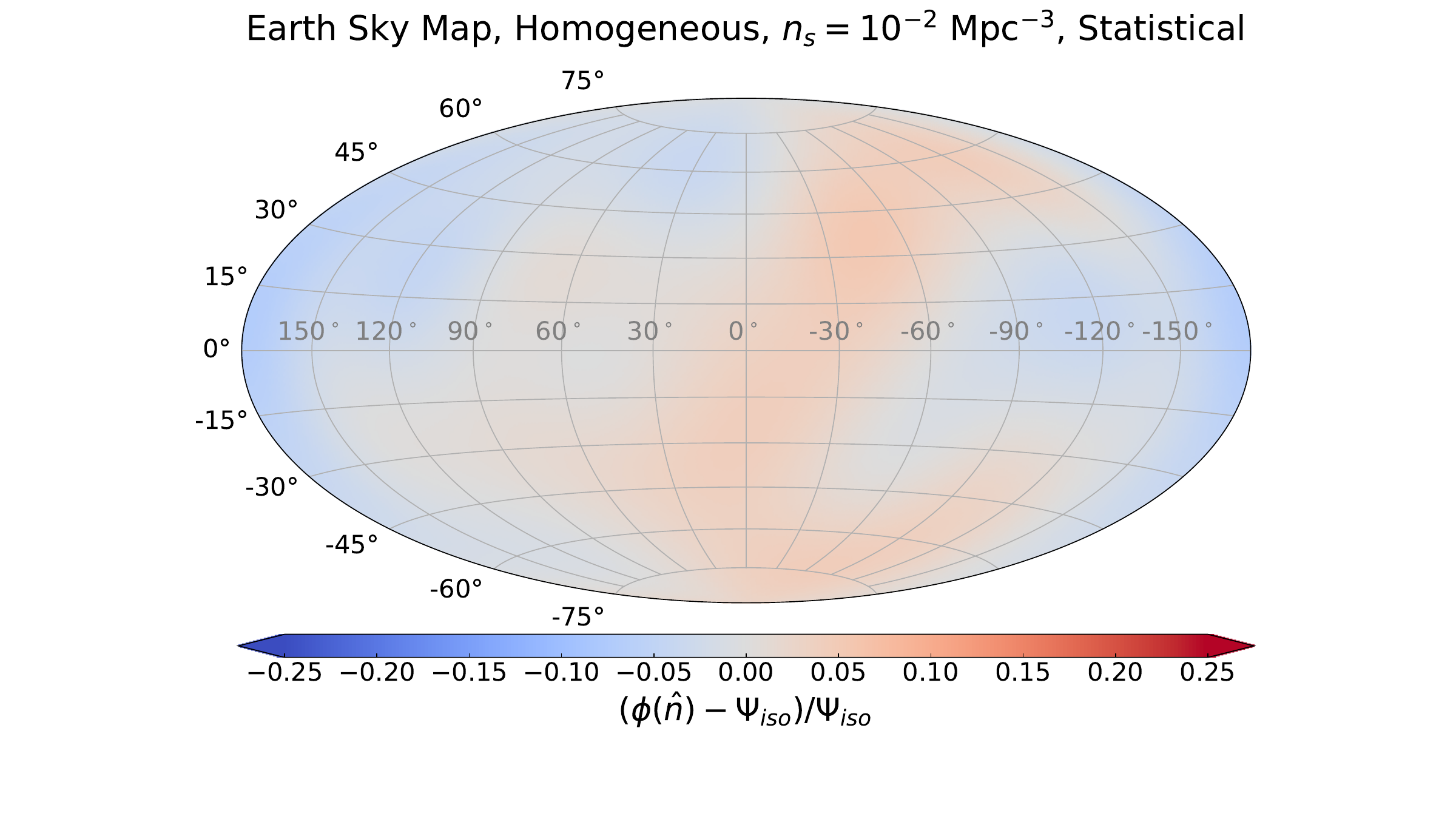}
\end{minipage}
\caption{Same as Fig.~\ref{sky_maps_den_lensed}, but for the \textit{homogeneous} scenarios.}
\label{sky_maps_homo_lensed}
\end{figure}

\begin{figure}[t]
\centering
\begin{minipage}{8.5cm}
\centering
\includegraphics[scale=0.5]{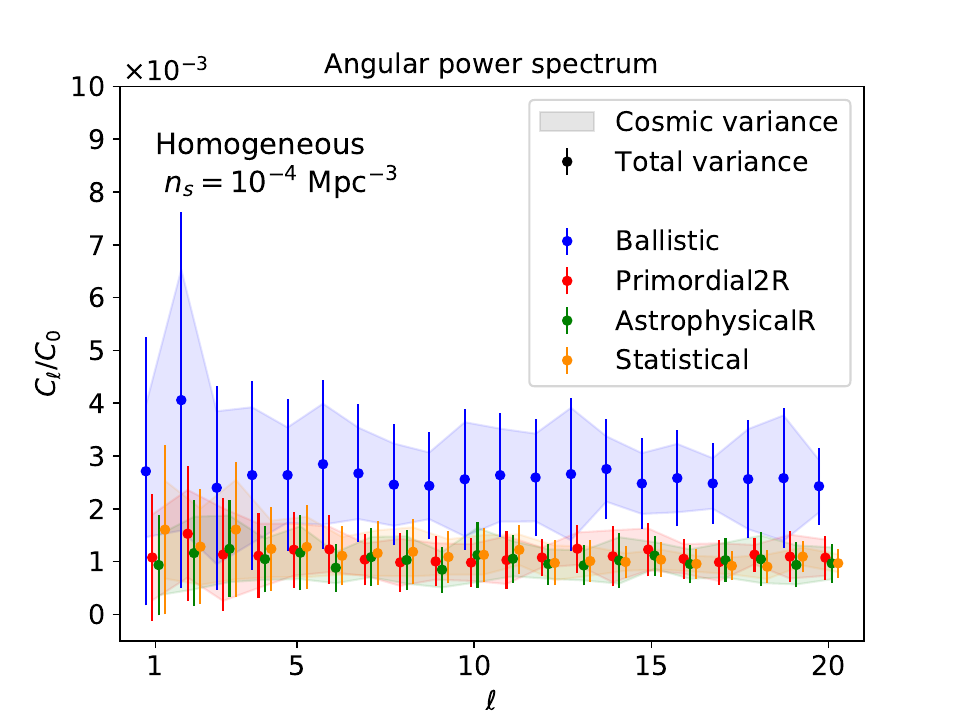}
\end{minipage}
\begin{minipage}{8.5cm}
\centering
\includegraphics[scale=0.5]{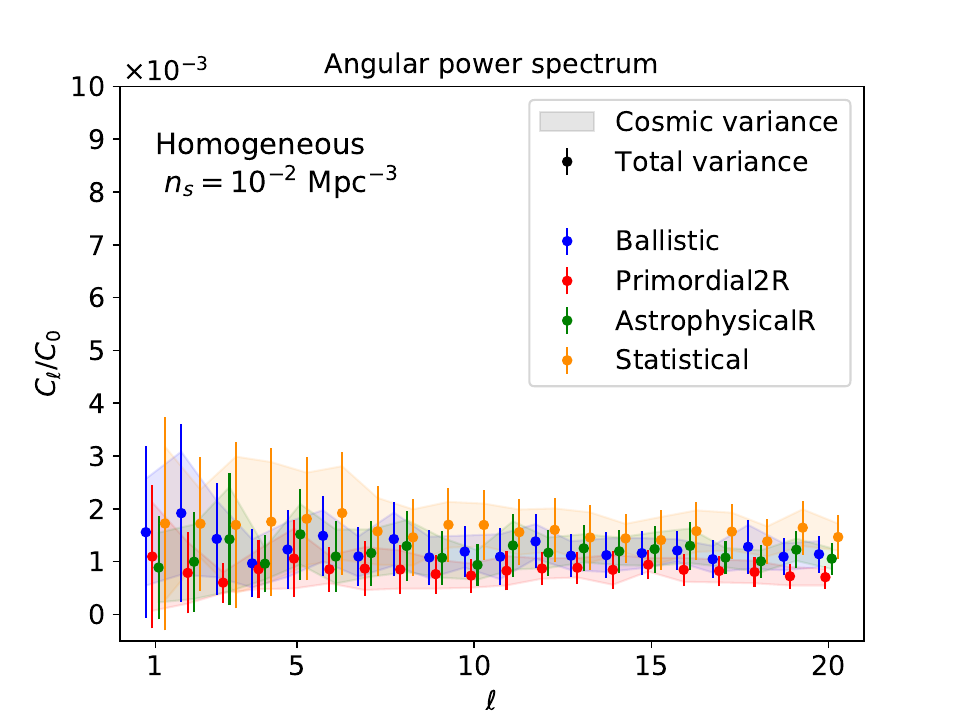}
\end{minipage}
\caption{Same as Fig.~\ref{angular_power_spectrum_den}, but for the \textit{homogeneous} scenarios.}
\label{angular_power_spectrum_homo}
\end{figure}

\begin{figure}[t]
\centering
\begin{minipage}{8.5cm}
\centering
\includegraphics[scale=0.5]{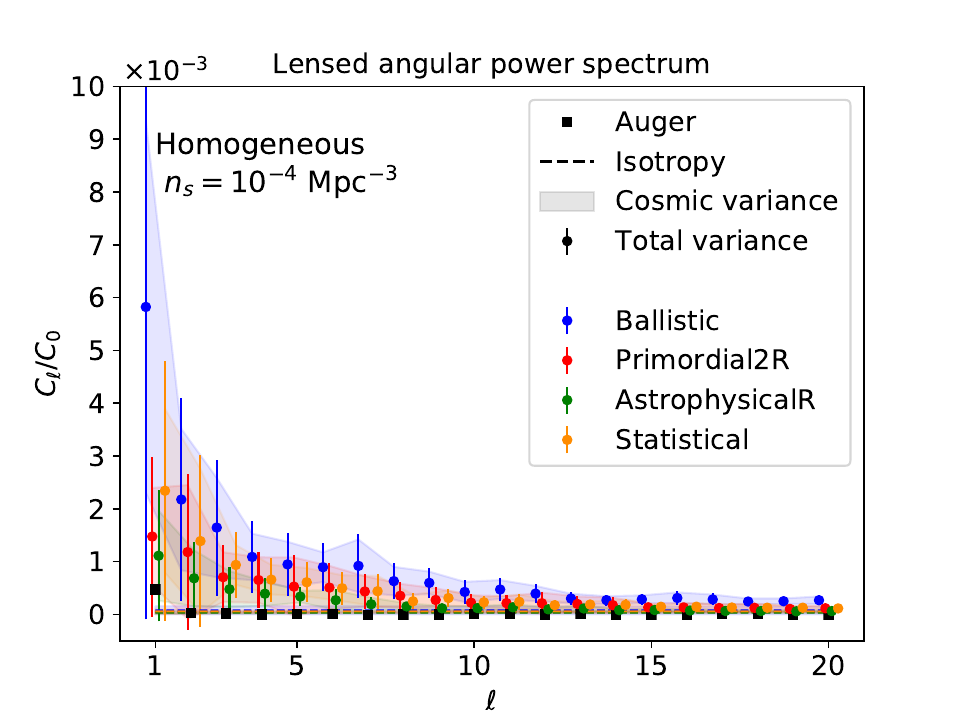}
\end{minipage}
\begin{minipage}{8.5cm}
\centering
\includegraphics[scale=0.5]{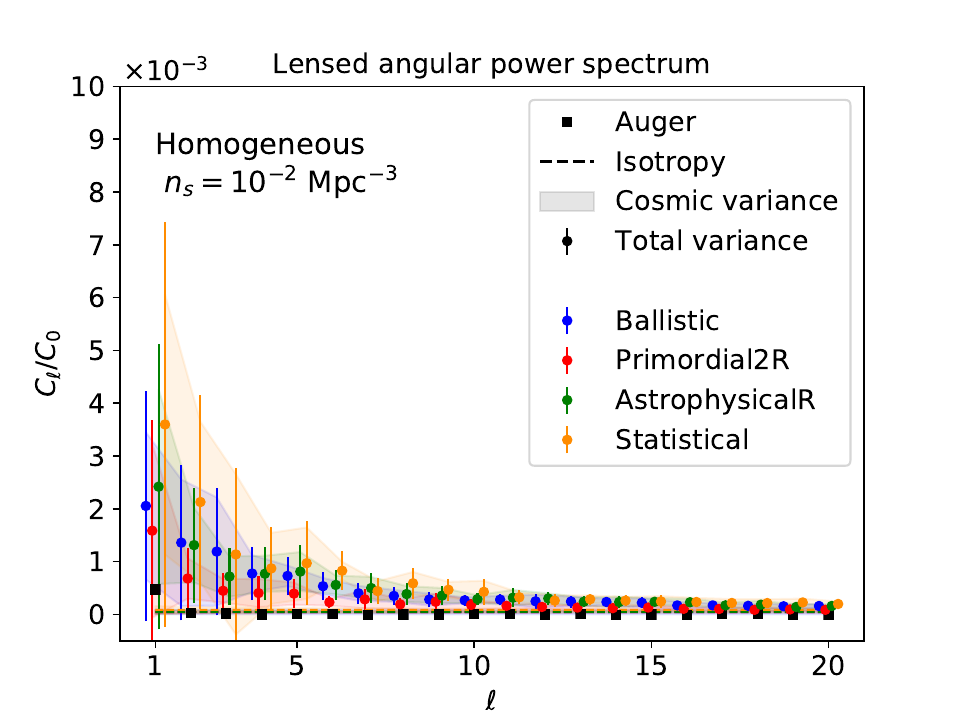}
\end{minipage}
\caption{Same as Fig.~\ref{angular_power_spectrum_den_lensed}, but for the \textit{homogeneous} scenarios.}
\label{angular_power_spectrum_homo_lensed}
\end{figure}

This appendix presents the results obtained for the \textit{homogeneous} source distribution scenarios following the same analysis discussed in Sec.~\ref{subsec_analysis}. The minimization results of Eq.~\eqref{chi_sq_def} for the \textit{homogeneous} scenarios are given in Table~\ref{table_homo} for both the source number densities considered. We find the same general results as for the \textit{mass density} scenarios: a hard spectral index up to $\gamma\simeq-3$ and a rigidity cutoff of $\log (R_\text{cut}/1\,\text{EV})\simeq0.2$ on average. The mass composition is again dominated by the nitrogen and silicon components. The propagated energy spectra and mass compositions are shown in Fig.~\ref{spectrum_composition_homo}: we do not observe any interesting effect due to the EGMF model considered. Fig.~\ref{deflection_homo} shows the distributions of the angular deflection of the propagated UHECRs from the injection position to the edge of the galaxy. We can see that the relative position of the sources with respect to the EGMF has no effect on the average extragalactic deflection pattern. Similar results are obtained for the flux ratios, shown in Fig.~\ref{number_sources_homo}. However, we observe an expected smaller cosmic variance in the case of the \textit{homogeneous} catalogues. Compared to the case of inhomogeneous source distributions, the sky distributions shown in Figs.~\ref{sky_maps_homo} and~\ref{sky_maps_homo_lensed} are more isotropic both at the edge of the galaxy and at Earth. We do not observe dominant hotspots of events in the sky. This can also be seen in the multipole distribution in Fig.~\ref{angular_power_spectrum_homo}, where no significant large dipole or quadrupole values are observed in any scenario. We observe that the angular power spectrum obtained in the low density scenario, without any EGMF, is larger than in the magnetised cases. This is due to the ballistic propagation of the UHECRs emitted by a small number of local sources. The effect of the GMF on the angular power spectra can be seen in Fig.~\ref{angular_power_spectrum_homo_lensed}, where we observe the same suppression as in the \textit{mass density} scenarios. However, the lensed dipole and quadrupole components are still small when \textit{homogeneous} source catalogues are used. We note that we still observe an anisotropy signal due to the cosmic variance of the source catalogues. In general, lower source densities still correspond to stronger anisotropy signals.

\clearpage
\twocolumngrid

\bibliography{apssamp}
\end{document}